\documentclass[11pt, a4paper]{article}
\usepackage[top=2.54cm,bottom=2.54cm,right=2.54cm,left=2.54cm]{geometry}
\usepackage[utf8]{inputenc}
\usepackage{amsmath}
\usepackage{dsfont}
\usepackage{amssymb}
\usepackage{booktabs}
\usepackage{epstopdf}
\usepackage{graphicx}
\usepackage[center]{caption}
\usepackage{subcaption}
\usepackage{verbatim}
\usepackage{bbm}
\usepackage{xcolor}
\usepackage{enumitem}
\usepackage[title]{appendix}
\usepackage{multirow}
\usepackage{xr-hyper}
\usepackage{hyperref}
\usepackage{apacite}

\renewcommand{\P}{\mathbb{P}}
\newcommand{\Q}{\mathbb{Q}}
\newcommand{\E}{\mathbb{E}}
\newcommand{\R}{\mathbb{R}}
\newcommand{\F}{\mathcal{F}}
\newcommand{\diff}{\mathrm{d}}
\newcommand{\dt}{\mathrm{d}t}
\newcommand{\ds}{\mathrm{d}s}

\newtheorem{proposition}{Proposition}
\newtheorem{assumption}{Assumption}
\newtheorem{lemma}{Lemma}

\title{ \vskip -1.0cm 
    Estimating Option Pricing Models Using a Characteristic Function-Based Linear State Space Representation\thanks{\scriptsize We are very grateful to 
Torben Andersen,
Kris Jacobs,
Frank Kleibergen,
Siem Jan Koopman, 
Olivier Scaillet,
George Tauchen,
Viktor Todorov,
Fabio Trojani,
and conference and seminar participants at 
the 2021 SoFiE Financial Econometrics Summer School at Northwestern University,
the 2022 Quantitative Finance and Financial Econometrics (QFFE) Conference at Aix-Marseille University,
the 2022 Annual SoFiE Conference at the University of Cambridge,
the 2022 Dynstoch meeting at the Institut Henri Poincar\'e in Paris,
the 74th European Meeting of the Econometric Society (ESEM) at Bocconi University in Milan,
the University of Amsterdam,
Kellogg School of Management at Northwestern University,
the Center for Econometrics and Business Analytics at St.~Petersburg State University,
and the Tinbergen Institute
for helpful comments and suggestions. 
{\tt Julia} code to implement the estimation procedure developed in this paper is available from \url{https://github.com/evladimirov/OptionModels-cKF-ccf}.
This research was funded in part by the Netherlands Organization for Scientific Research (NWO) under grant NWO-Vici 2019/2020 (Laeven).
Email addresses: 
\href{mailto:H.P.Boswijk@uva.nl}{\tt H.P.Boswijk@uva.nl}, 
\href{mailto:R.J.A.Laeven@uva.nl}{\tt R.J.A.Laeven@uva.nl}, and 
\href{mailto:E.Vladimirov@uva.nl}{\tt E.Vladimirov@uva.nl}.}
}

\author{
    H.~Peter Boswijk\\ \small{Amsterdam School of Economics}\\ \small{University of Amsterdam}\\ \small{and Tinbergen Institute} 
    \and Roger J.~A. Laeven\\ \small{Amsterdam School of Economics}\\ \small{University of Amsterdam, EURANDOM}\\ \small{and CentER}\\
    \and Evgenii Vladimirov\\ \small{Amsterdam School of Economics}\\ \small{University of Amsterdam}\\  \small{and Tinbergen Institute}\\
    \medskip
    }
    
\date{\vskip -0.4cm \today }

\begin{document}

\maketitle

\begin{abstract}
    We develop a novel filtering and estimation procedure for parametric option pricing models driven by general affine jump-diffusions. 
    Our procedure is based on the comparison between an option-implied, model-free representation of the conditional log-characteristic function and the model-implied conditional log-characteristic function, which is functionally affine in the model's state vector. 
    We formally derive an associated linear state space representation and establish the asymptotic properties of the corresponding measurement errors.  
    The state space representation allows us to use a suitably modified Kalman filtering technique to learn about the latent state vector and a quasi-maximum likelihood estimator of the model parameters, which brings important computational advantages. 
    We analyze the finite-sample behavior of our procedure in Monte Carlo simulations. 
    The applicability of our procedure is illustrated in two case studies that analyze S\&P~500 option prices and the impact of exogenous state variables capturing Covid-19 reproduction and economic policy uncertainty.

\end{abstract}

\noindent {\small \textit{Keywords:} Options; Characteristic Function; Affine Jump-Diffusion; State Space Representation.}\\
\noindent {\small \textit{JEL Classification:} Primary: C13; C58; G13; Secondary: C32; G01.}

\section{Introduction}
\label{sec:Introduction}

Over the past decades, explosive growth in the trading of option contracts has attracted the attention of academics and practitioners to the development and estimation of increasingly sophisticated option pricing models. 
The building blocks of many continuous-time option pricing models are semimartingale stochastic processes that govern the dynamics of the underlying asset. 
These processes are often latent with stochastic diffusive volatility as the prototypical example, as in the classical \citeA{heston1993} model. 
The literature also suggests the need to allow for a discontinuous jump component, both in the asset price dynamics and in its volatility process, potentially with a time-varying stochastic jump intensity. 

An important econometric challenge lies in estimating the parameters of these continuous-time models and in filtering their unobserved and time-varying components, since option prices are highly nonlinear functions of the state vector. 
This stands in contrast to, for instance, term structure models, where bond yields can be represented as a linear function of the states, at least within the affine framework (see, e.g., \citeNP{piazzesi2010affine}, for a review of the affine term structure literature). 
To evaluate option prices as a function of the state vector, one typically needs to apply either Fourier-based methods or simulation-based approaches, in both cases at a substantial computational cost. 
This is one of the reasons why in much of the empirical research on option pricing, only a subset of the available option price data is used, such as at-the-money contracts or weekly (typically Wednesday) options data.

In this paper, we develop a new latent state filtering and parameter estimation procedure for option pricing models governed by general affine jump-diffusion processes. 
Our procedure leverages the linear relationship between the logarithm of an option-implied, model-free spanning formula for the conditional characteristic function of the underlying asset return on the one hand, and the state vector induced by parametric model specification on the other hand. 
From this relationship, we formally derive a linear state space representation, and establish the asymptotic properties of the corresponding measurement errors.
Linearity of the measurement and state updating equations that make up the state space representation, with coefficient and variance matrices that are (semi-)closed-form functions of the parameters, allows us to exploit Kalman filtering techniques. 
The proposed estimation procedure is fast and easy to implement, circumventing the typical computational burden in conducting inference on option pricing models.

Exploiting the option-spanning formula of \citeA{carr2001optimal} for European-style payoff functions, we replicate the risk-neutral conditional characteristic function (CCF) of the underlying log-asset price at the expiration date in a completely model-independent way. 
In other words, we imply information about the CCF from the option prices without imposing any parametric assumptions on the underlying asset price dynamics. 
A similar option-spanning approach for the CCF is used by \citeA{todorov2019nonparametric} to develop an option-based nonparametric spot volatility estimator. 
On the other hand, a large stream of literature is devoted to parametric option pricing models belonging to the general affine jump-diffusion (AJD) family; canonical examples are \citeA{heston1993}, \citeA{DPS2000}, \citeA{pan2002}, and \citeA{bates2006maximum}.\footnote{See also, e.g., \citeA{broadie2007model}, \citeA{ait2015contagion},  \citeA{AFT2017} and the references therein.} 
The defining property of the AJD class is the exponential-affine joint CCF, which is available in semi-closed form. 
By comparing the two option pricing representations---model-free and model-implied---we can obtain a linear relation between the logarithm of the option-implied CCF and the model-dependent CCF within the affine framework.

The state vector in AJD option pricing models typically contains both observable processes and latent factors. 
We address the filtering of such latent factors by developing a linear state space representation for this model class. 
The development includes an asymptotic analysis of the measurement error components, consisting of observation, truncation and discretization errors, under a double asymptotic scheme in the moneyness dimension.
The state space representation allows us to employ suitably modified Kalman filtering techniques to learn about the unobserved intrinsic components of the model and estimate the model parameters using quasi-maximum likelihood (QML). 
QML approaches based on Kalman filtering are often used in the affine term structure literature, where the yields themselves are linear functions of the state vector (see, e.g., \citeNP{duffee1999estimating},  \citeNP{jong2000time}, \citeNP{driessen2005default}). 
Besides the possibility to exploit Kalman filtering and QML estimation techniques, another advantage of our approach is that, once the model-free CCF has been obtained from the data, no further numerical option pricing methods, such as the FFT approach of \citeA{CM1999} or simulation-based methods, are needed. 
Therefore, our method reduces computational costs considerably relative to many existing approaches in the option pricing literature. 
We note that, whereas the parametric CCF is used to price options in Fourier-based methods, here we use the CCF to directly learn about the latent factors and model parameters. 

We analyze the developed estimation procedure in Monte Carlo simulations based on several AJD specifications.
We consider a one-factor AJD option pricing model, with the stochastic volatility and jump intensity both being affine functions of a single latent process, and a two-factor AJD model specification with an observable exogenous factor. 
We find good finite-sample performance in both cases, notwithstanding the challenging nature of the econometric problem.

Finally, we illustrate our new filtering and estimation approach in an empirical application to S\&P~500 index options. 
In particular, we filter and estimate the latent volatility and jump intensity from a stochastic volatility model with co-jumps in returns and volatility. 
We also investigate the impact of the Covid-19 propagation rate on the stock market within this model, by embedding the associated reproduction number as an exogenous factor into the volatility and jump intensity dynamics. 
Our results show that while the reproduction number has only a mild effect on total diffusive volatility, it contributes substantially to the likelihood of jumps.
By contrast, when we consider an Economic Policy Uncertainty index as exogenous factor, the jump intensity process is not affected, but the exogenous factor contributes significantly to diffusive volatility.

Various estimation and filtering strategies for option pricing models have been developed in the literature. 
These include the (penalized) nonlinear least squares methods in, for instance, \citeA{bakshi1997empirical}, \citeA{broadie2007model}, \citeA{andersen2015parametric}; the efficient method of moments of \citeA{gallant1996moments} as applied in \citeA{chernov2000study} and \citeA{andersen2002empirical}; the implied-state methods initiated by \citeA{pan2002} and further analyzed by \citeA{santa2010crashes};  
the Markov Chain Monte Carlo method in \citeA{eraker2004stock} and \citeA{eraker2003impact}; and the particle filtering method, see \citeA{johannes2009optimal} and \citeA{bardgett2019inferring}. 
Most of these estimation methods use as inputs option prices or a monotonic transformation thereof, such as implied volatilities. 
By contrast, we propose an estimation procedure based on the prices of spanning option portfolios that by the bijection between CCFs and conditional distributions, in principle, contain \textit{all} probabilistic information about the stochastic process governing the dynamics of the underlying asset.

In general, estimation strategies based on the transform space of conditional characteristic functions are, of course, not new to the literature. 
For instance, \citeA{carrasco2000generalization} develop a generalized method of moments (GMM) estimator with a continuum of moment conditions based on the CCF; see also \citeA{singleton2001estimation}, \citeA{carrasco2007}. 
In applications to option prices, \citeA{BLL2015} and \citeA{boswijk2021jump} propose to imply the latent state vector from a panel of options and then estimate the model via GMM with a continuum of moments. 
\citeA{bates2006maximum} develops maximum likelihood estimation and filtering using CCFs. 
In particular, he proposes a recursive likelihood evaluation by updating the CCF of a latent variable conditional upon observed data. 
However, unlike our approach, these methods require numerical integration over the dimension of the state vector, thus suffering from a `curse of dimensionality'.

Our work is also related to \citeA{feunou2018risk}, who exploit the linear relation between the first four risk-neutral cumulants of the log-asset price and latent factors. 
They obtain these cumulants via a portfolio of options and employ the Kalman filter to estimate the latent factors. 
The main difference with our approach is that we exploit the CCF, and the corresponding state space representation we develop, instead of the first four moments. 
The CCF contains much richer information, leading to more efficient inference.
Another difference is in dimension reduction: \citeA{feunou2018risk} use a two-step principal components analysis (PCA) to reduce the dimension of the risk-neutral cumulants observed at different maturities. 
Instead, we use a modified version of a so-called collapsed Kalman filtering approach, originally developed by \citeA{jungbacker2015likelihood}, which does not suffer from information losses relative to the full-dimensional setting. 

The paper is organized as follows. 
Section~\ref{sec:Framework} provides the theoretical framework for aligning the option-implied and model-implied CCFs. 
In Section~\ref{sec:Estimation}, we develop the state space representation, and establish the main result about the orders of measurement errors, under a double asymptotic scheme. 
This allows us to next develop the filtering approach and corresponding estimation procedure. 
Section~\ref{sec:Simulation} presents the Monte Carlo simulation results. 
We describe the data in Section~\ref{sec:Data} and the empirical applications in Section~\ref{sec:Empirics}. 
Conclusions are in Section~\ref{sec:Conclusion}.
In supplementary material, four appendices provide details on 
($i$) the proof of Proposition~\ref{prop1},
($ii$) the computation of conditional moments, 
($iii$) the inter- and extrapolation scheme for option prices and the measurement errors in the CCF replication, and 
($iv$) additional simulation and empirical results.

\section{Theoretical Framework}
\label{sec:Framework}

In this section, we provide the theoretical framework for our approach. 
We start with extracting information about the CCF from option prices allowing for general underlying dynamics. 
Next, we consider the CCF within the AJD class, which is exponentially affine in the model's state variables. 
Finally, we discuss how to align the two CCFs---option-implied model-free and AJD model-implied---in order to conduct inference about the model parameters and the latent state variables.

\subsection{Option-implied CCF}
\label{sec:Framework1}

Throughout the paper, we fix a filtered probability space $(\Omega, \F, \{\F_t\}_{t\geq 0}, \P)$. 
On this probability space, we consider the dynamics of an arbitrage-free financial market. 
The no-arbitrage assumption guarantees the existence of a risk-neutral probability measure $\Q$, locally equivalent to $\P$. 
Since we are interested in exploiting information from options, we formulate the model dynamics under $\Q$.

Let us denote by $F_t$ the futures price at time $t$ for a stock or an index futures contract with some fixed maturity. 
The absence of arbitrage implies that the futures price process is a semimartingale. 
In this subsection, we assume the following general dynamics for $F_t$ under $\Q$:
    \begin{align}\label{GD}
        \frac{\diff F_t}{F_t} = v_t \diff W_t + \int_{\R} x \tilde{\mu}(\dt, \diff x),\qquad F_0>0,
    \end{align}
    where $v_t$ is an adapted, locally bounded, but otherwise unspecified stochastic volatility process; $W_t$ is a standard Brownian motion; $\mu$ is a counting random measure with compensator $\nu_t(\diff x)\dt$ such that $\tilde{\mu}(\dt, \diff x) := \mu(\dt, \diff x) - \nu_t(\diff x)\dt $ is the associated martingale measure and $\int (x^2 \wedge 1)\nu_t(\diff x) < \infty$. 

We further denote out-of-the-money (OTM) European-style option prices at time $t$ with time-to-maturity $\tau > 0$ and strike price $K>0$ by $O_t(\tau, K)$. 
Under the no-arbitrage assumption, the option prices equal the risk-neutral conditional expectations of the corresponding discounted payoff functions:
    \begin{align*}
        O_t(\tau, K) = 
        \begin{cases}
            \E^\Q[ e^{-r\tau}(F_{t+\tau} - K)^+ \rvert \F_t],  \quad \mbox{if } K>F_{t},\\
            \E^\Q[ e^{-r\tau}(K - F_{t+\tau})^+ \rvert \F_t],  \quad \mbox{if } K\leq F_{t}.
        \end{cases}
    \end{align*}
The OTM price $O_t(\tau, K)$ is a call option price if $K > F_{t}$ and a put option price if $K\leq F_{t}$. 
For simplicity, we assume a constant interest rate $r$. 

Following \citeA{carr2001optimal}, any twice continuously differentiable  European-style payoff function $g(F_{t+\tau})$, with first and second derivatives $g_{F}$ and $g_{FF}$, can be spanned via a position in risk-free bonds, futures (or stocks) and options with a continuum of strikes, as follows:
    \begin{align*}
        g(F_{t+\tau}) =\ & g(x) + g_{F}(x)(F_{t+\tau} - x) \\
        & + \int_0^x g_{FF}(K) (K-F_{t+\tau})^+ \diff K + \int_x^\infty g_{FF}(K) (F_{t+\tau} - K)^+ \diff K.
    \end{align*}
Here, $x\in\mathbb{R}^{+}$, the first and second terms on the right-hand side correspond to risk-free bonds and futures positions, and the third and fourth terms correspond to OTM options.
Taking conditional expectations under the risk-neutral measure for $x=F_t$, we find that the price at time $t$ of a contingent claim with payoff function $g(F_{t+\tau})$ can be expressed as a weighted portfolio of a risk-free bond and OTM options:
    \begin{align}\label{spanning}
        \E^{\Q}[e^{-r\tau} g(F_{t+\tau})\rvert \F_t] = e^{-r\tau} g(F_t) + \int_0^\infty g_{FF}(K) O_t(\tau, K) \diff K.
    \end{align}

This general spanning result lies behind the construction of one of most popular `fear' indices---the VIX index, when $g(F_{t+\tau}) = \log(F_{t+\tau}/F_t)$. 
Some other applications of the spanning formula \eqref{spanning} include the calculation of the option-implied skewness and kurtosis \cite{bakshi2003stock} and of the corridor implied volatility \cite{andersen2007construction}. 

Applying this result to the complex-valued payoff function $g(x) = e^{\mathrm{i }u \log(x/F_t)}$ yields that the discounted CCF of log returns can be spanned as 
    \begin{align}\label{CCF_spanning}
        \phi_t(u, \tau) &:= e^{-r\tau} \E^\Q[e^{\mathrm{i} u \log(F_{t+\tau}/F_t)} \rvert \F_t] \notag \\
        &= e^{-r\tau} - (u^2 + \mathrm{i}u) \int_0^\infty \frac{1}{K^2} e^{\mathrm{i}u(\log K - \log F_t)} \cdot O_t(\tau, K)  \diff K \notag \\
        &=e^{-r\tau} - (u^2 + \mathrm{i}u) \frac{1}{F_t} \int_\mathbb{R} e^{(\mathrm{i}u -1) m } \cdot O_t(\tau, m) \diff m,
    \end{align}
    where $m = \log (K/F_t)$ is the log-moneyness of an option with strike price $K$.\footnote{With slight abuse of notation, we use the same symbol $O_t$ for the option value as a function of $(\tau,K)$ and as a function of $(\tau,m)$.}

It is important to emphasize that the spanning of the CCF in equation~\eqref{CCF_spanning} is exact and is furthermore completely model independent akin to the VIX construction. 
Therefore, the CCF of log returns over a particular horizon $\tau$ can be replicated in a model-free way given a single cross-sectional slice of liquid option prices with all strikes (and the same maturity $\tau$). 
A similar approach of CCF spanning is taken by \citeA{todorov2019nonparametric} to nonparametrically estimate spot volatilities from option prices (considering the limit as $\tau \downarrow 0$).

The expression for $\phi_t(u,\tau)$ in~\eqref{CCF_spanning} cannot be computed in reality as we do not observe option prices for a continuum of strikes. 
Nevertheless, as we detail in Section~\ref{sec:Estimation-ss}, the expression
in~\eqref{CCF_spanning} is easy to approximate using a limited number of observable option prices.
When developing our estimation procedure, we take both the resulting approximation errors as well as the observation errors in option prices, and hence in the CCF approximation, into account. 
Henceforth, we denote by $\widehat{\phi}_t(u,\tau)$ the computationally feasible counterpart of the option-implied CCF; it is explicitly defined in~\eqref{phi-hat} below. 
In our simulation experiments and empirical applications, we further employ an interpolation-extrapolation scheme to improve the reliability of the approximation. 

\subsection{Affine jump-diffusion CCF}
\label{sec:Framework2}

Whereas the CCF in~\eqref{CCF_spanning} is model independent, the CCF of log returns of the underlying asset is often considered under some parametric assumptions on the return dynamics. 
A model-implied CCF depends on the model parameters, which we generally do not know, and potentially on the dynamics of other latent processes, which affect the distribution of returns. 
Therefore, by suitably aligning the model-free and parameter-dependent CCFs, we may learn about the model parameters and the unobservable state dynamics.    

We restrict our attention to the broad class of AJD models defined in~\citeA{DPS2000}. 
The main attraction of the AJD class is that the Laplace transform has a semi-closed-form expression and is of the exponential-affine form. 
Suppose that $X_t$ is a Markov process representing an $d_X$-dimensional state vector in $D \subset \R^{d_X}$ with the first component being the log price of an asset. 
We assume that under the physical and risk-neutral probability measures, the state vector $X_t$ solves the following stochastic differential equation:
    \begin{align}\label{AJD-DPS}
        \diff X_t = \mu(X_t; \theta) \dt + \sigma(X_t; \theta)\diff W_t + \sum_{i=1}^{d_J} J_{i,t} \diff N_{i,t},
    \end{align}
    where $W_t$ is a standard Brownian motion in $\R^{d_W}$; $\mu{:}\ D \to \R^{d_X}$ and $\sigma{:}\ D \to \R^{d_X \times d_W}$ are the drift and diffusion functions; $N_{i,t}$ is a pure jump process with intensity $\{\lambda^{i}(X_t;\theta){:}\ t\geq 0\}$, $\lambda^{i}{:}\ D \to \R^{+}$; $\{J_{i,t}\}_{t\geq 0}$ constitutes a sequence of jump sizes with generic conditional distribution $\nu^{i}$ on $\R^{d_X}$ for $i=1,\dots,d_J$; and $\theta$ is a vector of unknown parameters that governs the model for $X_t$.
We note that we allow for multiple jump types each arriving with their own intensity process as in the generalized AJD class in Appendix~B of \citeA{DPS2000}.  
The specification~\eqref{AJD-DPS} can be extended further, e.g., to include a time-dependent structure and infinite activity jumps;  
see \citeA{DPS2000}, in particular Appendix~B, and \citeA{duffie2003affine} for more details on the AJD class formulation.

Following \citeA{DPS2000}, the drift $\mu(x)$, diffusive variance $\sigma(x)\sigma(x)'$ and jump intensities $\lambda^{i}(x)$ are assumed to be affine on $D$:
    \begin{align*}
            \mu(x) &= K_0 + K_1 x, & K_0& \in \R^{d_X},\ K_1 \in \R^{d_X\times d_X},\\
            \sigma(x)\sigma(x)' &= H_0 + \textstyle{\sum_{j=1}^{d_X}} x_j H_1^{(j)}, &H_0& \in \R^{d_X\times d_X},\ H_1^{(j)} \in \R^{d_X\times d_X},\ j=1,\dots,d_X,\\
            \lambda^{i}(x) &= l_{i,0} + l_{i,1} \cdot x, & l_{i,0}& \in \R,\ l_{i,1} \in \R^{d_X},\ i=1,\dots,d_J,  
    \end{align*}
    where $x_{j}$ is the $j$-th element of a vector $x$ and $H_1^{(j)}$ for $j{=}1,\dots,d_X$ form a $d_X {\times} d_X {\times} d_X$ tensor $H_1$ by stacking matrices along a new dimension. 
The joint regularity conditions on $(D, \mu, \sigma, \lambda, \nu)$ that guarantee a unique solution to the SDE \eqref{AJD-DPS} are discussed in \citeA{duffie1996yield} and \citeA{dai2000specification}. 
These joint conditions put constraints on the parameter vector $\theta$. 
Therefore, we consider a model from the AJD class indexed by $\theta$ in a parameter space $\Theta$ containing such admissible parameter values, on which there is a unique solution to \eqref{AJD-DPS} that remains in $D$. 
For instance, in the case of the stochastic volatility component, the admissible parameter values in $\Theta$ ensure that the volatility process remains nonnegative, by satisfying Feller's condition; 
see also the discussion of the admissibility problem in~\citeA[Chapter 5]{singleton2009empirical}.

\citeA{DPS2000} show that the affine dependence of the functions $\mu(x)$, $\sigma(x)\sigma(x)'$ and $\lambda(x)$ implies an exponential-affine form of the CCF of the state vector $X_t$. 
Specifically, the discounted joint CCF of $X_{t+\tau}$ conditional on $\F_t$ with $\tau > 0$ is given by
    \begin{align}\label{AJD-CCF}
        \psi_X(\mathbf{u}, \tau) := e^{-r\tau} \E^\Q[ e^{ \mathrm{i} \mathbf{u} \cdot X_{t+\tau}} \rvert \F_t] = e^{\alpha(\mathbf{u},\tau;\theta) + \beta(\mathbf{u},\tau;\theta)\cdot X_t},
    \end{align}
    where $\mathbf{u} \in \mathbb{R}^{d_X}$ is an argument vector and $\alpha(\mathbf{u}, \tau; \theta)$ and $\beta(\mathbf{u}, \tau; \theta)$ are solutions to the following complex-valued system of ordinary differential equations (ODEs) in time:
    \begin{align}\label{ODEs}
        \begin{cases}
            \dot \beta(\mathbf{u},s) = K_1' \beta(\mathbf{u},s) + \frac{1}{2} \beta(\mathbf{u}, s)' H_1 \beta(\mathbf{u}, s) + \sum_{i=1}^{d_J} l_1^i(\chi^i(\beta(\mathbf{u}, s))-1), \\
            \dot \alpha(\mathbf{u}, s) = K_0' \beta(\mathbf{u}, s) + \frac{1}{2} \beta(\mathbf{u}, s)' H_0 \beta(\mathbf{u}, s) + \sum_{i=1}^{d_J} l_0^i(\chi^i(\beta(\mathbf{u}, s))-1) - r,
        \end{cases}
    \end{align}
    with initial conditions $\beta(\mathbf{u}, 0) = \mathrm{i} \mathbf{u}$ and $\alpha(\mathbf{u}, 0) = 0$. 
Here, $\chi^{i}(c) = \int_{\R^n}\exp(c \cdot z) \diff \nu^{i}(z),\ c\in \mathbb{C}^{d_X}$, are jump transforms, 
which determine the conditional jump-size distributions. 
The ODE for $\beta$ is known as a generalized Riccati equation, 
whereas the solution for the second ODE can be obtained by simply integrating the right-hand side expression over time. 

The affine dependence of the characteristic exponent $\alpha(\mathbf{u},\tau ; \theta) + \beta(\mathbf{u}, \tau; \theta){\cdot} X_t$ on the current state $X_t$ is even the defining property of the AJD class under some regularity conditions (see \citeNP{duffie2003affine}). 
In other words, the AJD class can be defined as a class in which characteristic exponents of $X_{t+\tau}$ given $X_{t}$ are affine functions of $X_t$. 
In fact, this is a key property in our estimation procedure. 
While it is also possible to obtain the CCF for some non-affine models, the exponential-affine form allows us to use linear Kalman filtering techniques in the estimation procedure. 
This is the main motivation why we restrict our attention to the parametric models of the AJD class.\footnote{The considered AJD class could, in principle, be broadened further to the linear-quadratic jump-diffusion class by augmenting the state vector (see \citeNP{cheng2007linear}, for more details).}

Unlike the option-implied CCF \eqref{CCF_spanning}, the CCF in \eqref{AJD-CCF} is fully parametric, that is, it requires parametric AJD model dynamics of the state vector $X_t$.   
Although the AJD class is more restrictive than the general dynamics of $F_t$ in \eqref{GD}, it includes a myriad of popular option pricing models such as those in~\citeA{heston1993}, \citeA{DPS2000}, \citeA{pan2002}, \citeA{bates2006maximum}, \citeA{broadie2007model}, \citeA{BLL2015}, and \citeA{AFT2017} among many others.

The state process $X_t$ often includes both observed and unobserved state variables that affect the dynamics of the log futures price $\log F_t$.  
In our empirical application, we consider the presence of both. 
Therefore, it is convenient to partition the state vector as $X_t' = (w_t', x_t')$, where $w_t$ represents the observable component and $x_t$ includes $d < d_X$ latent state variables. 
Then, the dynamics of $X_t$ given by equation~\eqref{AJD-DPS}, can be rewritten as
    \begin{align}\label{AJD-w}
        \diff w_t &= \mu^w(w_t, x_t)\dt + \sigma^w(w_t, x_t) \diff W_t + \sum_{i=1}^{d_J} J_{i,t}^w \diff N_{i,t},\\
        \label{AJD-x}
        \diff x_t &= \mu^x(w_t, x_t)\dt + \sigma^x(w_t, x_t) \diff W_t + \sum_{i=1}^{d_J} J_{i,t}^x \diff N_{i,t},
    \end{align}
    where $\mu^{w}{:}\ D \to \R^{d_X-d},\ \mu^{x}{:}\ D \to \R^{d},\ \sigma^w{:}\ D \to \R^{(d_X-d) \times d_J},\ \sigma^x{:}\ D \to \R^{d \times d_J}$ and $J_{i,t}^w$ and $J_{i,t}^x$ are marginal jump sizes of $J_{i,t}$ associated with $w_t$ and $x_t$, respectively. 
In the simplest case, the observable component includes only the log futures prices, that is, $w_t = \log F_t$. 
In more general settings, the stochastic volatility is often a main latent driver of the log returns dynamics, as e.g., in \citeA{heston1993}.

\subsection{Marrying the two CCFs}
\label{sec:Framework3}

Given the two CCFs~\eqref{CCF_spanning} and~\eqref{AJD-CCF}, we can now align them to conduct inference about the model parameters and the unobservable state variables. 
For that purpose, first note that the CCF in \eqref{AJD-CCF} is joint for the state vector $X_t$. 
We assume, without loss of generality, that the first component of the state vector $X_t$ is the log futures price. 
Therefore, we can easily obtain its marginal CCF by plugging in an argument vector of the form $\mathbf{u_1} := (u, 0,\dots, 0)' \in \mathbb{R}^{d_X}$ with $u\in \mathbb{R}$. 
To obtain the marginal CCF of log returns, we further subtract the term $\mathrm{i}u \log F_t$ in the exponent. 
That is, the marginal CCF of log returns under the AJD specification is aligned to that in~\eqref{CCF_spanning} as follows:
    \begin{align}\label{AJD-CCF-ret}
        \phi_t(u, \tau) = \psi_{X}(\mathbf{u_1}, \tau) e^{ - \mathrm{i}u \log F_t} 
        = e^{\alpha(\mathbf{u_1},\tau;\theta) + \tilde\beta(\mathbf{u_1},\tau; \theta) \cdot X_t},
    \end{align}
where $\tilde\beta(\mathbf{u_1},\tau; \theta) := \beta(\mathbf{u_1},\tau; \theta) - \mathrm{i}\mathbf{u_1}$; i.e., the first component of $\tilde\beta(\mathbf{u_1},\tau; \theta)$ differs from that of $\beta(\mathbf{u_1},\tau; \theta)$, since we are interested in the CCF of log returns rather than that of log prices.

Note that the log of the (joint) CCF (also known as cumulant generating function) is linear in the state vector $X_t$. 
Therefore, under a correctly specified AJD model we obtain a simple linear relation between the log of the option-spanned CCF\footnote{Although the logarithm of a complex number is a multivalued function, here, the ambiguity is resolved given the fact that $\phi(0) = 1$ and the CCF is a continuous function. 
In fact, in practice we ensure that the logarithm of the CCF does not have `jumps' by taking the logarithm sequentially with respect to $u$, starting from the origin.}
of log returns and the model's state vector:
    \begin{align}
        \log \phi_t(u, \tau) 
        = \alpha(\mathbf{u_1}, \tau; \theta) + \tilde\beta(\mathbf{u_1}, \tau; \theta) \cdot X_t.
    \end{align}
Replacing the cumulant generating function on the left-hand side with its computationally feasible counterpart $\widehat \phi_t(u, \tau)$, 
which we will explicitly define in Section~\ref{sec:Estimation-ss}, 
we obtain the following equation, which will play a central role in our estimation procedure:
    \begin{align}\label{key_relation}
        \log \widehat \phi_t(u, \tau) = \alpha(\mathbf{u_1}, \tau; \theta) + \tilde\beta(\mathbf{u_1}, \tau; \theta) \cdot X_t + \xi_t(u, \tau), \qquad u \in \R.
    \end{align}
Here, $\xi_t(u, \tau)$ is the measurement error, which is related to the observation,  truncation and discretization errors in the CCF-spanning option portfolios. 
We elaborate in detail on the relation between the computable counterpart of the CCF and the source of the measurement errors in the next section.    

Equation~\eqref{key_relation} is the key relation in our analysis and a few remarks shall be made here regarding it.
First, \eqref{key_relation} is essentially a \textit{functional} linear model since this equation holds for any argument variable of the CCF, $u \in \mathbb{R}$.
Furthermore, the functions $\alpha(\mathbf{u_1}, \tau; \theta)$ and $\tilde\beta(\mathbf{u_1}, \tau; \theta)$ are parameter-dependent and solutions to the system of Riccati ODEs \eqref{ODEs}. 
Therefore, if the state vector $X_t$ is observable, then the model parameters can be estimated by solving a continuum version of a non-linear least-squares problem.

Second, in the case in which the state vector is (partially) unobservable, \eqref{key_relation} represents a linear latent factor model with a continuum of linear relations. 
The factors are given by the state components of the AJD model. 
Therefore, one could apply, e.g., a (functional) principal component analysis to learn about the unobserved factors. 
In this paper, we utilize a (suitably modified) Kalman filtering technique to conduct inference about the model parameters and the latent factors.

In other words, \eqref{key_relation} reveals that, using the present approach, AJD models become amenable to filtering and estimation using approaches from the rich literature on linear factor and state space models. 
This is reminiscent of the term structure literature, 
where in affine term structure models (see \citeNP{piazzesi2010affine}, for a review of this class of models) bond yields themselves are assumed to be linear functions of the state vector. 
For instance, \citeA{duffee1999estimating}, \citeA{jong2000time}, \citeA{driessen2005default}  use the Kalman filter in their estimation of affine term structure models.

Furthermore, another advantage of this approach is that it does not require evaluating option prices given a certain parametric model. 
Therefore, our estimation procedure is computationally more appealing than many alternative approaches, which often involve the Carr-Madan FFT pricer (\citeNP{CM1999}) or the COS method (\citeNP{fang2008novel}) to price options. 
This also implies that the usage of the characteristic function is different: with the FFT or COS methods one needs a model-dependent CCF only to evaluate option prices, while here we use the CCF to directly learn about the latent factors and the model parameters. 

Finally, given the partition of the state vector into observable and unobservable components, the linear relation between the option-implied and model-implied CCFs in~\eqref{key_relation} can be rewritten as
    \begin{align}\label{key_relation2}
        \log \widehat \phi_t(u, \tau) = \alpha(\mathbf{u_1}, \tau; \theta) + \beta^w(\mathbf{u_1}, \tau; \theta) \cdot w_t + \beta^x(\mathbf{u_1}, \tau; \theta) \cdot x_t + \xi_t(u, \tau), \qquad u \in \R,
    \end{align}
    where $\beta^{w}(\mathbf{u_1}, \tau; \theta) \in \mathbb{C}^{d_X-d}$ and $\beta^{x}(\mathbf{u_1}, \tau; \theta) \in \mathbb{C}^{d}$ are such that  $\tilde\beta' = (\beta^{w\prime}, \beta^{x\prime})$ is the solution to the ODE system~\eqref{ODEs}. 
Representation~\eqref{key_relation2} serves as the basis for an observation (or measurement) equation in our estimation procedure.

\section{Estimation Procedure}
\label{sec:Estimation}

In this section, we develop our filtering approach and corresponding estimation procedure for the general class of AJD models under consideration. 
First, we provide the formal state space representation for the defined class of models. 
Then, we describe our estimation strategy, which uses the collapsed Kalman filter.
    
\subsection{State space representation}
\label{sec:Estimation-ss}

As discussed in the previous section, we restrict our attention to the parametric models of the AJD class due to their exponential-affine form of the characteristic function. 
This form will allow us to exploit a linear Kalman filter in the estimation procedure. 
In the following, we summarize the assumptions we impose on the parametric model:
    \begin{assumption}\label{ass-ajd}
        \begin{enumerate}[label=(\roman*)]
            \item The stochastic process $X_t$ is Markov and affine, with finite second moments under both the physical and risk-neutral probability measures $\P$ and $\Q$. 
            In particular, $X_t$ is the unique solution to the SDE \eqref{AJD-DPS} and its characteristic function is of the exponential-affine form \eqref{AJD-CCF};
            \item The true parameter vector $\theta_0$ lies in the interior of a compact parameter space $\Theta$ containing admissible parameter values.
        \end{enumerate}
    \end{assumption} 

Assumption~\ref{ass-ajd} guarantees the existence of a unique solution to the SDE \eqref{AJD-DPS} within the AJD class. 
As discussed in Section~\ref{sec:Framework2}, admissible values $\theta \in \Theta$ reflect the regularity conditions imposed on the model such that there is a unique solution to \eqref{AJD-DPS}, with, e.g., non-negative volatilities and jump intensities. 
Such admissibility conditions will need to be checked in a case-by-case model analysis.
Assumption~\ref{ass-ajd}(\textit{i}) also presumes the technical conditions required to represent the AJD process, defined via the affine dependence of its drift, diffusive variance and jump intensities on the state vector, through the exponential-affine characteristic function. 
For a detailed analysis of the AJD theory, we refer to \citeA{DPS2000} and \citeA{duffie2003affine}. Note that Assumption~\ref{ass-ajd} does not require the state process to be stationary. Stationarity of the latent state variables $x_t$ is reasonable but not essential for the results to follow; the observed state variables $w_t$ (often including the log-forward price) are typically non-stationary.
    
In our estimation procedure, we discretize the continuous-time model along two dimensions: with respect to time and with respect to the argument of the CCF. 
The former naturally follows from the discrete sampling times of financial data, which we denote by the integer indices $t=1,\dots,T$. 
The latter allows us to rely on the existing literature about filtering techniques. 
For that, let us denote the collection of discretely sampled arguments by a set $\mathcal{U} \subseteq \R$ with cardinality $q \in \mathbb{N}$. 
We further consider options with $k \in \mathbb{N}$ different maturities $\tau$ and $n \in \mathbb{N}$ different log-moneyness values $m$ on each day. 
    
Since the input of our estimation procedure is a portfolio of option prices, we need to take into account the measurement errors in these option portfolios. 
For that purpose, we assume an observation error scheme on the option prices that constitute the portfolios. 
The measurement errors will be defined on the common probability space $(\Omega, \F, \P)$,
but in what follows, the filtration $\{\F_t\}_{t\ge 0}$ is generated by the state process $\{X_t\}_{t\ge 0}$ only. 
Note that the theoretical option prices $O_t(\tau, m)$ are $\F_t$-measurable, 
and hence the same applies to functionals of the option prices such as the (theoretical) Black-Scholes implied volatility (BSIV) and vega. 
\begin{assumption}\label{ass:option_errors}
Option prices are observed with an additive error term: 
        \begin{equation}
            \widehat{O}_t(\tau_i, m_j) := O_t(\tau_i, m_j) + \zeta_t(\tau_i, m_j),\qquad t=1,\ldots,T, \quad i=1,\ldots,k, \quad j=1,\ldots, n, 
        \label{eq:additiveerror}\end{equation}
        where the observation errors $\zeta_t(\tau, m)$ are such that: 
\begin{enumerate}[label=(\roman*)]
\item $\zeta_t(\tau, m)$ are $\F_t$-conditionally independent along tenors $\tau$, moneyness $m$ and time $t$;
\item $\E[\zeta_t(\tau, m)|\F_t] = 0$; 
\item $\E[\zeta_t(\tau, m)^2|\F_t] = \sigma_{t}^2(\tau,m) < \infty$ with $\sigma_{t}(\tau,m) := \sigma_\varkappa \kappa_t(\tau, m) \nu_t(\tau, m)$, where $\sigma_\varkappa \in \R^+$, $\kappa_t(\tau, m)$ is the Black-Scholes implied volatility, and $\nu_t(\tau, m)$ is the Black-Scholes vega.
\end{enumerate}
\end{assumption}
    
The additive error assumption is commonly imposed in the option pricing literature. 
For instance, \citeA{andersen2015parametric} and \citeA{todorov2019nonparametric} use additive error assumptions for option prices quoted in terms of BSIV and dollar amount, respectively. 
Additive observation errors are also often implicitly assumed when calibrating an option pricing model to market-observed prices, since the calibration is often performed using non-linear least squares as in, e.g., \citeA{broadie2007model}. 
    
Assumption~\ref{ass:option_errors}(\textit{i}) excludes in particular dependence of the observation errors across strikes and is also often imposed in the literature (see, for instance, \citeNP{christoffersen2010volatility}, \citeNP{andersen2015parametric} and \citeNP{todorov2019nonparametric}). 
This assumption can be relaxed by introducing a spatial dependence as in \citeA{andersen2021spatial}. 
This would, however, result in more complex expressions for the covariance terms in the measurement errors that we derive below. 
Furthermore, \citeA{andersen2021spatial} find evidence of limited dependence in the observation errors for S\&P 500 index options. 
They also show that this dependence declined sharply for short-dated options in recent years, due to improved liquidity.
Since in our empirical application we consider S\&P 500 index options with short tenors focusing on the past three years, the independence assumption will play a secondary role for the estimation procedure.
    
The conditional mean zero Assumption~\ref{ass:option_errors}(\textit{ii}) is crucial for our main result.  
Assumption~\ref{ass:option_errors}(\textit{iii}) asserts the standard deviation of the observation errors to be proportional to the product of the option's BSIV and vega.  
The motivation for this structure is as follows. 
Let $\widehat\kappa(m_j)$ and $\kappa(m_j)$ denote the error-distorted and true BSIV of an option,
and assume that the relative volatility errors $\varkappa_j = (\widehat\kappa(m_j) - \kappa(m_j))/\kappa(m_j)$ are homoskedastic across the strikes,
such that $\E[\varkappa_j^2|\F_t] = \sigma_\varkappa^2$. 
A Taylor-series expansion of the Black-Scholes pricing function $O^{BS}(\widehat\kappa(m_j), m_j)$ around $\kappa(m_j)$ 
then gives $\widehat{O}(m_j) = O^{BS}(\widehat\kappa(m_j), m_j) \approx O(m_j) + \nu(m_j) \kappa(m_j) \varkappa_j $, 
with $\nu(m_j) = \partial O^{BS}(\kappa(m_j), m_j) / \partial \kappa(m_j)$ the theoretical Black-Scholes vega.
Homoskedastic errors in relative implied volatilities are also assumed by \citeA{christoffersen2012dynamic} and \citeA{du2019pricing} in their MLE based on the particle filter and the unscented Kalman filter, respectively.
    
Finally, to assess the error sizes of the CCF approximation specified below, we impose the following assumption on the existence of moments for the underlying asset and on the log-moneyness grid that allows nonequidistant sampling in the moneyness dimension:
\begin{assumption}\label{ass-moments}
    \begin{enumerate}[label=(\roman*)]
        \item The underlying process and its reciprocal process have finite second moments under the risk-neutral measure: $\E^{\Q}[F_{t+\tau}^2|\F_t] < \infty$ and $\E^{\Q}[F_{t+\tau}^{-2}|\F_t] < \infty$ with $\tau >0$;
        \item For the log-moneyness grid $\underline{m} := m_1 < \ldots < m_n =: \overline{m}$, there exists a deterministic sequence $\Delta m$ depending on $n$ such that $\Delta m \to 0$ as $n \to \infty$ and
            \begin{align*}
                \eta \Delta m \leq \inf_{j=2,\dots,n} \Delta m_j \leq \sup_{j=2,\dots,n} \Delta m_j \leq \Delta m,
            \end{align*}
            where $\Delta m_j := m_j - m_{j-1}$ and $\eta \in (0,1]$ is some constant.
    \end{enumerate}
\end{assumption}
    
Using $n > 1$ observable option prices with time-to-maturity $\tau >0$ and log-moneyness values $\{m_j\}_{j=1}^n$, 
we may approximate the CCF $\phi_t(u,\tau)$ given in~\eqref{CCF_spanning} by replacing the theoretical option prices by their observed counterparts,
and the integral by a Riemann sum:
    \begin{align}\label{phi-hat}
        \widehat{\phi}_t(u, \tau) &= e^{-r\tau} - u_t \sum_{j=2}^n e^{(\mathrm{i}u-1)m_j} \cdot \widehat{O}_t(\tau, m_j) \Delta m_j,
    \end{align}
where we use the notation $u_t := (u^2 + \mathrm{i} u)/F_t$,
and where $\widehat{O}_t(\tau, m_j)$ satisfies Assumption~\ref{ass:option_errors}. 

The deviation of the option-spanned CCF from its theoretical counterpart, $\zeta_t^{\phi}(u,\tau) := \widehat{\phi}_t(u, \tau) - \phi_t(u, \tau)$,
stems from \textit{observation}, \textit{truncation} and \textit{discretization} errors, 
where truncation refers to the fact that the integration interval $[\underline{m},\overline{m}]$ does not cover the entire real line.
The truncation and discretization errors also arise in VIX calculations and depend on the availability of option prices. 
They will be shown to be of smaller order than the observation errors, and can further be efficiently reduced by using an interpolation-extrapolation scheme (see, e.g., \citeNP{jiang2005model,jiang2007extracting}, \citeNP{chang2012option}, and Appendix~\ref{sec:Appendix-Extrapolation}). 
Appendix~\ref{sec:Appendix-CCFReplication} illustrates the impact of the three different types of measurement errors on the CCF approximation, and the effectiveness of the interpolation-extrapolation scheme.\footnote{The interpolation-extrapolation scheme may induce some cross-sectional dependence in the observation errors $\zeta_t(\tau,m)$. This is in deviation from Assumption~\ref{ass:option_errors}, which is only realistic when referring to the errors before application of the interpolation-extrapolation scheme. We will not consider this effect explicitly in Proposition~\ref{prop1} that follows; it would lead to a more complicated expression for the covariance matrix of the measurement errors, but, importantly, would not affect the main result otherwise.} 
    
From the preceding analysis, the functional measurement equation~\eqref{key_relation2} is then obtained using the following log-linearization:
    \begin{align}\label{loglinear}
        \xi_t(u,\tau) := \log \widehat{\phi}_t(u, \tau) - \log \phi_t(u, \tau) = \log \left ( 1 + \frac{\zeta_t^{\phi}(u,\tau)}{\phi_t(u, \tau)}\right ) = \xi_t^{(1)}(u,\tau) + r_t(u,\tau),
    \end{align}
where the log-linearized observation errors $\xi_t^{(1)}(u,\tau)$ are defined by $\zeta_t^{(1)}(u,\tau)/\phi_t(u, \tau)$, with 
    \begin{align*}
        \zeta_t^{(1)}(u,\tau) := - u_t \sum_{j=2}^n e^{(\mathrm{i}u-1)m_j} \cdot \zeta_t(\tau, m_j) \Delta m_j,
    \end{align*}
and where $r_t(u,\tau)$ is a remainder term that collects the log-linearized truncation and discretization errors as well as the higher-order terms from the required Taylor-series expansion.
(The superscript $^{(1)}$ refers to the first, and prime, source of the measurement errors, the observation errors; see also the detailed decomposition in equation~\eqref{eq:measdec}.)

To formulate the main result, we turn the complex-valued functional measurement equation \eqref{key_relation2} into a real vector measurement equation, as usual in state space model formulations.
First, we stack the log CCF and the corresponding measurement errors along $q$ values $u_1,\ldots, u_q$ for the CCF argument $u \in \mathcal{U}$, for a fixed expiration period $\tau_i$:
    \begin{align*}
        \log \widehat{\phi}_{t, i} :=
        \begin{pmatrix}
            \log \widehat{\phi}_{t}(u_1,\tau_i)\\
            \log \widehat{\phi}_{t}(u_2,\tau_i)\\
            \vdots\\
            \log \widehat{\phi}_{t}(u_q,\tau_i)
        \end{pmatrix},
        \quad
        r_{t,i} := 
        \begin{pmatrix}
            r_t(u_1, \tau_i)\\
            r_t(u_2, \tau_i)\\
            \vdots\\
            r_t(u_q, \tau_i)
        \end{pmatrix},
        \quad
        \xi^{(1)}_{t,i} := 
        \begin{pmatrix}
            \xi^{(1)}_t(u_1, \tau_i)\\
            \xi^{(1)}_t(u_2, \tau_i)\\
            \vdots\\
            \xi^{(1)}_t(u_q, \tau_i)
        \end{pmatrix}.
    \end{align*}
In a similar way, we denote by $a_{t,i}$, $b^w_{t,i}$ and $b^x_{t,i}$ the stacked outputs\footnote{Here, we attribute these elements (and system matrices $\tilde{d}_t, W_t$ and $Z_t$ in equation \eqref{stacking}) with an additional time index although the coefficient functions are assumed to be time-invariant in the exposition. 
This is because in practice we can have different expiration periods for different days.} of the functions
$\alpha(u,\tau_i)$, $\beta^w(u,\tau_i)$ and $\beta^x(u,\tau_i)$, respectively.
Next, to tackle the complex-valued measurement equation~\eqref{key_relation2},
we stack the real and imaginary parts, as well as $k$ maturities: 
    \begin{align}\label{stacking}
        \underbrace{
        \begin{pmatrix}
            \Re(\log \widehat{\phi}_{t,1}) \\
            \Im(\log \widehat{\phi}_{t,1}) \\
            \vdots\\
            \Re(\log \widehat{\phi}_{t,k}) \\
            \Im(\log \widehat{\phi}_{t,k})          
        \end{pmatrix}}_{\strut \textstyle =: y_t \in \R^{p}} 
        &=
        \underbrace{
        \begin{pmatrix}
            \Re(a_{t,1}) \\
            \Im(a_{t,1}) \\
            \vdots\\
            \Re(a_{t,k}) \\
            \Im(a_{t,k})
        \end{pmatrix}}_{\strut \textstyle =: \tilde d_t }
        +
        \underbrace{
        \begin{pmatrix}
            \Re(b^w_{t,1}) \\
            \Im(b^w_{t,1}) \\
            \vdots\\
            \Re(b^w_{t,k}) \\
            \Im(b^w_{t,k})
        \end{pmatrix}}_{\strut \textstyle =: W_t} w_t
        +
        \underbrace{
        \begin{pmatrix}
            \Re(b^x_{t,1}) \\
            \Im(b^x_{t,1}) \\
            \vdots\\
            \Re(b^x_{t,k}) \\
            \Im(b^x_{t,k})
        \end{pmatrix}}_{\strut \textstyle =: Z_t} x_t
        +
        \underbrace{
        \begin{pmatrix}
            \Re(r_{t,1}) \\
            \Im(r_{t,1}) \\
            \vdots\\
            \Re(r_{t,k}) \\
            \Im(r_{t,k})
        \end{pmatrix}}_{\strut \textstyle =: r_{t,n} }
        +
        \underbrace{
        \begin{pmatrix}
            \Re(\xi^{(1)}_{t,1}) \\
            \Im(\xi^{(1)}_{t,1}) \\
            \vdots\\
            \Re(\xi^{(1)}_{t,k}) \\
            \Im(\xi^{(1)}_{t,k})
        \end{pmatrix}}_{\strut \textstyle =: \varepsilon_t},
    \end{align}
    where $p=2qk$. 
Stacking the real and imaginary parts of the measurements is a natural approach when the state vector is real-valued;\footnote{See \citeA{singleton2001estimation} and \citeA{chacko2003spectral}, who use this approach in a GMM estimation setting based on the empirical characteristic function.} a complex-valued state vector would have required a complex Kalman filter based on the so-called widely linear complex estimator, as in~\citeA{dini2012class}.
The stacked observation equation \eqref{stacking} links all available information from option prices with several tenors at time $t$ to the state vectors $w_t$ and $x_t$ in a linear way.

To complete the state space model, we need to augment the measurement equation~\eqref{stacking} by a transition equation for the unobservable state vector $x_t$.
This is a linear, discrete-time dynamic system, to be derived from the continuous-time stochastic differential equation. 
An Euler discretization of the state process~\eqref{AJD-x} would converge to the true transition dynamics as the discretization step $\Delta t \to 0$. 
However, the maximum likelihood (ML) estimator based on the Euler discretization is, in general, inconsistent for fixed non-zero $\Delta t$ \cite{lo1988maximum}, because the discretization has conditional moments different from those of the true process \cite{piazzesi2010affine}. 
Fortunately, the AJD assumption under $\P$ implies that the first and second conditional moments of $x_{t+1}$ given $\F_t$ are linear and available in semi-closed form (possibly requiring the solution of a system of ODEs):
    \begin{align}
        \label{discrmean}
        \E[x_{t+1}|\F_t] &= c_t + T_t x_t,\\
        \label{discrvar}
        \mbox{Var}(x_{t+1}|\F_t) &= Q_t(x_t),
    \end{align}
    where $Q_t{:}\ \R^d \to \R^{d\times d}$ is an affine function in $x_t$. 
The finiteness of the conditional moments is ensured by Assumption~\ref{ass-ajd}(\textit{i}). 
Both conditional moments will in general be linear in both the observed state $w_t$ and the latent state $x_t$; but because the former does not need filtering,
we absorb its effect in the time-varying intercept $c_t$, and similarly in the intercept of the affine function $Q_t$.\footnote{The transition matrix $T_t$ will not be time-varying in stationary AJD processes with equidistant observations, but we do not impose this time-constancy in the notation, also to avoid confusion with the sample size $T$.} 
    
In Appendix~\ref{sec:Appendix-CM}, we show how these transition coefficients can be computed for the AJD model.
Using this approach, which will in principle be model-dependent and hence has to be applied case by case, we obtain a discrete-time transition equation with the same conditional mean and variance as the true continuous-time process (but possibly different higher-order moments). 
Quasi-maximum likelihood (QML) estimation based on conditionally normally distributed measurement and transition errors in the state space representation yields consistent estimation results \cite{fisher1996estimating}.
A similar approach has been adopted in the term structure literature (see, e.g., \citeNP{jong2000time}, \citeNP{duffee2002term}).

We summarize the development of the state space representation, and analyze properties of the errors, in the following proposition. 
The main result contains a remainder term in the measurement equation that collects the truncation and discretization errors in the construction of $\log \widehat{\phi} (u,\tau)$ and higher-order terms in the log-linearization. 
This term vanishes under an asymptotic scheme, where $\overline{m} = \max_{1\le j \le n} m_j \to \infty$, $\underline{m} = \min_{1\le j \le n} m_j \to -\infty$ and $\Delta m \to 0$. 
We also denote the corresponding smallest and largest strike prices by $\underline{K}$ and $\overline{K}$, and express the asymptotic orders with respect to the number of option prices $n$ with fixed maturity.

    \begin{proposition} \label{prop1}
        Suppose Assumptions~\ref{ass-ajd},~\ref{ass:option_errors} and~\ref{ass-moments} hold, and in addition $\underline{K} \asymp n ^{-\underline{\alpha}}$ and $\overline{K} \asymp n^{\overline{\alpha}}$ with $\underline{\alpha} >0$ and $\overline{\alpha} >0$. 
        Then $\{ (y_t,x_t), t=1,\ldots,T \}$ satisfy the linear state space representation
        \begin{align}
        \label{ss-observ}
        y_t &= d_t + Z_t x_t + r_{t,n} + \varepsilon_t,  &\E[\varepsilon_t | \F_t] &= 0,
        &\E[\varepsilon_t \varepsilon_t' | \F_t] &= H_t,\\
        \label{ss-state}
        x_{t+1}&= c_t + T_t x_t + \eta_{t+1},  &\E[\eta_{t+1} | \F_t] &=0, &\E[\eta_{t+1} \eta_{t+1}' | \F_t] &= Q_t(x_t),
    \end{align}
    where $r_{t,n} = \mathcal{O}_p\left( n^{-2(\underline{\alpha} \wedge \overline{\alpha})} \vee n^{-1}\log n \right) $ and $\varepsilon_t = \mathcal{O}_p\left( \sqrt{n^{-1}\log n} \right)$;
    $d_t=\tilde{d}_t + W_t w_t$ and $Z_t$ are defined in \eqref{stacking} and $c_t$, $T_t$ and $Q_t$ are as given in
    \eqref{discrmean}--\eqref{discrvar}; and $H_t = \mathrm{blkdiag}\{H_{t,1}, \dots, H_{t,k}\}$,
    with $H_{t,i} = \sigma_\varkappa^2 \cdot \widetilde{H}_{t,i}$, 
    where
    \begin{align}\label{complex_covar}
        \widetilde{H}_{t,i} =
        \begin{pmatrix}
            \frac{1}{2} \Re(\widetilde\Gamma_{t,i} + \widetilde C_{t, i}) & \frac{1}{2} \Im(-\tilde\Gamma_{t,i} + \widetilde C_{t, i})\\
            \frac{1}{2} \Im(\widetilde\Gamma_{t,i} + \widetilde C_{t, i}) & \frac{1}{2} \Re(\widetilde\Gamma_{t, i} - \widetilde C_{t,i})
        \end{pmatrix},\quad i=1,\dots,k,
    \end{align}
    and $\widetilde\Gamma_{t,i}$ and $\widetilde C_{t,i}$ are covariance and pseudo-covariance matrices of $\xi_{i,t}/\sigma_{\varkappa}$,
    with elements
    \begin{align*}
        (\widetilde\Gamma_{t,i})_{kl} &= \frac{u_{k,t} \overline{u_{l,t}} \sum_{j=2}^n e^{(\mathrm{i}(u_k - u_l)-2)m_j} \kappa^2_{t}(\tau_i, m_j) \nu^2_{t}(\tau_i, m_j) (\Delta m_j)^2}
        {\phi_t(u_k,\tau_i)\phi_t(-u_l,\tau_i)}, \quad k,l =1,\dots,q,
        \\
        (\widetilde C_{t,i})_{kl} &= \frac{u_{k,t} u_{l,t} \sum_{j=2}^n e^{(\mathrm{i}(u_k + u_l)-2)m_j} \kappa^2_{t}(\tau_i, m_j) \nu^2_{t}(\tau_i, m_j) (\Delta m_j)^2}
        {\phi_t(u_k,\tau_i)\phi_t(u_l,\tau_i)}, \quad k,l =1,\dots,q.        
    \end{align*}
    Furthermore, 
        \begin{enumerate}[label=(\roman*)]
            \item $\E[\varepsilon_t \varepsilon_{s}'] = 0$ and $\E[\eta_t \eta_s'] = 0$ for $s \neq t = 1,\ldots, T$;
            \item $\E[\varepsilon_t \eta_{s}'] = 0$ for all $s,t=1,\dots, T$;
            \item $\E[\varepsilon_t x_1'] = 0$ and $\E[\eta_{t+1} x_1'] = 0$ for $t=1,\ldots,T$.
        \end{enumerate}
    \end{proposition}

The proof is given in Appendix~\ref{sec:Appendix-Proofs}. 
The orders indicate that the remainder term goes to zero faster than the observation term given some minimum non-zero requirements for $\underline{\alpha}$ and $\overline{\alpha}$. 
In the sequel, we assume that $(\underline{\alpha} \wedge \overline{\alpha}) > \frac{1}{4}$ and neglect the remainder term in the estimation and filtering procedures.  
The system matrices $Z_t, T_t, Q_t(x_t)$ and system vectors $d_t$ and $c_t$ are known up to a parameter vector $\theta$, assumed to lie in the interior of a compact parameter space $\Theta$ by Assumption~\ref{ass-ajd}($ii$). 
Similarly, the system matrix $H_t$ depends on the data and $\theta$ (via $u_t, \phi_t, \kappa_t$ and $\nu_t$), and an additional unknown parameter $\sigma_{\varkappa}^2$. 
Note that $(d_t,Z_t,H_t)$ are derived from the $\Q$-dynamics of~\eqref{AJD-DPS}, whereas $(c_t, T_t, Q_t(\cdot))$ correspond to the $\P$-dynamics. 
Therefore, possible deviations between $\P$ and $\Q$, reflecting the presence of factor risk premia, will require an extension of the parameter vector; we discuss this possibility further in Section~\ref{sec:Simulation} and Appendix~\ref{sec:Appendix-SimulationEmpirical}.
Estimation of $\theta$ and filtering of the latent state vector via (versions of) the Kalman filter is considered in the next sub-section.

\subsection{Modified and collapsed Kalman filter}
\label{sec:Estimation-cKF}

Consider the state space representation \eqref{ss-observ}--\eqref{ss-state}, where from now on we will ignore the remainder term $r_{t,n}$, and hence assume that
the set of strike prices $\{m_j\}_{j=1}^n$ on each day is rich enough to make this term negligible.
Define the dataset $Y_t = \{y_1, \ldots, y_t\}$, and linear projections (denoted by $\widehat \E$) of the latent state vector conditional on the data:
$\widehat x_{t|t} = \widehat \E[x_t|Y_t]$ and $\widehat{x}_{t|t-1} = \widehat \E[x_{t}| Y_{t-1}]$, with corresponding mean square error matrices
$P_{t|t} = \E [(x_t-\widehat{x}_{t|t})(x_t-\widehat{x}_{t|t})']$ and $P_{t|t-1} = \E [(x_t-\widehat{x}_{t|t-1})(x_t-\widehat{x}_{t|t-1})']$.
Then a modified version of the Kalman filter reads as follows:
    \begin{align*}
        \omega_t &= y_t - (d_t + Z_t \widehat{x}_{t|t-1}),
        &G_t =&\ Z_t P_{t|t-1} Z_t' + H_t, \\
        \widehat{x}_{t|t} &= \widehat{x}_{t|t-1} + P_{t|t-1} Z_t' G_t^{-1} \omega_t, 
        &P_{t|t} =&\ P_{t|t-1} - P_{t|t-1} Z_t' G_t^{-1} Z_t P_{t|t-1},\\
        \widehat{x}_{t+1|t} &= c_t + T_t \widehat{x}_{t|t},
        &P_{t+1|t} =&\ T_t P_{t|t} T_t' + Q_t(\widehat{x}_{t|t}), 
    \end{align*}
    for $t=1,\dots,T$. 
If the latent state process $x_t$ is stationary, the initial conditions $\widehat{x}_{1|0}$ and $P_{1|0}$ for the filter can be set to the unconditional mean and variance, respectively.

In traditional homoskedastic Gaussian state space models, where the distribution of the vector $(\varepsilon_t^{\prime}, \eta_{t+1}^{\prime})^{\prime}$, conditional on $x_t$, is Gaussian with a constant variance matrix, the filtered state $\widehat{x}_{t|t}$ is the conditional expectation of the true process $x_t$ given the observations up to time $t$. 
When the errors are non-Gaussian homoskedastic, the filtered state represents the linear projection (or minimum mean square error linear predictor) instead of the conditional expectation. 
This property can be used to prove that quasi-maximum likelihood (QML) estimation based on the Gaussian likelihood still yields consistent and asymptotically normal parameter estimates \cite[Chapter 13]{hamilton1994time}. 
In general AJD models, on the other hand, the distribution of the errors will be non-Gaussian with a conditional variance $Q_t(x_t)$ that is an affine function of the true latent state vector $x_t$. 
Therefore, the Kalman filter recursions have been modified by using $Q_t(\widehat{x}_{t|t})$ instead of the unobserved $Q_t(x_{t})$. 
A similar modification is used in, e.g.,~\citeA{jong2000time}, \citeA{monfort2017staying} and \citeA{feunou2018risk}. 
Although consistency of QML based on this modification has not been proved, Monte Carlo simulation results in these articles suggest that the method works well in practice. 

Given the large dimension of the observation vector $p=2qk$, the Kalman filter and its QML estimation will be computationally challenging if not infeasible. 
In fact, an important caveat with this approach is that one needs a non-singular innovation variance matrix $G_t$. 
Since our CCF approximation is based on common option price data for $q$ different arguments $u$ and fixed time-to-maturity $\tau$, this matrix is likely to be (near-)singular for large $q$. 
Furthermore, with large cross-sectional dimension, the computation of the inverse matrix for each time $t$ adds a significant computational burden to the estimation procedure. 
To overcome these issues, we consider the collapsed Kalman filter, originally developed by \citeA{jungbacker2015likelihood}, which we describe below.
We modify their method to allow for a (near-)singular variance matrix $H_{t}$, using generalized inverses.

The idea of the collapsed Kalman filter is to transform the observation vector $y_t$ into an uncorrelated pair of vectors $y_t^*$ and $y_t^+$ such that $y_t^*$ depends on the state vector $x_t$ and has dimension $d \times 1$, whereas $y_t^+$ does not depend on $x_t$ and has dimension $(p-d) \times 1$. 
Such a transformation can be done using, for instance, the projection matrices $A_t^* = (Z_t'H_t^- Z_t)^{-1} Z_t'H_t^-$ and $A_t^+ = L_t H_t^-(I_p - Z_t A_t^*)$, where $L_t$ is chosen such that $A_t^+$ has full row rank and where $H^-$ is the generalized inverse of $H$ and $I_p$ is the identity matrix of size $p$. 
Since $A_t^* Z_t = I_p$ and $A_t^+ Z_t = 0$, the observation equation is then transformed into
    \begin{align}\label{collapsing}
        \begin{pmatrix}
            y_t^* \\
            y_t^+ 
        \end{pmatrix} 
        :=  
        \begin{bmatrix}
            A_t^* \\
            A_t^+ 
        \end{bmatrix} y_t
        =
        \begin{pmatrix}
            d_t^* \\
            d_t^+ 
        \end{pmatrix} +     
        \begin{pmatrix}
            x_t \\
            0 
        \end{pmatrix} 
        +   \begin{pmatrix}
            \varepsilon_t^* \\
            \varepsilon_t^+ 
        \end{pmatrix},
    \end{align}
    with 
    $d_t^* = A_t^* d_t$,  $d_t^+ = A_t^+ d_t$, $\varepsilon_t^* = A_t^* \varepsilon_t$ and $\varepsilon_t^+ = A_t^+ \varepsilon_t$. 
Using $H^-HH^-=H^-$, we have
    \begin{align*}
        \mbox{Var} (\varepsilon_t^*) &= A_t^* H_t A_t^{*\prime} = (Z_t'H_t^-Z_t)^{-1} =: H_t^*,\\
        \mbox{Var} (\varepsilon_t^+) &= A_t^+ H_t A_t^{+\prime} =: H_t^+ ,\\
        \mbox{Cov} (\varepsilon_t^*, \varepsilon_t^+) &= A_t^* H_t A_t^{+\prime} 
        = A_t^*H_t (I_p - A_t^{*\prime} Z_t)H_t^-L' \\
        &= A_t^*H_t H_t^- L' - (Z_t'H_t^-Z_t)^{-1} Z_t'H_t^-L' = A_t^* L' - A_t^* L' = 0.
    \end{align*}

In the preceding display, it has been assumed that $\mathrm{rank}(Z_t'H_t^{-}Z_t) =d$; this is not very restrictive, given that the dimension $d$ of the state vector will typically be much smaller than the dimension $p$ of the observation vector.
We also require that the matrix $A_t = [A_{t}^{*\prime},A_{t}^{+\prime} ]'$ is non-singular, such that the transformation $A_t y_t$ does not lead to a loss of information.

The representation~\eqref{collapsing} shows that information about the state vector $x_t$ is contained in the observation equation for $y_t^*$;
thus we may ignore the second equation with $y_t^+$ and focus on the collapsed state space model:
    \begin{align}\label{collapsed_ss1}
        y_t^* &= d_t^* +  x_t + \varepsilon_t^*,  &\E[\varepsilon_t^*|\F_t] &= 0,\ \mbox{Var} (\varepsilon_t^*|\F_t) = H_t^*,\\
        \label{collapsed_ss2}
        x_{t+1} &= c_t + T_t x_t + \eta_{t+1},  &\E[\eta_{t+1}|\F_t] &= 0,\  \mbox{Var} (\eta_{t+1}|\F_t) = Q_t(x_t).
    \end{align}

Let us emphasize that the collapsing transformation into a lower-dimensional state space form is also valid for the Moore-Penrose inverse covariance matrix $H_t^{-}$. 
Therefore, we can collapse a high-dimensional data vector into a lower-dimensional vector even when the covariance system matrix of disturbances is (near-)singular. 

The logarithm of the Gaussian likelihood function of the data vector $Y_T=(y_1', \dots, y_T')'$ is given by
    \begin{align*}
        l(Y_T; \theta) = \sum_{t=1}^T \log p_{\theta}(y_t|Y_{t-1}),
    \end{align*}
    where $p_{\theta}(y_t|Y_{t-1})$ is the (misspecified) Gaussian distribution of $y_t$ conditional on $Y_{t-1}$ (and $w_1,\ldots,w_{t-1}$), which can be evaluated via the prediction error decomposition based on the original state space representation \eqref{ss-observ}--\eqref{ss-state}. 
Given the assumption of a full rank transformation matrix $|A_t|$,
the collapsed transformation allows to decompose the log-likelihood function $l(Y_T; \theta)$ into three parts to ease computation:
    \begin{align}\label{loglike}
        l(Y_T; \theta) = l(Y^*_T; \theta) + l(Y_T^+; \theta) + \sum_{t=1}^T \log |A_t|,
    \end{align}
    where $Y_T^*$ and $Y_T^+$ are stacked vectors of $y_t^*$ and $y_t^+$ over $t=1,\dots,T$, respectively.

The first term in~\eqref{loglike} is the quasi-loglikelihood evaluated by the Kalman filter applied to the collapsed state space system \eqref{collapsed_ss1}--\eqref{collapsed_ss2}:
    \begin{align*}
        l(Y_T^*; \theta) = -\frac{dT}{2} \log 2\pi - \frac{1}{2} \sum_{t=1}^T \log |G_t^*| - \frac{1}{2} \sum_{t=1}^T \omega_t^{*\prime}G_t^{*-1} \omega_t^{*},
    \end{align*}
    where $\omega_t^{*}$ are the prediction errors and $G_t^*$ are their mean square error matrices from the Kalman filter.

Since $y_t^+$ does not depend on the state vector $\alpha_t$ and $|H_t^+|=1$ may be imposed without loss of generality, 
the second term in~\eqref{loglike} is given by
    \begin{align*}
        l(Y_T^+; \theta) = -\frac{(p-d)T}{2} \log 2 \pi - \frac{1}{2} \sum_{t=1}^T (y_t^{+} - d_t^+)'(H_t^{+})^{-1} (y_t^{+} - d_t^+).
    \end{align*}
Fortunately, the last term in the expression above can be calculated without construction of the matrix $A_t^+$: 
    \begin{align*}
        (y_t^{+} - d_t^+)'(H_t^{+})^{-1} (y_t^{+} - d_t^+) 
        &= (y_t - d_t)'A_t^{+\prime}(A_t^{+}H_t A_t^{+})^{-1} A_t^{+}(y_t - d_t)\\
        &= (y_t - d_t)' J_t^{+} H_t^{-} (y_t - d_t)\\
        &= (y_t - d_t)' J_t^{+} H_t^{-} J_t^{+\prime}(y_t - d_t)\\
        &= (y_t - d_t)' M_Z' H_t^{-} M_Z (y_t - d_t)\\
        &= e_t 'H_t^{-} e_t,
    \end{align*}
    where $M_Z = I - Z_t(Z_t'H_t^-Z_t)^{-1}Z_t'H_t^- = I - Z_t A_t^*$,\ $J_t^+ = A_t^{+\prime}(A_t^{+}H_t A_t^{+})^{-1} A_t^{+} H_t$ and $e_t = M_Z (y_t - d_t)$, that is, these are the generalized least squares (GLS) residuals from the observation vector $y_t$ with the covariate matrix $Z_t$ and variance matrix $H_t$. For derivation details,\footnote{ 
    The derivation in~\citeA{jungbacker2015likelihood} is based on the invertible covariance matrix $H_t$, but the same result and the same derivation are valid when using the pseudo-inverse matrix $H_t^-$.
    } see \citeA{jungbacker2015likelihood}.

 Finally, the third term in~\eqref{loglike}, $|A_t|$, can be found from the relation
    \begin{align}\label{detA}
        |A_t|^2 \cdot |H_t| = |A_t H_t A_t'| = |H_t^*|\cdot |H_t^+| = |H_t^*|,
    \end{align}
    which follows from the fact that the covariance matrix $A_t H_t A_t'$ is block diagonal given the uncorrelated error terms $\varepsilon_t^*$ and $\varepsilon_t^+ $ and using again $|H_t^+|=1$. 

Given the measurement error structure as implied by Proposition~\ref{prop1}, the single scale parameter $\sigma_\varkappa^2$ of the covariance matrix can be factored out as
$H_t = \sigma_\varkappa^2 \cdot \widetilde{H}_t$. 
The matrix $\widetilde{H}_t$ has a block-diagonal structure; although its blocks depend on the state vector and parameters via the theoretical BSIV $\kappa_t(\tau, m)$ and vega $\nu_t (\tau,m)$, we estimate these quantities directly from the data, hence they are not updated as we optimize over $\theta$. 
Thus, we have from \eqref{detA} that 
    \begin{align*}
        \log |A_t| 
        &= \frac{1}{2}\left(\log |H_t^*| - \log |H_t| \right)\\
        &= \frac{1}{2}\left(\log |H_t^*| - \log \sigma_\varkappa^{2p} - \log |\widetilde{H}_t| \right)\\
        &\propto \frac{1}{2} \log |H_t^*| - p \log \sigma_\varkappa.
    \end{align*}
Therefore, the log-likelihood~\eqref{loglike} is proportional to 
    \begin{align}\label{loglike_proprto}
        l(Y_T; \theta) \propto \frac{1}{2}
        \sum_{t=1}^T \left( -\log |G_t^*| - \omega_t^{*\prime}G_t^{*-1} \omega_t^{*}
        -  e_t 'H_t^{-} e_t
        + \log |H_t^*| \right) - pT \log \sigma_\varkappa.
    \end{align}
Note that the inversions and determinants of the matrices $G_t^*$ and $H_t^*$ can be computed efficiently since they have small dimensions $d\times d$. 
This eases maximization of the log-likelihood function~\eqref{loglike_proprto} substantially. 

The quasi maximum-likelihood parameter estimates $\widehat{\theta}$ are obtained by maximizing~\eqref{loglike_proprto} over the model parameter space $\Theta$, where we implicitly assume that the parameter vector $\theta$ has been extended to include the additional parameter $\sigma_{\varkappa}^2$. 
Its asymptotic properties are analogous to QML estimation based on the (modified) Kalman filter, as discussed at the beginning of this sub-section. 
In cases in which the conditional covariance matrix $Q_t$ does not depend on the latent state vector $x_t$, and the latent state process $x_t$ is stationary, 
QML based on the Kalman filter will yield consistent and asymptotically normal estimators. 
When $Q_t$ is affine in $x_t$, 
then QML based on the modified Kalman filter appears to have comparable properties in Monte Carlo simulations, 
but no formal consistency proof is available.

\section{Monte Carlo Study}
\label{sec:Simulation}

In this section, we study the finite-sample performance of our estimation procedure. 
In particular, we consider two AJD specifications: a one-factor model and two versions of an option pricing model with two factors.

\subsection{SVCDEJ}
\label{sec:Simulation-svcdej}

As a starting point, we illustrate the developed estimation procedure based on a modification of the widely used `double-jump' stochastic volatility model of~\citeA{DPS2000}. 
The modification is due to using double-exponential (rather than Gaussian) jump sizes in returns as in~\citeA{kou2002jump} and \citeA{andersen2015parametric}, and a stochastic (rather than constant) jump intensity that is a multiple of the stochastic variance as in~\citeA{pan2002}. 
We label this specification as `SVCDEJ' for stochastic volatility model with co-jumps in volatility and double-exponential jumps in returns.
    
In particular, we assume the following data-generating process for the log forward price under both the $\P$ and $\Q$ probability measures:
    \begin{align}
        \label{SVCDEJ-y}
        \diff \log F_t &= ( -\tfrac{1}{2}v_t - \mu \lambda_t) \dt + \sqrt{v_t} \diff W_{1,t} + J_t \diff N_t,\\
        \label{SVCDEJ-v}
        \diff v_t &= \kappa(\bar{v} - v_t) \dt + \sigma \sqrt{v_t} \diff W_{2,t} + J_t^v \mathbf{1}_{\{J_t <0\}} \diff N_t,
    \end{align}
    where the two standard Brownian motions $W_1$ and $W_2$ are assumed to be correlated with coefficient $\rho\in[-1,1]$, and $N_t$ is a Poisson jump process with jump intensity proportional to the stochastic variance, $\lambda_t = \delta v_t$, $\delta>0$. 
We further assume that $J_t$ is a double-exponentially distributed jump size with generic probability density function 
    \begin{equation*}
        f_J(x) = p^+ \frac{1}{\eta^+} e^{-x/\eta^{+}} \mathbf{1}_{\{x \geq 0\}} + p^- \frac{1}{\eta^-} e^{x/\eta^{-}} \mathbf{1}_{\{x < 0\}},   
    \end{equation*}
    where $p^+$ and $p^-$ are probabilities of positive and negative jumps, respectively, and $\eta^+$ and $\eta^-$ are the corresponding conditional means of the jump sizes. 
We assume that all of these parameters are positive, $p^+ + p^- =1$ and $\eta^+<1$. 
Given the jump size distribution, the expected relative jump size in returns is 
    \begin{equation*}
    \mu := \E[e^{J} {-} 1] =  \frac{p^+}{1-\eta^{+}} + \frac{p^-}{1+\eta^{-}} - 1.
    \end{equation*}
We allow the volatility to co-jump only with negative jumps in returns, with exponentially distributed jump sizes $J_t^v$ with mean $\mu_v>0$. 
Finally, we assume $\kappa$, $\bar{v}$ and $\sigma$ to be positive
and impose Feller's condition $2 \kappa \bar{v} > \sigma^2$ and the covariance stationarity condition $\kappa > p^- \delta \mu_v$.

The model in~\eqref{SVCDEJ-y}--\eqref{SVCDEJ-v} belongs to the AJD class and exhibits all important ingredients of option pricing models: stochastic volatility, jump components in returns and volatility, time-varying jump intensity and a self-excitation feature (because a negative jump in returns is associated with a positive jump in volatility, which increases the volatility and hence the jump intensity). 
Furthermore, this specification assumes a double-exponential jump size distribution in returns, which has recently been  advocated in the literature (see, e.g., \citeNP{kou2002jump}, \citeNP{ait2015contagion}, \citeNP{andersen2015parametric} and \citeNP{bardgett2019inferring}). 

Our developed estimation and filtering approach uses information from option prices, and is agnostic about equity risk premia. 
Indeed, the measurements are constructed as portfolios of options rather than the underlying asset. 
On the other hand, since the transition equation in the state space representation reflects the dynamics of the latent components (under $\P$), it is, in principle, possible to learn about the risk premia associated with the latent processes (for instance, the variance risk premium). 
However, additional simulation results, reported in Appendix~\ref{sec:Appendix-SimulationEmpirical}, suggest that the $\Q$-information in the option prices largely dominates the $\P$-information, making the identification of risk premium parameters weak. 
A similar difficulty of identifying the physical dynamics arises in the term structure literature (see, e.g., \citeNP{kim2012term}). 
Therefore, we assume no variance (or state related) risk premia, that is, the latent components have the same dynamics under both probability measures. 
Importantly, the results in 
Appendix~\ref{sec:Appendix-SimulationEmpirical} suggest that estimation of the $\Q$-parameters is hardly affected by imposing this (possibly invalid) restriction.

The discounted marginal CCF of the log forward prices in the SVCDEJ model can be derived using the results in \citeA{DPS2000} and is given by
    \begin{align}\label{SVCDEJ-CCF}
        \psi_X(\mathbf{u}_1, \tau) = e^{-r\tau} \E^{\Q}[ e^{ \mathrm{i} u \log F_{t+\tau}} \rvert \F_t] = e^{\alpha(u,\tau) + \beta_1(u,\tau) \log F_t + \beta_2(u,\tau) v_t},
    \end{align}
    where $\alpha(u, \tau)$ and $\beta(u, \tau)$ are solutions to the complex-valued ODE system in time:
    \begin{align*}
        \begin{cases}
            \dot \beta_1(u,s) &= 0, \\
            \dot \beta_2(u,s) &= {-}\left(\frac{1}{2} + \mu \delta \right)\beta_1(u,s) - \kappa \beta_2(u,s) + \frac{1}{2}\beta_1(u,s)  + \rho \sigma \beta_1(u,s) \beta_2(u,s) \\ 
            &\quad + \frac{1}{2} \sigma^2 \beta_2^2(u,s) 
             + \delta (\chi(\beta_1(u,s), \beta_2(u,s)) - 1), \\
            \dot \alpha(u,s) &= \kappa \bar{v} \beta_2(u, s) - r,
        \end{cases}
    \end{align*}
    with initial conditions $\beta_1(u,0) = \mathrm{i} u,\ \beta_2(u,0) = 0$ and $\alpha(u,0) = 0$. 
Here the `jump transform' takes the form
    \begin{align*}
        \chi(\beta_1, \beta_2) = \frac{p^+}{1-\beta_1 \eta^+} + \frac{p^-}{(1 + \beta_1 \eta^-)(1 - \beta_2 \mu_v)}.
    \end{align*}

The CCF of the log price in~\eqref{SVCDEJ-CCF} is used to price options. 
For the state space representation, we turn it into the CCF of log returns as described in Section~\ref{sec:Framework3}.
Using the fact that the solution to the ODE system satisfies $\beta_1(u,\tau)=\mathrm{i}u$, the linear relation between the log of the option-implied CCF and the state vector is given by 
    \begin{align*}
        \log \widehat \phi_t(u, \tau) = \alpha(u, \tau) + \beta_2(u, \tau) v_t + \xi_t(u, \tau), \qquad u \in \R,
    \end{align*} 
    where $\widehat \phi_t(u, \tau)$ is the option-implied CCF, $\tau > 0$ is the time-to-maturity of available options and $\xi_t(u, \tau)$ is the measurement error term due to observation and approximation errors in the option-implied CCF. 
We use this linear relation to construct the measurement equation as discussed in Section~\ref{sec:Estimation-ss}. 

Following Appendix~\ref{sec:Appendix-CM}, the conditional mean and variance of the latent stochastic volatility process are given by
    \begin{align}\label{mean_svcdej}
        \E[v_{t+1}| \F_t] &= e^{g_1\Delta t} v_t + \frac{g_0}{g_1}\left(e^{g_1\Delta t} - 1 \right),\\
        \label{var_svcdej}
        \mbox{Var}(v_{t+1} | \F_t) &= - \frac{\sigma^2 + 2 p^-\delta  \mu_v^2}{2g_1^2}  \left[ 2 g_1\left(e^{g_1\Delta t} - e^{2g_1\Delta t} \right) v_t - g_0\left( 1 - e^{g_1\Delta t} \right)^2\right],
    \end{align}
    with $g_0 = \kappa \bar{v}$ and $g_1 = -\kappa + p^-\delta \mu_v$. 
Equations~\eqref{mean_svcdej}--\eqref{var_svcdej} are then used to define the state updating equation:
    \begin{align}
        v_{t+1} = c_t + T_t v_t + \eta_{t+1},
    \end{align}
    where $c_t =\frac{g_0}{g_1}\left(e^{g_1\Delta t} - 1 \right),\ T_t = e^{g_1\Delta t}$ and $\mbox{Var}(\eta_{t+1}| \F_t) =\mbox{Var}(v_{t+1}|\F_t) =: Q_t(v_t)$. 

The model specification has nine parameters of interest and one additional parameter that characterizes the observation errors. 
We note that the parameter $\delta$ often enters as a multiple of $p^-$, which can possibly cause identification issues in the estimation procedure. 
Therefore, to avoid these identification issues, we fix the probability of negative jumps to be $p^- = 0.7$.
This is consistent with findings in~\citeA{ait2015contagion} and our empirical results for the unrestricted model provided in Appendix~\ref{sec:Appendix-Empirical}, 
where we also assess the robustness of our empirical results to fixing $p^- = 0.7$.

In the simulation study, we use $T=500$ time points with $\Delta t =1/250$. 
The time-series of the log prices and true spot volatilities are simulated using an Euler scheme applied to the specification~\eqref{SVCDEJ-y}--\eqref{SVCDEJ-v}. 
The initial values are set to $F_0 = 100$ and $v_0=0.015$.
The options data are generated using the COS method of~\citeA{fang2008novel} based on the CCF, specified in~\eqref{SVCDEJ-CCF}. 
The true model parameters are displayed in Table~\ref{tab:svcdej-mc}.   

In the simulations, we consider three tenors for options equal to 10, 30 and 60 days. 
For each tenor, we simulate a finite number of options with log-moneyness between $\underline{m} = -10 \cdot \sigma_{ATM, \tau} \sqrt{\tau}$ and $\overline{m} = 4 \cdot \sigma_{ATM, \tau} \sqrt{\tau}$, where $\sigma_{ATM,\tau}$ is the BSIV of the ATM option with time-to-maturity $\tau$. 
Furthermore, the strikes are generated equidistantly with $\Delta K = 0.01\cdot F_t$. 
Finally, we distort the options data by adding the observation errors to the option prices for each tenor $\tau$ and each log-moneyness level $m$ as specified in Assumption~\ref{ass:option_errors}, i.e.,
    \begin{align*}
        \widehat{O}_t(\tau, m) = O_t(\tau, m) + \sigma_{\varkappa}\cdot \kappa_t(\tau, m) \nu_t(\tau, m) \cdot \epsilon,
    \end{align*}
    where $\epsilon$ is an i.i.d.\ standard normal random variable and $\sigma_{\varkappa}=0.02$. 
The distorted option prices, in terms of total implied variance, are then interpolated using a cubic spline and extrapolated linearly in log-moneyness, as described in Appendix~\ref{sec:Appendix-Extrapolation}.  

    \begin{table}[htbp]
        \centering
        \scriptsize
        \caption{Monte Carlo results for the SVCDEJ model}
          \begin{tabular}{lccccccccc}
            \toprule
            parameter & $\sigma$ & $\kappa$ & $\bar{v}$ & $\rho$ & $\delta$  & $\eta^+$ & $\eta^-$ & $\mu_v$ & $\sigma_{\varkappa}$ \\ [1ex]
          \multicolumn{10}{c}{$u=1,\ldots,5$} \\
          \midrule
          true value & 0.450 & 8.000 & 0.0150 & -0.9500 & 100.000 & 0.020  & 0.050  & 0.050  & 0.020 \\
          mean  & 0.498 & 8.213 & 0.0151 & -0.8983 & 105.906 &  0.021 & 0.048 & 0.048 & 0.032 \\
          std dev   & 0.007 & 0.257 & 0.0005 & 0.0088 & 7.567 &    0.001 & 0.001 & 0.001 & 0.005 \\
          q10   & 0.494 & 7.965 & 0.0146 & -0.9097 & 96.930 &     0.021 & 0.047 & 0.046 & 0.026 \\
          q50   & 0.498 & 8.198 & 0.0151 & -0.8956 & 105.936 &    0.021 & 0.048 & 0.048 & 0.032 \\
          q90   & 0.504 & 8.503 & 0.0158 & -0.8904 & 114.125 &    0.022 & 0.049 & 0.049 & 0.038 \\ [1ex]
          \multicolumn{10}{c}{$u=1,\ldots,10$} \\
          \midrule
          true value & 0.450 & 8.000 & 0.0150 & -0.9500 & 100.000 &    0.020  & 0.050  & 0.050  & 0.020 \\
          mean  & 0.440 & 8.779 & 0.0136 & -0.9968 & 136.200 &    0.023 & 0.045 & 0.043 & 0.035 \\
          std dev  & 0.017 & 0.310 & 0.0007 & 0.0119 & 13.316 &       0.001 & 0.002 & 0.002 & 0.006 \\
          q10   & 0.427 & 8.409 & 0.0130 & -1.0000 & 118.640 &      0.022 & 0.043 & 0.041 & 0.027 \\
          q50   & 0.437 & 8.836 & 0.0134 & -1.0000 & 139.079 &      0.023 & 0.045 & 0.043 & 0.036 \\
          q90   & 0.451 & 9.102 & 0.0142 & -1.0000 & 150.730 &      0.024 & 0.047 & 0.046 & 0.042 \\ [1ex]
          \multicolumn{10}{c}{$u=1,\ldots,15$} \\
          \midrule
          true value & 0.450 & 8.000 & 0.0150 & -0.9500 & 100.000 & 0.020  & 0.050  & 0.050  & 0.020 \\
          mean  & 0.440 & 8.723 & 0.0139 & -0.9942 & 128.617 &   0.022 & 0.046 & 0.045 & 0.027 \\
          std dev  & 0.013 & 0.316 & 0.0005 & 0.0151 & 10.260 &     0.001 & 0.001 & 0.002 & 0.005 \\
          q10   & 0.428 & 8.330 & 0.0134 & -1.0000 & 113.412 &   0.021 & 0.044 & 0.043 & 0.022 \\
          q50   & 0.438 & 8.741 & 0.0138 & -1.0000 & 130.575 &   0.022 & 0.045 & 0.044 & 0.028 \\
          q90   & 0.455 & 9.102 & 0.0144 & -0.9785 & 139.999 &   0.023 & 0.048 & 0.047 & 0.033 \\ [1ex]
          \multicolumn{10}{c}{$u=1,\ldots,20$} \\
          \midrule
          true value & 0.450 & 8.000 & 0.0150 & -0.9500 & 100.000 &  0.020  & 0.050  & 0.050  & 0.020 \\
          mean  & 0.455 & 8.143 & 0.0147 & -0.9558 & 110.734 &   0.022 & 0.048 & 0.046 & 0.023 \\
          std dev  & 0.007 & 0.205 & 0.0003 & 0.0126 & 4.847 &      0.001 & 0.001 & 0.001 & 0.006 \\
          q10   & 0.449 & 7.919 & 0.0144 & -0.9707 & 105.192 &    0.021 & 0.047 & 0.045 & 0.018 \\
          q50   & 0.454 & 8.142 & 0.0147 & -0.9563 & 110.689 &      0.021 & 0.048 & 0.046 & 0.022 \\
          q90   & 0.461 & 8.389 & 0.0150 & -0.9430 & 116.719 &     0.022 & 0.049 & 0.047 & 0.027 \\ [1ex]
          \multicolumn{10}{c}{$u=1,\ldots,25$} \\
          \midrule
          true value & 0.450 & 8.000 & 0.0150 & -0.9500 & 100.000 & 0.020  & 0.050  & 0.050  & 0.020 \\
          mean  & 0.460 & 7.918 & 0.0150 & -0.9404 & 105.489 &   0.021 & 0.049 & 0.047 & 0.023 \\
          std dev  & 0.008 & 0.188 & 0.0003 & 0.0126 & 3.881 &      0.000 & 0.001 & 0.001 & 0.008 \\
          q10   & 0.453 & 7.697 & 0.0148 & -0.9492 & 103.098 &   0.021 & 0.048 & 0.046 & 0.017 \\
          q50   & 0.459 & 7.958 & 0.0150 & -0.9449 & 105.744 &   0.021 & 0.049 & 0.047 & 0.020 \\
          q90   & 0.468 & 8.082 & 0.0152 & -0.9229 & 108.351 &   0.022 & 0.050 & 0.048 & 0.032 \\ [1ex]
          \multicolumn{10}{c}{$u=1,\ldots,30$} \\
          \midrule
          true value & 0.450 & 8.000 & 0.0150 & -0.9500 & 100.000 & 0.020  & 0.050  & 0.050  & 0.020 \\
          mean  & 0.457 & 7.815 & 0.0149 & -0.9432 & 111.115 &   0.022 & 0.049 & 0.044 & 0.026 \\
          std dev  & 0.011 & 0.209 & 0.0003 & 0.0190 & 5.794 &     0.001 & 0.001 & 0.002 & 0.010 \\
          q10   & 0.448 & 7.523 & 0.0147 & -0.9596 & 104.771 &    0.021 & 0.048 & 0.042 & 0.019 \\
          q50   & 0.453 & 7.857 & 0.0148 & -0.9514 & 111.854 &   0.022 & 0.048 & 0.044 & 0.023 \\
          q90   & 0.472 & 8.016 & 0.0151 & -0.9105 & 116.199 &     0.023 & 0.049 & 0.046 & 0.038 \\
          \bottomrule
          \end{tabular}%
        \label{tab:svcdej-mc}%
          
        \medskip
        \begin{minipage}{0.85\textwidth}\scriptsize
            Note: This table provides Monte Carlo simulation results for the SVCDEJ model, based on 300 replications. 
            Six settings with different ranges of the argument $u$ are considered. 
            Each panel lists, for each parameter,
            the true value, the Monte Carlo mean and standard deviation, and
            the 10th, 50th and 90th Monte Carlo percentiles, respectively.
            We use $T=500$ time points with $\Delta t =1/250$. 
            The initial values are set to $F_0 = 100$ and $v_0=0.015$. 
            The threshold for singular values is set to $\bar{s} = 10^{-7}$. 
            The probability of negative jumps is fixed to $p^-=0.7$.
        \end{minipage}
      \end{table}%

The covariance matrix of the errors in the measurement equation is calculated according to equation~\eqref{complex_covar}. 
To calculate the pseudo-inverse of the $2q \times 2q$ covariance matrix $\widetilde{H}_{t,i}$ for each of the maturities $i=1,\dots,k$, we set the following level of the threshold for the singular values:
    \begin{align*}
        \mbox{tol} := \bar{s} \cdot 2q \cdot \max_j s_j,
    \end{align*}
    with $\bar{s} = 10^{-7}$ and where the maximum is taken over all singular values $s_j$ of $\widetilde{H}_{t,i}$. 
We also analyze the robustness of our results to the choice of $\bar{s}$.

With these specifications, we take the number of replications $N$ to be $N=300$,
thus running the estimation procedure of Section~\ref{sec:Estimation} 300 times.
Table~\ref{tab:svcdej-mc} provides the Monte Carlo results for the SVCDEJ model, for six different ranges of the argument set $\mathcal{U}$. 
The results in general show a good finite-sample performance. 
We notice that for smaller ranges of the CCF argument, the estimates exhibit biases for some model parameters. 
This is expected since the smaller ranges provide coarser information on which we build the filtering and parameter estimation procedures. 
On the other hand, we also notice that the variance of some parameter estimates increases when using a very large range of arguments (in particular, $u=1,\dots,30$). 
This is likely due to an increased variance in the CCF approximation for large arguments $u$.

\begin{figure}
        \centering
        \caption{RMSPE for different levels of $\bar{s}$}
        \vspace*{-0.45cm}
        \includegraphics[scale=0.5]{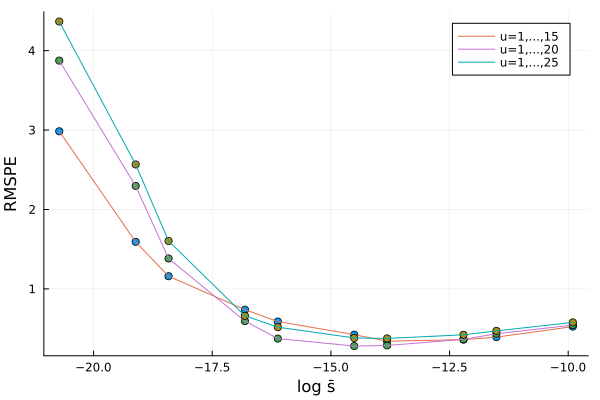}

        \label{fig:tols}

        \medskip
        
        \begin{minipage}{0.75\textwidth}\scriptsize
            Note: This figure plots the RMSPE for different levels of $\bar{s}$, considering three different ranges of the argument $u$. 
            In particular, we estimate $N=300$ replications for each $\bar{s} = \{10^{-9}, 5{\cdot}10^{-9}, 10^{-8}, 5{\cdot}10^{-8}, 10^{-7}, 5{\cdot}10^{-7}, 10^{-6}, 5{\cdot}10^{-6}, 10^{-5}, 5{\cdot}10^{-5}\}$ and plot the resulting RMSPEs against $\bar{s}$ in log-scale.
        \end{minipage}
    \end{figure}

To explore the robustness to the choice of the truncation level in the pseudo-inversion of the covariance matrix, we also consider other values of $\bar{s}$. 
In particular, we run $N=300$ simulations for each level of $\bar{s}$ using the same parameter values as in Table~\ref{tab:svcdej-mc}, and construct the root mean square percentage error (RMSPE) metrics, defined as the square root of
$N^{-1} \sum_{i=1}^N \sum_{j=1}^{d_{\theta}} \left ( (\widehat{\theta}_{i,j} - \theta_{0,j})/\theta_{0,j} \right ) ^2 $, with $d_{\theta}$ the dimension of $\theta$.
Figure~\ref{fig:tols} plots the resulting RMSPEs for different levels of $\bar{s}$ and three different ranges of the argument $u$. 
As we can see, the levels $\bar{s}$ in between $10^{-7}$ and $10^{-6}$ yield the smallest RMSPE. 
In the following simulations and empirical applications, we therefore set $\bar{s} = 10^{-7}$.

We end this subsection by noting that we have also conducted simulation studies for some related alternative one-factor specifications. 
In particular, in Appendix~\ref{sec:Appendix-SimulationEmpirical}, we provide additional simulation results for the `SVCJ' model with a Gaussian jump size distribution, 
and the `SVCEJ' model with two separate counting processes for positive and negative jumps. 
The former shows a very good finite-sample performance, while the latter, a richer model specification, shows reasonable results, gradually reaching the limits of what can be identified using the present input data and design.
    
\subsection{SVCDEJ with external factors}
\label{sec:Simulation-svcdejex}

Now we extend the one-factor specification by adding an external factor. 
This modification can be seen as a two-factor specification, but we will assume that the second factor is observable. 
The motivation comes from the fact that in some situations we might have an understanding of possible drivers of the risks in the market. 
Therefore, we would like to embed exogenous variables into the model's risk factors and quantify their impact.

In particular, next to the stochastic volatility component we introduce the exogenous factor $h_t$, which affects the intensity of jumps and the diffusive component. 
The model reads as follows:
    \begin{align}
        \label{SVCDEJex-y}
        \diff \log F_t &= ( -\tfrac{1}{2}V_t - \mu \lambda_t ) \dt + \sqrt{v_t} \diff W_{1,t} + q\sqrt{h_t} \diff W_{3,t} + J_t \diff N_t,\\
        \label{SVCDEJex-v}
        \diff v_t &= \kappa(\bar{v} - v_t) \dt + \sigma \sqrt{v_t} \diff W_{2,t} + J_t^v \mathbf{1}_{\{J_t <0\}} \diff N_t,\\ \label{SVCDEJex-h}
        \diff h_t &= \kappa_h(\bar{h} - h_t) \dt + \sigma_h \sqrt{h_t} \diff W_{4,t},
    \end{align}
    where $V_t = v_t + q^2 h_t$ is the total diffusive variance of the process and the jump intensity process $\lambda_t$ is also affected by $h_t$ with $\lambda_t = \delta v_t + \gamma h_t$, $q,\delta,\gamma>0$. 
We assume that $W_{3,t}$ and $W_{4,t}$ are independent standard Brownian motions, jointly independent of $(W_{1,t},W_{2,t})$. 
The process $h_t$ is exogenous to the SVCDEJ dynamics, meaning that the dynamics of $\log F_t$ and $v_t$ do not affect the dynamics of $h_t$. 
In turn, the exogenous factor $h_t$ affects the intensity of jumps and the diffusive component of the log return dynamics. 
This specification is similar to the two-factor model in~\citeA{andersen2015parametric}, which includes short- and long-term stochastic volatility components. 
The difference is that here the exogenous process $h_t$ is observable, although its parameters are unknown.

In the Monte Carlo simulations, we consider two possible estimation approaches. 
In the first approach, we assume a correct specification of the dynamics of $h_t$ with known true parameters $\kappa_h,\ \bar{h}$ and $\sigma_h$. 
In practice, these parameters can be pre-estimated given the observed path of the exogenous process. 
In the second approach, we estimate the misspecified model in which the contribution of $h_t$ is constant throughout the maturity of an option. 
In other words, under this approach we ignore the dynamics of $h_t$ when pricing options, but let $h_t$ still affect the level of the jump intensity and of the total variance. 
The motivation is that when the exogenous process is persistent and smooth relative to $v_t$, its dynamics can be neglected when pricing options with short expiration periods. 
In a similar way, interest rates are often assumed to enter option prices in a deterministic way.
Moreover, the true parametric specification for an exogenous variable is likely unknown in practice, but if its dynamics are persistent and smooth, we can find its effect on option prices via this approach. 
Therefore, in the Monte Carlo experiment, we simulate $h_t$ with a mean-reversion rate that is smaller than that in $v_t$, mimicking the specification we will explore in the empirical application.

For the rest, the Monte Carlo setting for the SVCDEJ model with an external factor is the same as for the SVCDEJ specification in the previous subsection. 
The parameters of the external factor are set to $\kappa_h = 1$, $\bar{h}=1$ and $\sigma_h=0.1$. 
The simulation results are provided in Table~\ref{tab:svcdejex_mc}. 
The parameters of the SVCDEJ model exhibit similar good performance under both estimation approaches. Importantly, the parameters related to the external factors, $\gamma$ and $q$, also show similar good performance in the correctly specified model as in the misspecified setting. 
We emphasize that this is achieved due to simulating a relatively smooth and persistent exogenous process $h_t$ and using short-dated options in the estimation procedure. 

    \begin{table}
        \centering
        \scriptsize
        \caption{Monte Carlo results for the SVCDEJ model with external factor}
        \begin{tabular}{lccccccccccc}

            \toprule
            parameter & $\sigma$ & $\kappa$ & $\bar{v}$ & $\rho$ &  $\delta$ &  $\eta^+$ & $\eta^-$ & $\mu_v$ & $\gamma$ & $q$ & $\sigma_{\varkappa}$ \\ [1ex]
            \multicolumn{12}{c}{(a) Estimation with known true parameters for $h_t$} \\
            \midrule
            true value & 0.450  & 8.00  & 0.0150 & -0.950 & 100.00 &  0.0200 & 0.050  & 0.050  & 1.500 & 0.050 & 0.020 \\
            mean  & 0.452 & 8.15  & 0.0150 & -0.939 & 116.21 &  0.0211 & 0.049 & 0.046 & 1.480 & 0.046 & 0.024 \\
            std dev  & 0.034 & 0.36  & 0.0019 & 0.047 & 11.37 &  0.0008 & 0.001 & 0.002 & 0.167 & 0.005 & 0.011 \\
            q10   & 0.423 & 7.65  & 0.0140 & -0.977 & 102.67 &  0.0206 & 0.048 & 0.044 & 1.348 & 0.042 & 0.016 \\
            q50   & 0.442 & 8.22  & 0.0145 & -0.954 & 117.07 &  0.0209 & 0.048 & 0.045 & 1.531 & 0.048 & 0.019 \\
            q90   & 0.502 & 8.52  & 0.0167 & -0.857 & 128.02 &  0.0218 & 0.050 & 0.047 & 1.565 & 0.049 & 0.043 \\ [1ex]     
            \multicolumn{12}{c}{(b) Estimation of misspecified model} \\
            \midrule
            true value & 0.450  & 8.00  & 0.0150 & -0.950 & 100.00 &  0.0200 & 0.050  & 0.050  & 1.500 & 0.050 & 0.020 \\
            mean  & 0.450 & 8.16  & 0.0150 & -0.939 & 116.93 &  0.0210 & 0.049 & 0.046 & 1.473 & 0.046 & 0.024 \\
            std dev  & 0.034 & 0.37  & 0.0022 &  0.044 & 11.47 &    0.0016 & 0.001 & 0.002 & 0.165 & 0.006 & 0.012 \\
            q10   & 0.425 & 7.70  & 0.0140 & -0.973 & 106.71 &  0.0206 & 0.048 & 0.044 & 1.380 & 0.043 & 0.016 \\
            q50   & 0.441 & 8.23  & 0.0146 & -0.954 & 117.78 &  0.0209 & 0.048 & 0.045 & 1.516 & 0.047 & 0.019 \\
            q90   & 0.492 & 8.55  & 0.0159 & -0.861 & 128.68 &  0.0214 & 0.049 & 0.047 & 1.546 & 0.048 & 0.043 \\       
            \bottomrule
        \end{tabular}%
        \label{tab:svcdejex_mc}%

        \medskip
        \begin{minipage}{0.9\textwidth}\scriptsize
            Note: This table provides Monte Carlo simulation results for the SVCDEJ model with an exogenous factor, based on 300 replications. 
            Each panel lists, for each parameter, the true value, the Monte Carlo mean and standard deviation, and the 10th, 50th and 90th Monte Carlo percentiles, respectively. 
            We use $T=500$ time points with $\Delta t =1/250$. 
            The range for the arguments is set to $u=1,\dots,20$ and the threshold to $\bar{s} = 10^{-7}$. 
            The initial values are set to $F_0 = 100$ and $v_0=0.015$. 
            The probability of negative jumps is fixed to $p^-=0.7$. 
            The parameters of the external factor are set to $\kappa_h = 1$, $\bar{h}=1$ and $\sigma_h=0.1$.
        \end{minipage}
    \end{table}
    
\section{Data}
\label{sec:Data}

This section describes the data and the data selection process, which we use in our empirical application. 
Since our estimation procedure utilizes option-implied CCFs, we also pay attention to the construction of these objects in this section.
Further details are in Appendix~\ref{sec:Appendix-InterExtraSpanning}.

\subsection{Data description}

In this paper, we use options data on the S\&P 500 stock market index obtained from the Chicago Board Options Exchange (CBOE).
We focus on the period from May 1, 2017, to April 1, 2021, covering in particular the turbulent period in the stock market due to the outbreak of the Covid-19 pandemic. 
The CBOE provides end-of-day option quotes and a snapshot at 3:45 pm ET, 15 minutes prior to the market closure. 
We use the latter to calculate mid-quotes since it is considered to be a more accurate representation than the former in view of market liquidity. 
The data contain both the `standard' AM-settled SPX options and Weeklys and End-of-Months PM-settled SPXW products. 
The settlement value for the SPX options is based on the opening level of the S\&P 500 index on the settlement day, 
whereas for the SPXW options it is based on the closing prices of the index.

    \begin{table}[h]
        \centering
        \caption{Descriptive statistics for S\&P 500 index options}
        \footnotesize
        \begin{tabular}{lcccccc}
            \toprule
             &  $k \leq 0.8$ &  $0.8 < k \leq 0.95$ &  $0.95< k \leq 1.03$ &  $1.03 <k\leq 1.1$ &  $1.1 < k$ &  Total \\
            \midrule
            & \multicolumn{6}{c}{Panel A: Total volume of option contracts (in millions)} \\
            \midrule
            $\tau \in (2, 9]$     &        4.52 &        52.61 &        167.09 &        13.27 &        1.25 &     238.73 \\
            $\tau \in (9, 30]$    &       25.47 &        74.58 &        168.06 &        29.93 &        3.96 &     302.00 \\
            $\tau \in (30, 60] $  &       25.85 &        61.97 &        109.53 &        28.80 &        4.22 &     230.38 \\
            $\tau \in (60, 90]  $ &       11.68 &        21.70 &         40.82 &        11.05 &        3.05 &      88.30 \\
            $\tau \in (90, 180]$  &       19.93 &        22.83 &         27.23 &        10.59 &        4.57 &      85.15 \\
            $\tau \in (180, 365]$ &       10.74 &        10.91 &         10.82 &         4.54 &        5.12 &      42.12 \\
            Total  &       98.19 &       244.60 &        523.55 &        98.18 &       22.16 &     986.68 \\
            \midrule
            & \multicolumn{6}{c}{Panel B: Volume of OTM option contracts (in millions)} \\
            \midrule
            $\tau \in(2, 9]$     &        4.12 &        52.03 &        139.33 &        11.94 &        0.96 &     208.37 \\
            $\tau \in(9, 30]$    &       24.04 &        73.54 &        133.21 &        27.87 &        3.36 &     262.03 \\
            $\tau \in(30, 60]$   &       23.82 &        61.11 &         83.72 &        27.28 &        3.85 &     199.79 \\
            $\tau \in(60, 90] $  &       10.62 &        21.21 &         26.72 &        10.44 &        2.78 &      71.77 \\
            $\tau \in(90, 180]$  &       18.77 &        22.15 &         18.23 &         9.92 &        4.30 &      73.35 \\
            $\tau \in(180, 365]$ &       10.34 &        10.29 &          6.96 &         4.06 &        4.88 &      36.53 \\
            Total  &       91.71 &       240.32 &        408.17 &        91.51 &       20.14 &     851.85 \\
            \midrule
            & \multicolumn{6}{c}{Panel C: Average OTM option price (\$)} \\
            \midrule
            $\tau \in(2, 9]$     &        0.43 &         1.38 &          8.69 &         2.69 &        1.73 &       4.32 \\
            $\tau \in(9, 30]$    &        0.92 &         5.53 &         23.53 &         5.47 &        2.68 &      10.07 \\
            $\tau \in (30, 60]$   &        2.42 &        13.28 &         40.10 &        11.84 &        4.68 &      16.73 \\
            $\tau \in (60, 90]$   &        5.05 &        24.89 &         63.56 &        22.00 &        6.17 &      24.93 \\
            $\tau \in (90, 180]$  &       10.97 &        50.74 &        104.11 &        47.11 &       11.76 &      43.14 \\
            $\tau \in(180, 365]$ &       18.80 &        90.42 &        155.53 &        85.29 &       19.67 &      53.05 \\
            Total  &        7.97 &        22.29 &         47.73 &        22.35 &        9.64 &      24.02 \\
            
            \bottomrule
            \end{tabular}
            \label{descriptive}

            \medskip
            \begin{minipage}{\textwidth}\scriptsize
                Note: Descriptive statistics for filtered option data on the S\&P 500 stock market index. 
                The sample contains daily option data from 1 May 2017 to 1 April 2021.
                Observations are bucketed into six categories based on the time-to-maturity, $\tau$, and into five categories with respect to the moneyness level, defined as strike-to-forward ratio $k = K/F$.
            \end{minipage}
    \end{table}

Given that we need a reliable and wide coverage of option prices for each tenor, we use a fairly generous set of filters. 
In particular, we retain option observations that satisfy the following criteria: 
($i$) bid price is strictly positive and ask-to-bid ratio is less than a factor 10; 
($ii$) the maturity is larger than or equal to 2 calendar days, but less than or equal to 365 calendar days; 
($iii$) it is not an early-closure day.
The first criterion filters out illiquid observations and the second one limits our consideration in terms of options' maturity. 
The third criterion rules out shortened trading sessions, which in total constitute 10 days in our sample.

\begin{figure}
        \centering
        \caption{Stacked bar chart of time-to-maturity frequency}
        \vspace*{-0.45cm}
        \includegraphics[scale=0.7]{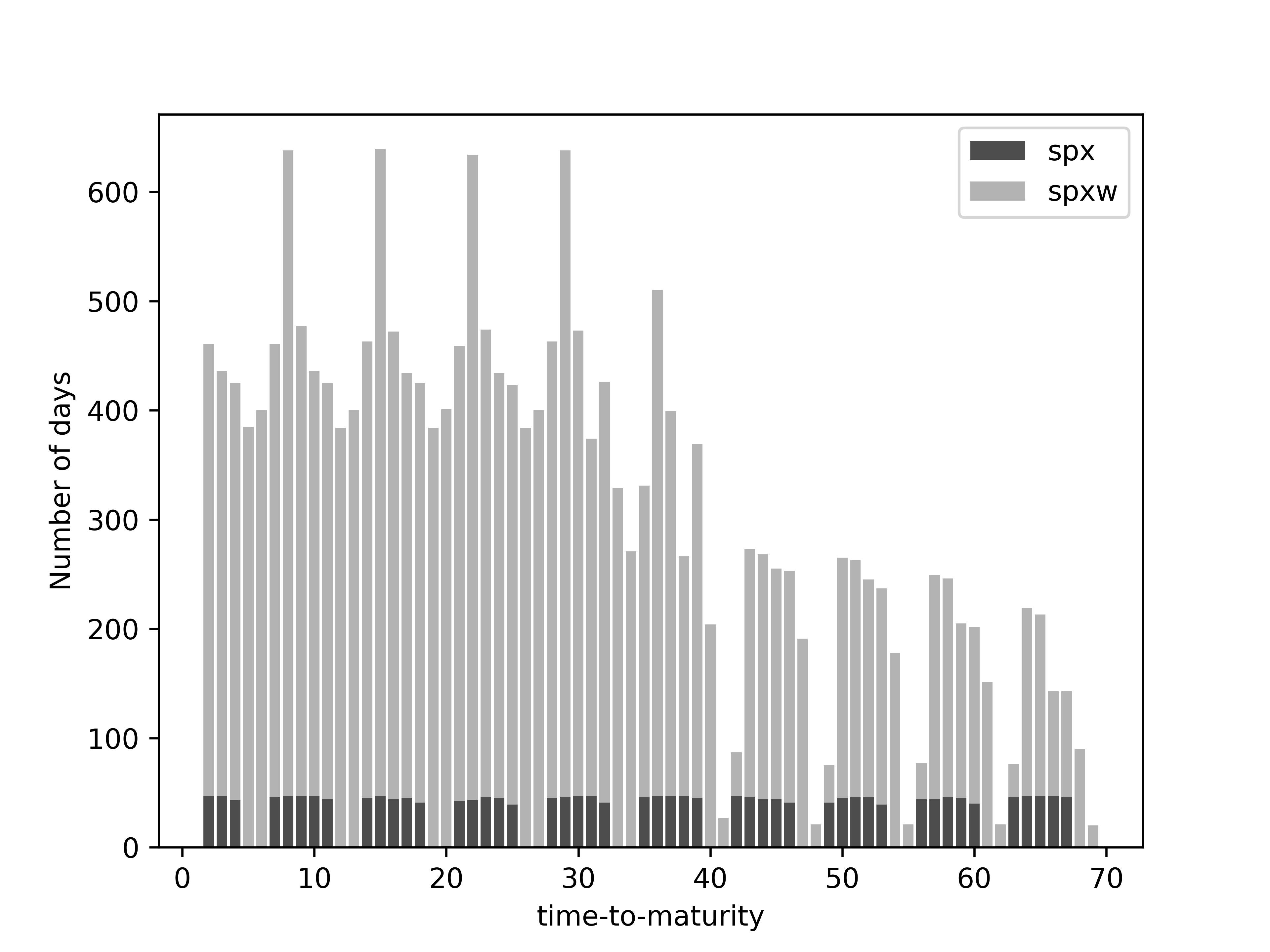}

        \label{fig:tenors}

        \medskip
        \begin{minipage}{0.75\textwidth}\scriptsize
            Note: This figure plots a stacked bar chart for the frequency of tenors in S\&P 500 index options.  
            The sample contains daily option data from 1 May 2017 to 1 April 2021 (constituting 978 trading days). 
            The indicators `spx' and `spxw' correspond to AM-settled `standard' and PM-settled `weeklys' and end-of-month contracts, respectively.
        \end{minipage}
    \end{figure}

For each tenor, we determine the moneyness based on the forward index level, $F_t(\tau)$. 
For that, we use the put-call parity to calculate the forward price for close to at-the-money (ATM) options. 
Specifically, we use up to 5 option pairs with the smallest absolute difference between the call and put prices. 
The median of their forward-implied prices is taken as the forward index level for the corresponding tenor. 
The risk-free rates are obtained by interpolating the LIBOR rates to any particular tenor. 
Finally, given the calculated forward prices and moneyness levels, we retain only out-of-the-money (OTM) options for further exploitation. 
Descriptive statistics of the S\&P 500 index options data sample are provided in Table~\ref{descriptive}. 
We observe that the largest portion of the trading volume is due to trades of OTM contracts and options with time-to-maturity less than 60 calendar days. 
Figure~\ref{fig:tenors} plots the frequency of tenors up to 70~calendar days. 

\subsection{CCF-spanning option portfolios}\label{sec:CCF-spanning}

The construction of the CCF-spanning option portfolios requires reliable option slices with wide coverage of strikes. 
Given that most of the trading volume is concentrated in option contracts with time-to-maturity of less than 60 days, our empirical application relies on the use of short-dated option slices with expiration period of no more than 2 months. 
In particular, on each trading day, we keep the six tenors closest to 8, 15, 22, 29, 36 and 61 days\footnote{The first five of these tenors are the most representative in the sample, see Figure~\ref{fig:tenors}.} from below with the largest trading volume and number of quoted OTM option contracts. 
Specifically, starting with the option slices closest to the indicated tenors, we compare them with every next shorter maturity option slice, and prefer the next one if it has a larger trading volume and a larger number of quoted contracts for OTM options. 
Table~\ref{descriptive_six} provides the descriptive statistics for each of the six selected tenors over the considered time span. 
We notice the wide coverage of strikes, since the average minimum put and call prices are close to the tick size of \$0.05, especially for very short-dated options. 
We also mention that in the selected option sample, each option slice at each trading day contains at least 55 different quoted contracts. 
Therefore, no additional filters on the minimum number of contracts are imposed. 
In total, we have 978~trading days, with six different tenors at each one of them, resulting in a total number of 1,158,059~contracts in the sample.

    \begin{table}
        \centering
        \caption{Descriptive statistics for the selected sample}
        \footnotesize
        \begin{tabular}{lccccccc}
            \toprule
            Number &       1 &       2 &       3 &       4 &       5 &       6 &   Total \\
            \midrule
            avg.\ tenor   &    6.97 &   12.10 &   18.95 &   26.08 &   33.95 &   53.67 &   25.29 \\
            avg.\ min put &    0.08 &    0.09 &    0.10 &    0.11 &    0.11 &    0.14 &    0.10 \\
            avg.\ min call &    0.09 &    0.09 &    0.10 &    0.11 &    0.12 &    0.14 &    0.11 \\
            avg.\ max price  &   22.48 &   31.65 &   40.24 &   47.56 &   54.83 &   71.05 &   44.64 \\
            avg.\ \# options  &  133.75 &  183.64 &  204.47 &  206.90 &  216.50 &  238.84 &  197.35 \\
            avg.\ min $K/F$   &    0.79 &    0.69 &    0.60 &    0.54 &    0.49 &    0.42 &    0.59 \\
            avg.\ max $K/F$   &    1.06 &    1.09 &    1.12 &    1.14 &    1.16 &    1.21 &    1.13 \\
            avg.\ ATM BSIV &    0.147 &    0.151 &    0.151 &    0.152 &    0.154 &    0.158 &    0.152 \\
            \bottomrule
            \end{tabular}           
        \label{descriptive_six}

        \medskip
        \begin{minipage}{0.8\textwidth}\scriptsize
            Note: Descriptive statistics for the selected data sample of options on the S\&P 500 stock market index. 
            The sample contains daily option data from 1 May 2017 to 1 April 2021.
            For each trading day, we select the six option tenors closest to 8, 15, 22, 29, 36 and 61 days from below with the largest trading volume and number of quoted OTM option contracts. 
            The table provides the descriptive statistics for each of the six tenors over the sample.
        \end{minipage}
    \end{table}

The inputs of our estimation procedure are option portfolios representing CCFs rather than BSIVs that are commonly used in the literature. 
Therefore, we pay careful attention to the construction of the option-implied CCF. 
As discussed in Section~\ref{sec:Estimation-ss}, we use a Riemann sum approximation to obtain a computationally feasible counterpart of the CCF spanning~\eqref{CCF_spanning}.
However, in order to reduce the truncation and discretization errors, we further employ an interpolation-extrapolation technique. 
In particular, we interpolate option prices using cubic splines with carefully selected knot sequences and extrapolate beyond the observable range of strike prices using a parametrization that satisfies the asymptotic results of~\citeA{lee2004moment}. 
The details of the interpolation-extrapolation scheme are provided in Appendix~\ref{sec:Appendix-Extrapolation}.

The calculation of the option-implied CCF then uses the Riemann sum approximation~\eqref{phi-hat} applied to the result of the interpolation-extrapolation scheme. 
The construction is conducted for each day and for each maturity separately. 
In particular, for equation~\eqref{phi-hat}, we set $\Delta m = 0.0001$ with a sufficiently wide range of log-moneyness between $\underline{m} = -6$ and $\overline{m} = 2$.

To conclude this section we emphasize again that, contrary to what is common in many existing approaches, the option prices---or a monotonic transformation thereof---are not used as inputs in our developed estimation procedure. 
Instead, we use the option portfolios that replicate the CCF of log returns. 
Furthermore, unlike in many other papers, our option dataset is daily and utilizes the information from short-dated options with maturities between two days and two months. 

\section{Empirical Applications}
\label{sec:Empirics}

Having thus constructed the dataset of option-implied CCFs for  S\&P 500 index options, we now illustrate our estimation procedure in two empirical applications, without and with an external factor.

\subsection{SVCDEJ}

We start with estimating the SVCDEJ model specified in Section~\ref{sec:Simulation-svcdej}, \eqref{SVCDEJ-y}--\eqref{SVCDEJ-v}, using the CCF-spanning option portfolios with six short-term tenors described in Section~\ref{sec:CCF-spanning}. 
Table~\ref{tab:svcdej_res} provides the parameter estimates.
Informed by the Monte Carlo results, the estimates are based on the range of CCF arguments $u=1,\dots,20$, a singular value threshold $\bar{s}=10^{-7}$, and a fixed parameter $p^- = -0.7$.
Standard errors are calculated using the familiar sandwich form covariance matrix.

  \begin{table}
      \centering
      \caption{SVCDEJ estimation results}
      \footnotesize
        \begin{tabular}{lccccccccc}
        \toprule
          & $\sigma$ & $\kappa$ & $\bar{v}$ & $\rho$ &  $\delta$ & $\eta^+$ & $\eta^-$ & $\mu_v$ & $\sigma_{\varkappa}$ \\
          \midrule
          $\widehat{\theta}$ & 0.5051 & 8.325 & 0.0153 & -0.997 & 157.51 & 0.0204 & 0.0424 & 0.0519 & 0.253 \\
          s.e. & 0.0075 & 0.207 & 0.0005 & 0.012 & 7.28  &  0.0005 & 0.0007 & 0.0009 &  0.004 \\
          \bottomrule
        \end{tabular}%
      \label{tab:svcdej_res}
            
        \medskip
        \begin{minipage}{0.8\textwidth}\scriptsize
            Note: This table provides the parameter estimates and standard errors for the SVCDEJ model. 
            Descriptive statistics of the options data are in Table~\ref{descriptive_six}.
            The model is estimated based on $u=1,\dots,20$ and $\bar{s} = 10^{-7}$, and with $p^- = 0.7$.
        \end{minipage}
    \end{table}%

The parameter estimates in Table~\ref{tab:svcdej_res} are meaningful, intuitive and broadly consistent with the literature. 
For instance, \citeA{andersen2015parametric} find the mean jump sizes to be 1.71\% and 5.33\% for positive and negative jumps in their three-factor model specification. 
(They use, however, only the Wednesday options with a different sample period, from 1996 to 2010.) 

We note that the leverage parameter $\rho$ is estimated close to its boundary value of $-1$,
implying almost perfectly correlated diffusive components in returns and volatility. 
The empirical literature suggests that $\rho$ is negative and large in absolute value. 
The estimate of $\rho$ being nearly equal to its boundary value might be due to the use of short-dated options that typically exhibit steep implied volatility slopes. 
Indeed, \citeA{AFT2017} also find this correlation to be close to $-1$ in their dataset dominated by option contracts with maturities of less than 2 months.
  
We also note that the estimated measurement standard error $\sigma_{\varkappa}$ corresponds to a standard deviation of about 25\% of the implied volatility. 
This is somewhat larger than what one might expect of measurement errors in option prices only, and might be interpreted to indicate e.g., missing state variables. 
In agreement with this, some of the extensions of the SVCDEJ model considered below and in Appendix~\ref{sec:Appendix-Empirical} show a slightly lower estimate of $\sigma_{\varkappa}$.

\begin{figure}
    \centering
    \caption{SVCDEJ filtered volatility}
    \vspace*{-0.45cm}
    \includegraphics[scale=0.10]{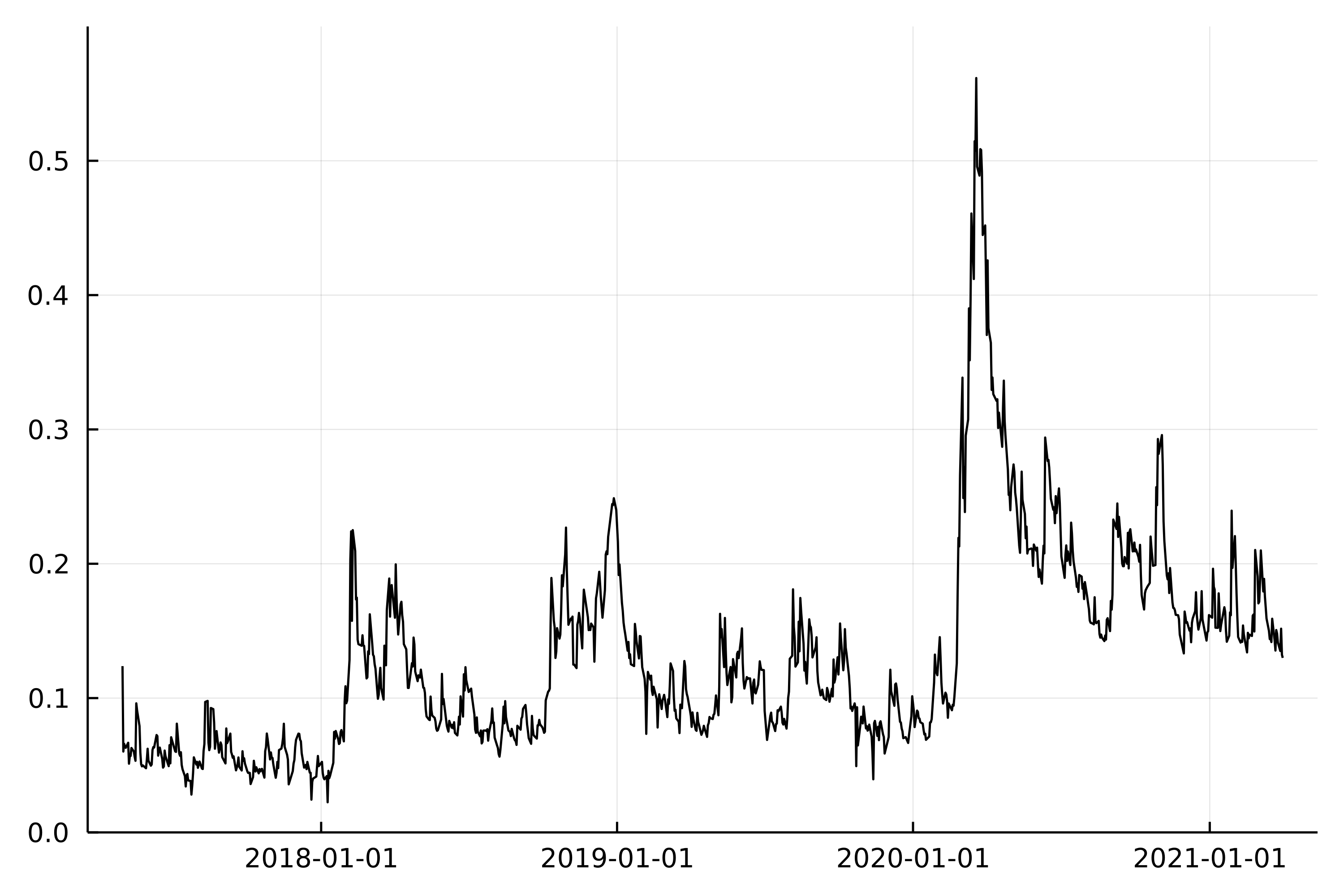}
    \label{fig:svcdej_vol}

    \medskip
    \begin{minipage}{0.8\textwidth}\scriptsize
      Note: This figure plots the filtered volatility (i.e., the square root of the filtered state $\widehat{x}_{t+1|t}$) given the parameter estimates of the SVCDEJ model using Kalman filter recursions.
    \end{minipage}
  \end{figure}

Figure~\ref{fig:svcdej_vol} plots the filtered volatility (i.e., the square root of the filtered state $\widehat{x}_{t+1|t}$) given the parameter estimates of the SVCDEJ model. 
As is clearly visible, the filtered volatility exhibits a relatively stable volatility regime prior to 2020 and jumps up in March 2020 at the outbreak of the Covid-19 pandemic.

\subsection{SVCDEJ with external factors}

Now we turn to model specifications with embedded external factors. 
In some situations, we might have specific information on possible drivers of the risks in the market, and would like to quantify their effect.  

An example is the recent Covid-19 crisis. 
The Covid-19 pandemic has dramatically affected our lives. 
It has also had a tremendous impact on the world's economy and financial markets.
The beginning of the pandemic, in particular, was associated with a spike in uncertainty. 
This uncertainty surrounded many aspects: 
the contagiousness and lethality of the virus, the time required to develop vaccines, the effectiveness of measures, the work-from-home policies, travel bans, and so on.
In this application, we explore the impact of the Covid-19 pandemic on the stock market through the lens of option prices. 
In particular, we consider how the spread of the virus affected the likelihood of jump events and the volatility in the U.S. stock market.

  \begin{figure}
        \centering
        \caption{Covid-19 daily cases and reproduction numbers}
        \begin{subfigure}{0.45\textwidth}
            \hspace*{-1.0cm}
            \includegraphics[scale=0.125]{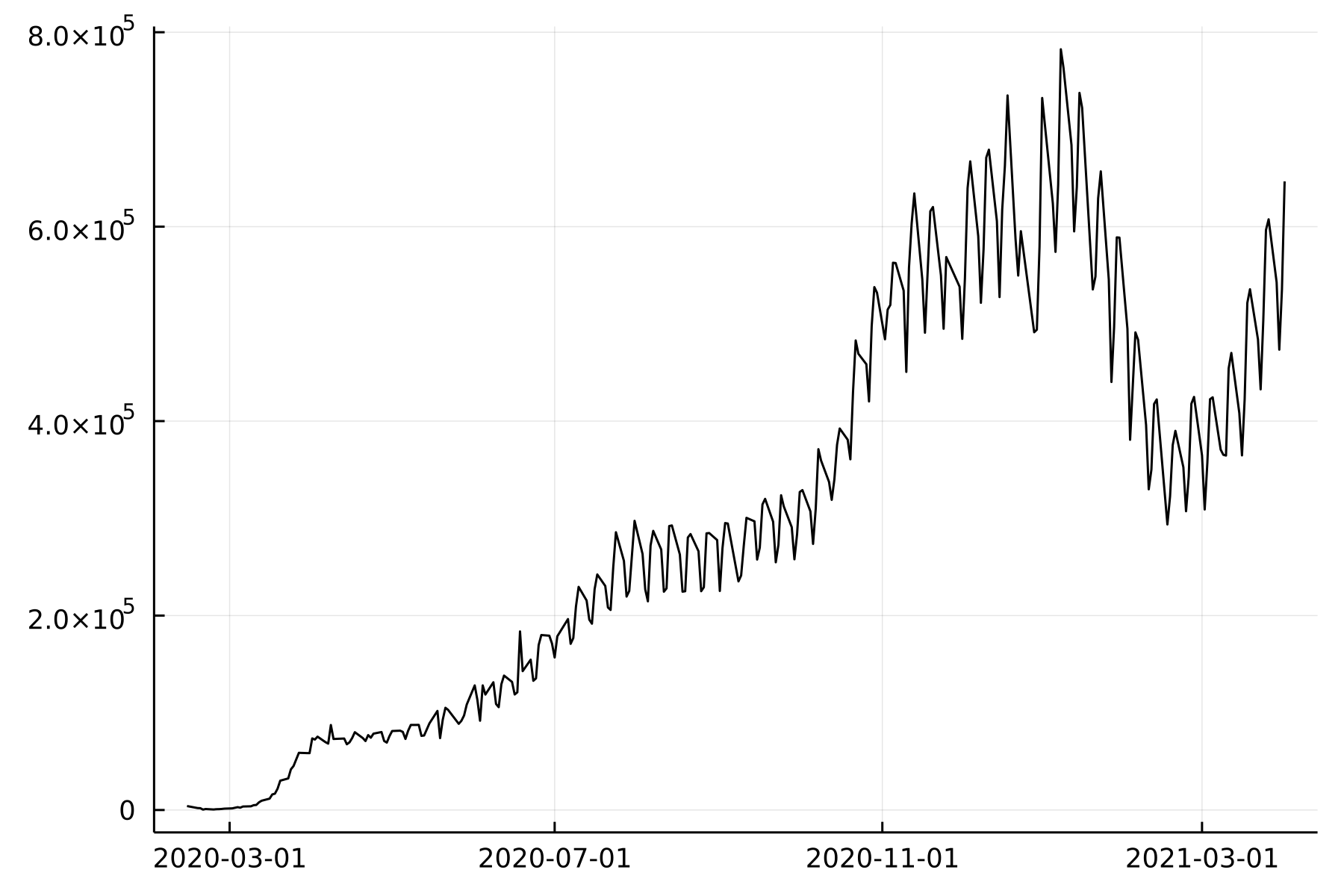}
            \caption{Daily cases}
        \end{subfigure}
        \begin{subfigure}{0.45\textwidth}
            \includegraphics[scale=0.125]{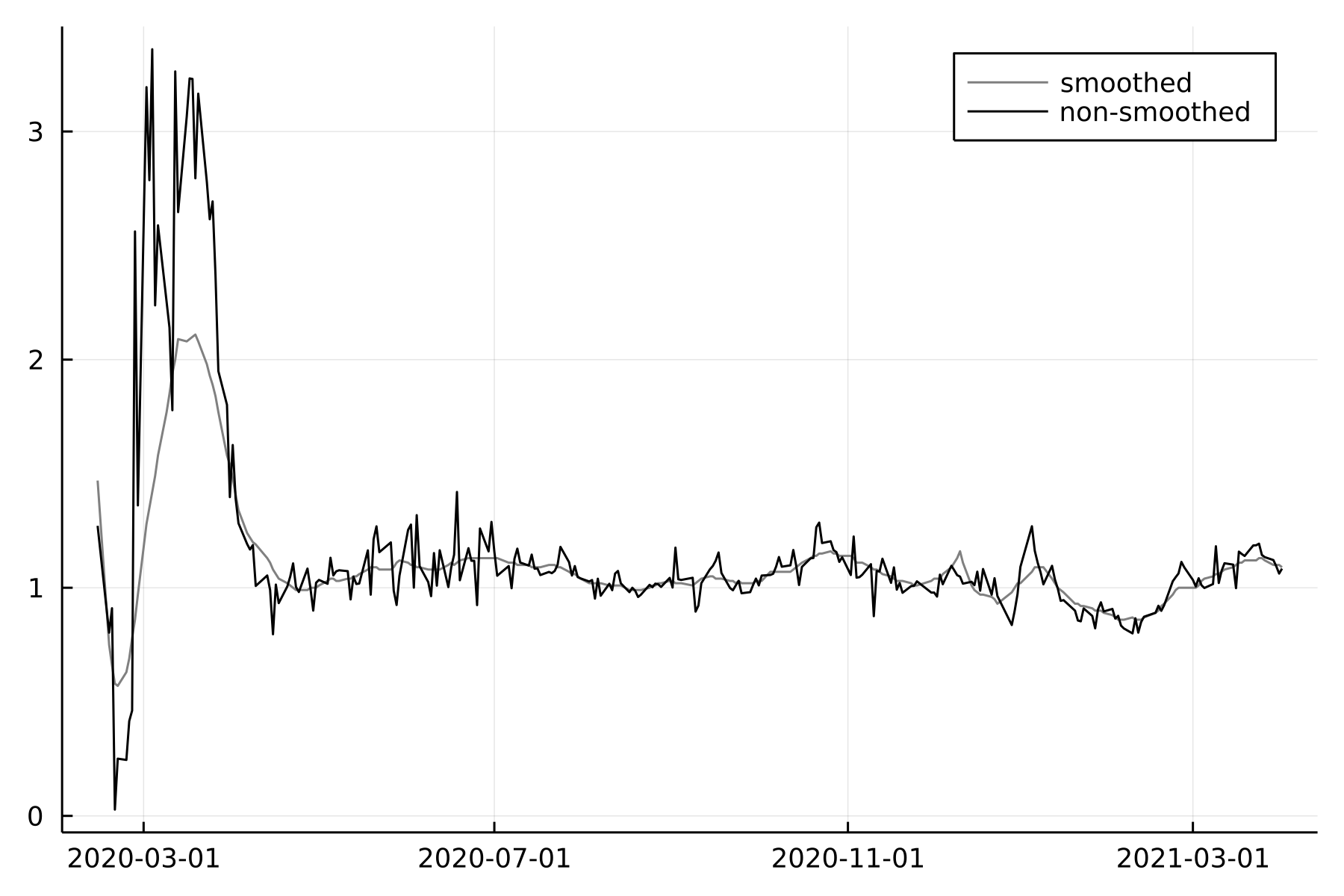}
            \caption{Reproduction numbers}
        \end{subfigure}
        \label{fig:Covid}

        \medskip
        \begin{minipage}{\textwidth}\scriptsize
            Note: This figure plots the daily Covid-19 cases (worldwide) and two reproduction numbers: 
            the first is taken from the website `Our World in Data' (using methodology of \citeA{arroyo2021tracking}); the second is calculated as the ratio $I_t/I_{t-7}$, where $I_t$ is number of infected people in day $t$ and 7 is the reported serial interval for Covid-19.
            The sample period runs from 14 February 2020 to 1 April 2021.
        \end{minipage}
    \end{figure}

Figure~\ref{fig:Covid}, Panel~(a), plots the daily cases of Covid-19 infections around the world obtained from the World Health Organization (WHO). 
The reported number of daily cases, however, does not represent well the contagiousness of the virus. 
Therefore, Panel~(b) of Figure~\ref{fig:Covid} displays the so-called reproduction number $R_t$, according to two measures: 
the first one is taken from the website `Our World in Data' and the second one is calculated as the ratio $R_t = I_t/I_{t-7}$, where $I_t$ is the number of infected people in day $t$ and $7$ is the reported serial interval for Covid-19. 
The former is based on the parametric methodology of~\citeA{arroyo2021tracking} and is smoothed over time.\footnote{In fact, \citeA{arroyo2021tracking} use a Kalman smoother.} 
The latter is non-smoothed and based on the assumption that the serial interval is $7$ days, which is consistent with the recent epidemiology literature (see, e.g., \citeNP{maier2020effective}, \citeNP{prem2020effect}, \citeNP{flaxman2020estimating}, \citeNP{arroyo2021tracking}). 
We will use the latter non-parametric and non-smoothed measure as the reproduction number in our application.

To quantify the effect of Covid-19 propagation on the financial market, we embed the reproduction number as an external factor into the (time-varying) levels of the stochastic volatility and jump intensity processes, as described in Section~\ref{sec:Simulation-svcdejex},
equations~\eqref{SVCDEJex-y}--\eqref{SVCDEJex-h}, with $h_t$ replaced by $R_t$. 
Given that the reproduction number constitutes a relatively persistent process, we will treat it as a deterministic process when pricing options; in other words, we follow the second estimation approach described in Section~\ref{sec:Simulation-svcdejex}.
In a similar way, the risk-free rate and dividend yields are often assumed to be deterministic in the option pricing literature. 
This allows us to be agnostic about the parametric dynamics of the reproduction number. 
Furthermore, given the short-dated options under consideration, the errors due to this deterministic treatment are likely to be negligible.\footnote{Similarly, \citeA{AFT2017} and \citeA{boswijk2021jump} consider an approximation of the return process with `freezed' spot volatility when estimating their option pricing models with short-dated options.}  

    \begin{table}[h]
        \centering
        \caption{SVCDEJ estimation results with Covid-19 reproduction numbers as external factor}
        \footnotesize
        \begin{tabular}{lccccccccccc}
        \toprule
            & $\sigma$ & $\kappa$ & $\bar{v}$ & $\rho$ &  $\delta$ &  $\eta^+$ & $\eta^-$ & $\mu_v$ & $\gamma$ &  $q$ & $\sigma_{\varkappa}$ \\
            \midrule
            $\widehat \theta$  & 0.5678 & 11.549 & 0.0140 & -1.000 & 130.12 & 0.0181 & 0.0413 & 0.0667 & 2.64  & 0.0003 & 0.245 \\
            s.e.  & 0.0176 & 0.646 & 0.0007 & 0.023 & 11.93 & 0.0007 & 0.0012 & 0.0029 & 0.25  & 0.0001 & 0.004 \\
            \bottomrule
          \end{tabular}%
        \label{tab:svcejex_res}%
                
        \medskip
        \begin{minipage}{0.95\textwidth}\scriptsize
            Note: This table provides the parameter estimates and standard errors for the SVCDEJ model with Covid-19 reproduction numbers as external factor.
            The reproduction numbers are set to zero before 14 February 2020, and are taken to be the ratios $R_t=I_t/I_{t-7}$ starting from 14 February 2020.
            Descriptive statistics of the options data are in Table~\ref{descriptive_six}.
            The model is estimated based on $u=1,\dots,20$ and $\bar{s} = 10^{-7}$, and with $p^- = 0.7$.
        \end{minipage}
    \end{table}%
      
    \begin{figure}[h]
        \centering
        \caption{SVCDEJ jump intensity with Covid-19 reproduction numbers as external factor}
        \vspace*{-0.45cm}
        \includegraphics[scale=0.1]{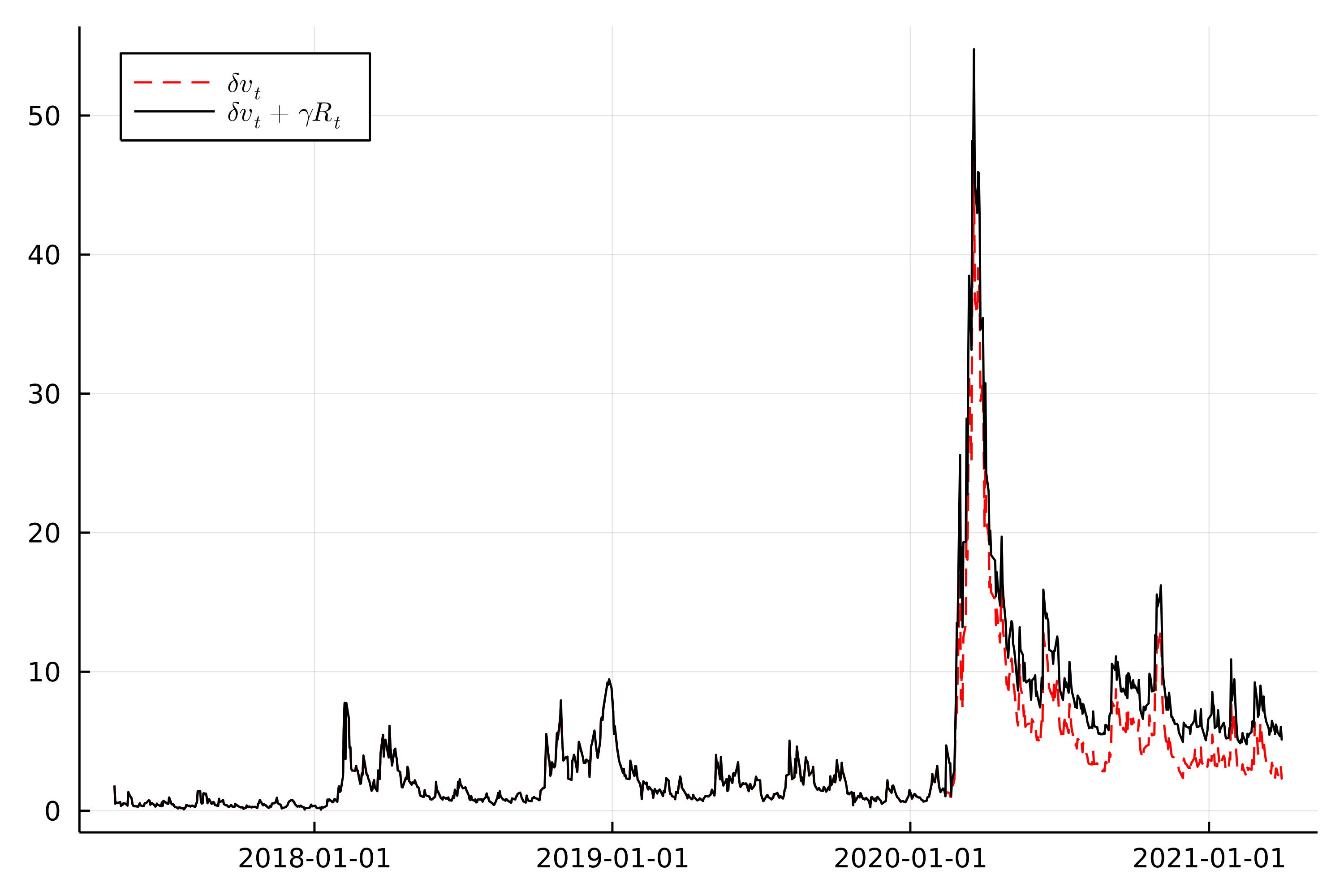}
        \label{fig:svcejex_intens}

        \medskip
        \begin{minipage}{0.8\textwidth}\scriptsize
            Note: This figure plots the filtered and total jump intensity given the parameter estimates of the SVCDEJ model with Covid-19 reproduction number dynamics.
        \end{minipage}
    \end{figure}

Table~\ref{tab:svcejex_res} provides the parameter estimates for the SVCDEJ model with Covid-19 reproduction numbers as an exogenous factor. 
With $q$ estimated at $0.0003$, the results indicate that the reproduction number dynamics have no substantial effect on the total diffusive volatility. 
A one unit increase in reproduction numbers, however, leads to an increase in the intensity of jumps by $\gamma$ which is estimated at $2.64$. 
In other words, the reproduction number contributes substantially to the likelihood of jumps. 
Figure~\ref{fig:svcejex_intens} illustrates the dynamics of the jump intensity without and with the added effect of reproduction numbers.
    
\begin{figure}[h]
        \centering
        \caption{SVCDEJ diffusive volatility with the EPU index as external factor}
        \vspace*{-0.45cm}
        \includegraphics[scale=0.1]{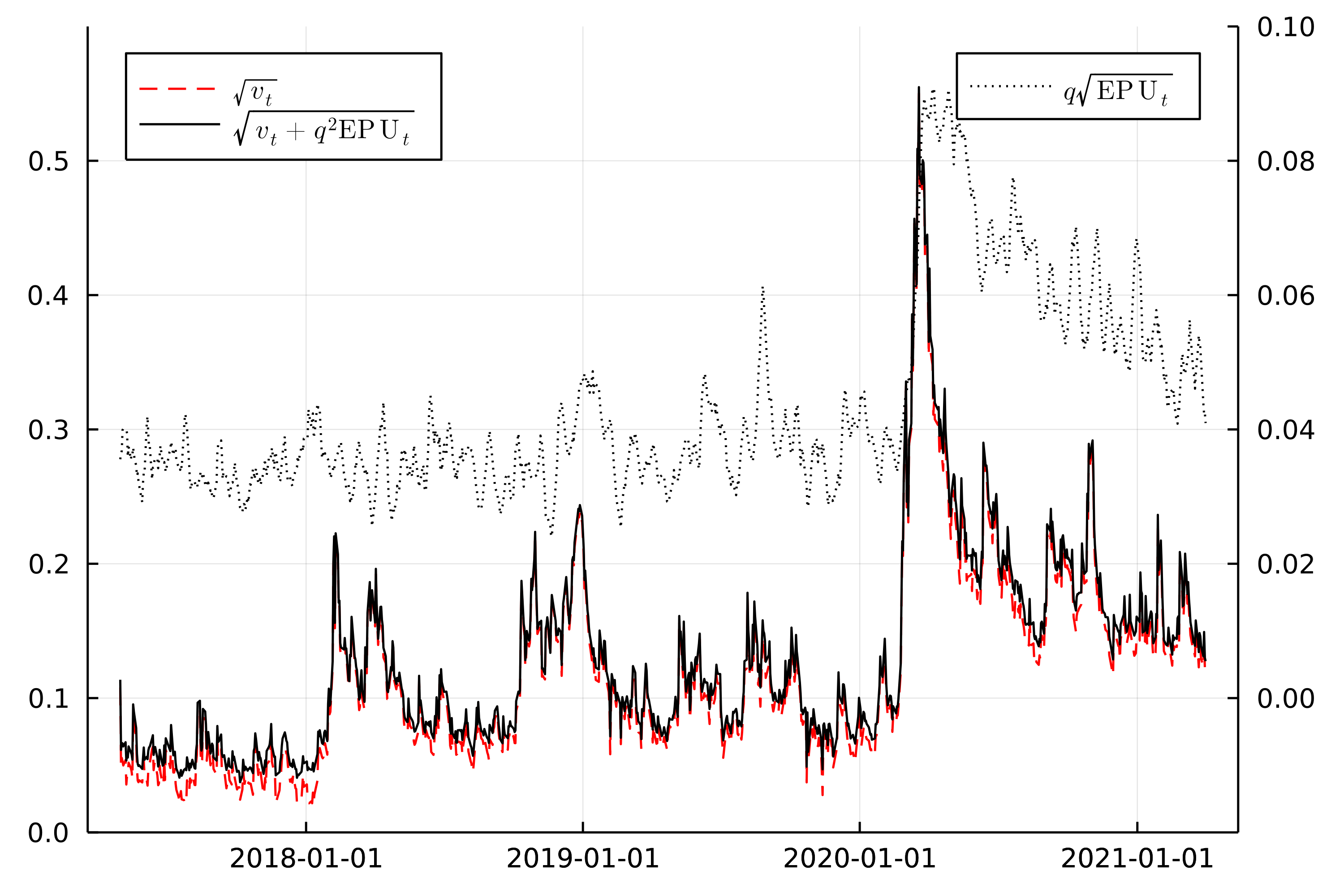}
        \label{fig:EPU}

        \medskip
        \begin{minipage}{0.8\textwidth}\scriptsize
            Note: This figure plots the filtered and total diffusive volatility given the parameter estimates of the SVCDEJ model with EPU index dynamics. 
            The total diffusive volatility equals $\sqrt{v_t + q^2 h_t}$, where $h_t$ is the external factor given by the EPU index.
            (The secondary vertical axis shows the values of $q\sqrt{\mathrm{EPU}_{t}}$.)
        \end{minipage}
    \end{figure}

It is also possible to investigate the contribution of other external factors to the diffusive volatility and jump intensity. 
As an example, we provide in Table~\ref{tab:svcejex_res_epu} estimation results for the SVCDEJ model with the Economic Policy Uncertainty (EPU) index embedded as an external factor. 
The EPU index, developed by~\citeA{baker2016measuring}, reflects policy-related economic uncertainty based on newspaper coverage frequency. 
The estimation results indicate that, unlike the reproduction number, the EPU index has no effect on the jump intensity process, but contributes significantly to the total diffusive volatility of the model, with $q$ estimated at $0.0369$; see also Figure~\ref{fig:EPU}.
Thus, whereas Covid-19 reproduction numbers contribute substantially to the jump intensity dynamics, the policy uncertainty index EPU contributes significantly to the total diffusive volatility.

\begin{table}[h]
        \centering
        \caption{SVCDEJ estimation results with the EPU index as external factor}
        \footnotesize
        \begin{tabular}{lccccccccccc}
        \toprule
        & $\sigma$ & $\kappa$ & $\bar{v}$ & $\rho$ &  $\delta$ &  $\eta^+$ & $\eta^-$ & $\mu_v$ & $\gamma$ &  $q$ & $\sigma_{\varkappa}$ \\
            \midrule
            $\widehat \theta$  & 0.4887 & 10.254 & 0.0116 & -1.000 & 223.94 & 0.0144 & 0.0410 & 0.0462 & 0.00  & 0.0369 & 0.249 \\
            s.e.  & 0.0186 & 0.783 & 0.0006 & 0.037 & 7.96 &  0.0011 & 0.0008 & 0.0024 & 0.13  & 0.0027 & 0.003 \\
            \bottomrule
          \end{tabular}%
        \label{tab:svcejex_res_epu}%
        
        \medskip
        \begin{minipage}{0.95\textwidth}\scriptsize
            Note: This table provides the parameter estimates and standard errors for the SVCDEJ model with the EPU index as external factor. 
            Descriptive statistics of the options data are in Table~\ref{descriptive_six}.
            The model is estimated based on $u=1,\dots,20$ and $\bar{s} = 10^{-7}$, and with $p^- = 0.7$.
        \end{minipage}
    \end{table}%

\section{Conclusion}
\label{sec:Conclusion}

We have proposed a novel state filtering and parameter estimation procedure for option pricing models that belong to the affine jump-diffusion class. 
Our procedure utilizes the log of the option-implied and model-free conditional characteristic function and the model-implied conditional log-characteristic function, which is functionally affine in the model's state vector. 
We have developed a linear state space representation for the considered class of option pricing models, which allows us to exploit suitably modified collapsed Kalman filtering techniques. 
Our estimation procedure is fast and easy to implement, circumventing the typical computational burden when working with option pricing models. 
We have demonstrated the applicability of our procedure in two empirical illustrations that analyze S\&P~500~index options and the impact of exogenous variables capturing Covid-19 reproduction and economic policy uncertainty data.
    
Although we have focused on Gaussian QML estimation based on Kalman filtering techniques, which delivers good results in our Monte Carlo simulations, the same state space formulation can also be analyzed by more refined methods such as those based on particle filters; see, e.g., \citeA{johannes2009optimal}, \citeA{christoffersen2014particle} and \citeA{bardgett2019inferring}. 
Such methods could exploit the non-Gaussianity and heteroskedasticity in the data to obtain more efficient estimates, at the cost of some increased computational complexity. 
We note that such extensions would still not require option price evaluation by the FFT or COS methods, and thus retain an important advantage of our approach.
    
Our proposed estimation procedure in principle allows for identification and estimation of factor risk premium parameters, by combining the risk-neutral parameters entering the measurement equation with the objective parameters entering the transition equation. 
Monte Carlo simulation results suggest, however, that option price data are not very informative about such risk premia, which is why we have concentrated on the case where the objective and risk-neutral measures coincide. 
Fortunately, the simulation results also suggest that inference on the risk-neutral parameters is quite robust with respect to deviations from this assumption. 
For more focused inference on (volatility) risk premium parameters, it may be possible to combine the information in daily option prices as considered in this paper with realized measures based on high-frequency returns on the underlying. 
We intend to explore this in future research.

    \clearpage
    \begin{appendices}
        \numberwithin{equation}{section}
        \numberwithin{assumption}{section}
        \numberwithin{figure}{section}
        \numberwithin{table}{section}
        
\section{Proofs}
\label{sec:Appendix-Proofs}

In this appendix, we provide the proof of Proposition~\ref{prop1}. 
First, we state and prove some preliminary results.

\subsection{Preliminary results}
\label{sec:Appendix-Proofs-Prelim}

We start by formally defining the measurement errors in the CCF approximation. 
Under the observation error structure specified in Assumption~\ref{ass:option_errors} and the CCF approximation given by equation~\eqref{phi-hat}, 
the total measurement error in the option-spanned CCF may be written as 
    \begin{align*}
        \zeta^{\phi}_t(u, \tau):=&\ \widehat{\phi}_t(u, \tau) - \phi_t(u, \tau) \\
        = &- u_t \sum_{j=2}^n e^{(\mathrm{i}u-1)m_j} \cdot \widehat{O}_t(\tau, m_j) \Delta m_j 
        + u_t \int_{-\infty}^{\infty} e^{(\mathrm{i}u -1) m } \cdot O_t(\tau, m) \diff m  \\
        = &- u_t \left[ \sum_{j=2}^n e^{(\mathrm{i}u-1)m_j} \cdot O_t(\tau, m_j) \Delta m_j + \sum_{j=2}^n e^{(\mathrm{i}u-1)m_j} \cdot \zeta_t(\tau, m_j) \Delta m_j \right] \\
        &+ u_t \int_{-\infty}^{\infty} e^{(\mathrm{i}u -1) m } \cdot O_t(\tau, m) \diff m ,
    \end{align*}
which after reordering of terms can be decomposed into the following three components:
    \begin{align}
        \zeta^{\phi}_t(u, \tau)=& \underbrace{- u_t \sum_{j=2}^n e^{(\mathrm{i}u-1)m_j} \cdot \zeta_t(\tau, m_j) \Delta m_j}_{=:\zeta^{(1)}_t(u, \tau)} \nonumber \\
        & + \underbrace{u_t \int_{-\infty}^{m_1} e^{(\mathrm{i}u -1) m } \cdot O_t(\tau, m) \diff m + u_t \int_{m_n}^{\infty} e^{(\mathrm{i}u -1) m } \cdot O_t(\tau, m) \diff m}_{=:\zeta^{(2)}_t(u, \tau)} \nonumber \\
        &+ \underbrace{u_t \sum_{j=2}^n  \int_{m_{j-1}}^{m_j} \left[ e^{(\mathrm{i}u-1) m } \cdot O_t(\tau, m) - e^{(\mathrm{i}u-1)m_j} \cdot O_t(\tau, m_j) \right] \diff m}_{=:\zeta^{(3)}_t(u, \tau)} \nonumber \\
        =&\ \zeta^{(1)}_t(u, \tau) + \zeta^{(2)}_t(u, \tau) + \zeta^{(3)}_t(u, \tau).\label{eq:measdec}
    \end{align}
    
The error terms $\zeta^{(1)}_t(u, \tau)$, $ \zeta^{(2)}_t(u, \tau)$ and $\zeta^{(3)}_t(u, \tau)$ represent observation, truncation and discretization errors, respectively. 
In order to characterize the asymptotic orders of these errors, we make use of the following auxiliary result: 

    \begin{lemma} \label{lemma1}
        Let $f_{t+\tau} = \frac{F_{t+\tau}}{F_t}$ be the futures price normalized to its value at time $t$ for $\tau >0$. 
        For all $m >0$, we have the call price bounds 
        \begin{align}\label{call-bound}
            \frac{O_t(\tau, m)}{F_t} \leq \left(\frac{p}{p+1}\right)^p \frac{e^{-r \tau} \E^\Q[f_{t+\tau}^{p+1}|\F_t]}{p+1} e^{-pm},
        \end{align} 
        for each $p >0$. 
        Similarly, for all $m <0$, we have the put price bounds 
        \begin{align}\label{put-bound}
            \frac{O_t(\tau, m)}{F_t} \leq \left(\frac{q}{q+1}\right)^q \frac{e^{-r \tau} \E^\Q[f_{t+\tau}^{-q}|\F_t]}{q+1} e^{(1+q)m},
        \end{align} 
        for each $q >0$. 
    \end{lemma}
    \textit{Proof:} The result is a straightforward adaptation of Theorem~2.1 in~\citeA{lee2004moment}.\hfill $\square$ \\

\noindent Lemma~\ref{lemma1} relates moments of the underlying process and of its reciprocal to bounds on option prices. 
Similar to \citeA{qin2019nonparametric} and \citeA{todorov2019nonparametric}, we assume the existence of at least the second order moment of the underlying process and of its reciprocal, formally stated in Assumption~\ref{ass-moments}($i$). 
If higher moments exist, then we can obtain even tighter bounds for the remainder term in Proposition~\ref{prop1} due to~\eqref{call-bound} and~\eqref{put-bound}.

The following lemma establishes the order of magnitude of the truncation and discretization errors under the joint asymptotic scheme, expressed with respect to the number of option prices $n$ with fixed maturity. 
As in the main text, we denote the smallest and largest log-moneyness by $\underline{m} = \min_{1\leq j \leq n} m_j $ and $\overline{m}=\max_{1\leq j \leq n} m_j $, and the corresponding strike prices by $\underline{K}$ and $\overline{K}$. 
In the proofs, we denote by $C_t$ an $\F_t$-adapted random variable that does not depend on $m$ and that may change from line to line.
    
    \begin{lemma}\label{lemma2}
        Suppose $\E^{\Q}[F_{t+\tau}^{p+1} \rvert \F_t] < \infty$ and $\E^{\Q}[F_{t+\tau}^{-q} \rvert \F_t] < \infty$ for some $p>0$ and $q>0$, Assumption \ref{ass-moments}(ii) holds, and in addition $\underline{K} \asymp n ^{-\underline{\alpha}}$ and $\overline{K} \asymp n^{\overline{\alpha}}$ with $\underline{\alpha} >0$ and $\overline{\alpha} >0$. 
        Then, as $n \to \infty$, we have
        \begin{align}
            \zeta_t^{(2)}(u, \tau) &= \mathcal{O}_p\left( n^{-(q \underline{\alpha} \wedge (1+p) \overline{\alpha})}\right),
        \end{align}
        and
        \begin{align}
            \zeta_t^{(3)}(u, \tau) &= \mathcal{O}_p\left(\frac{\log n}{n^{1 + q\underline{\alpha} \wedge (p+1) \overline{\alpha}}} \right),
        \end{align}
        for a fixed $u \in \mathcal{U}$ and $\tau>0$.
    \end{lemma}

    \textit{Proof:}
    We start with the truncation errors. 
    For $\underline{m} = m_1 < \ldots < m_n = \overline{m}$, with $\underline{m}<0$ and
    $\overline{m}>0$, and using Lemma~\ref{lemma1}, we can bound the upper and lower truncation parts as follows:
    \begin{align*}
        \Big\rvert \frac{1}{F_t} \int_{\overline{m}}^{\infty} e^{(\mathrm{i}u -1) m } \cdot O_t(\tau, m) \diff m \Big\rvert
        &\leq \int_{\overline{m}}^{\infty} \Big\rvert  e^{(\mathrm{i}u -1) m } \Big\rvert \cdot \Big\rvert \frac{ O_t(\tau, m)}{F_t} \Big\rvert \diff m  \leq C_t  e^{-(1+p) \overline{m}}, \\
        \Big\rvert \frac{1}{F_t} \int_{-\infty}^{\underline{m}} e^{(\mathrm{i}u -1) m } \cdot O_t(\tau, m) \diff m \Big\rvert
        &\leq \int_{-\infty}^{\underline{m}} \Big\rvert  e^{(\mathrm{i}u -1) m } \Big\rvert \cdot \Big\rvert \frac{ O_t(\tau, m)}{F_t} \Big\rvert \diff m  \leq  C_t e^{q \underline{m}},
    \end{align*}
    where, as mentioned before, $C_t$ is independent of $m$ and may vary from line to line.
    Therefore, as $\underline{m} \to -\infty$ and $\overline{m} \to \infty$, we have
    \begin{align*}
        \zeta_t^{(2)}(u, \tau) &= u_t \int_{-\infty}^{\underline{m}} e^{(\mathrm{i}u -1) m } \cdot O_t(\tau, m) \diff m + u_t \int_{\overline{m}}^{\infty} e^{(\mathrm{i}u -1) m } \cdot O_t(\tau, m) \diff m \\
        &= \mathcal{O}_p\left(e^{-q |\underline{m}|}\right) + \mathcal{O}_p\left(e^{-(1+p)|\overline{m}|}\right)
        = \mathcal{O}_p\left(e^{-(q |\underline{m}| \wedge (1+p)|\overline{m}|)}\right)
        = \mathcal{O}_p\left(n^{-(q \underline{\alpha} \wedge (1+p)\overline{\alpha})}\right).
    \end{align*}
    %

    For the discretization errors, we use the following decomposition:
    \begin{align*}
        &\int_{m_{j-1}}^{m_j} \left[ e^{(\mathrm{i}u-1) m } \cdot O_t(\tau, m) - e^{(\mathrm{i}u-1)m_j} \cdot O_t(\tau, m_j) \right] \diff m \\
        &= \int_{m_{j-1}}^{m_j} \left[ \left( e^{(\mathrm{i}u-1) m } - e^{(\mathrm{i}u-1) m_j } \right)\cdot O_t(\tau, m_j) + e^{(\mathrm{i}u-1)m} \left( O_t(\tau, m) - O_t(\tau, m_j) \right) \right] \diff m.
    \end{align*}
    By applying the mean value theorem twice, we have 
    \begin{align*}
        \Big \rvert e^{(\mathrm{i}u-1) m } - e^{(\mathrm{i}u-1) m_j } \Big \rvert
        \leq  |\mathrm{i}u-1|  \Big \rvert e^{(\mathrm{i}u-1) \tilde{m} }  \Big \rvert \Delta m_j   
        \leq e^{-m_{j-1}} (|u|\vee 1) \Delta m_j,
    \end{align*}
    and 
    \begin{align*}
        \Big \rvert e^{(\mathrm{i}u-1)m} \left( O_t(\tau, m) - O_t(\tau, m_j) \right)  \Big \rvert 
        &\leq e^{-m_{j-1}} \Bigg \rvert \frac{\partial O_t(\tau, m)}{\partial m} \Big \rvert_{m=\tilde m} \Bigg \rvert \Delta m_j \\
        &\leq e^{-m_{j-1}} \Bigg \rvert \frac{\partial O_t(\tau, m)}{\partial K}  \Big \rvert_{K=\tilde K} \Bigg \rvert e^{\tilde m} F_t \Delta m_j\\
        &\leq  \Bigg \rvert \frac{\partial O_t(\tau, m)}{\partial K} \Big \rvert_{K=\tilde K} \Bigg \rvert e^{\Delta m_{j}} F_t \Delta m_j,
    \end{align*}
    where $\tilde{m} = \log \frac{\tilde{K}}{F_t}$ lies between $m$ and $m_j$. 
    
    For the first term in the decomposition above, we use that Lemma~\ref{lemma1} implies that, for all $m$,
    \begin{align*}
        \frac{ O_t(\tau, m)}{F_t} \leq C_t e^{-(pm \vee -(1+q)m)}.
    \end{align*}
    Furthermore, for the second term in the decomposition, we exploit the fact that the derivative with respect to the strike price is the risk neutral distribution or survival function, which can be bounded using the Markov inequality. 
    In particular, for $m > 0$,
    \begin{align*}
        \Bigg \rvert \frac{\partial O_t(\tau, m)}{\partial K} \Bigg \rvert 
        &= e^{-r \tau} \Q(F_{t+\tau} > K)
        = e^{-r \tau} \Q\left(f_{t+\tau} > e^{m}\right)\\
        &\leq e^{-r \tau} \E^{\Q}[f_{t+\tau}^{p+1}|\F_t] e^{-(p+1)m},
    \end{align*}
    and, for $m<0$,
    \begin{align*}
        \Bigg \rvert \frac{\partial O_t(\tau, m)}{\partial K} \Bigg \rvert 
        &= e^{-r \tau} \Q(F_{t+\tau} < K)
        = e^{-r \tau} \Q\left(f_{t+\tau}^{-1} > e^{-m}\right)\\
        &\leq e^{-r \tau} \E^{\Q}[f_{t+\tau}^{-q}|\F_t] e^{qm}.
    \end{align*}
    Therefore, 
    \begin{align*}
        \Bigg \rvert \frac{\partial O_t(\tau, m)}{\partial K} \Bigg \rvert 
        \leq C_t e^{-((p+1)m \vee -qm)}.
    \end{align*}
    Combining all these inequalities together, we obtain
    \begin{align*}
        &\Big\rvert  \frac{1}{F_t} \int_{m_{j-1}}^{m_j} \left[ e^{(\mathrm{i}u-1) m } \cdot O_t(\tau, m) - e^{(\mathrm{i}u-1)m_j} \cdot O_t(\tau, m_j) \right] \diff m \Big\rvert  \\
        &=\Big\rvert  \frac{1}{F_t} \int_{m_{j-1}}^{m_j} \left[ \left( e^{(\mathrm{i}u-1) m } - e^{(\mathrm{i}u-1) m_j } \right)\cdot O_t(\tau, m_j) 
        + e^{(\mathrm{i}u-1)m} \left( O_t(\tau, m) - O_t(\tau, m_j) \right) \right] \diff m \Big\rvert \\
        &\leq \left(C_t (|u|\vee 1) \Delta m_j  e^{-m_{j-1}} e^{-[pm_{j-1} \vee -(1+q)m_{j-1}]} +  C_t e^{\Delta m}e^{-[(p+1)m_{j-1} \vee -qm_{j-1}]} \Delta m_j \right)\Delta m_j\\
        &\leq C_t e^{-[(p+1)m_{j-1} \vee -q m_{j-1}]} (\Delta m_j)^2.
    \end{align*}
    Then, for fixed $\overline{m}$ and $\underline{m}$, and $\Delta m \to 0$, we have 
    \begin{align}\label{zeta3_bounds}
        \zeta_t^{(3)}(u, \tau) = u_t \sum_{j=2}^n  \int_{m_{j-1}}^{m_j} \left[ e^{(\mathrm{i}u-1) m } \cdot O_t(\tau, m) - e^{(\mathrm{i}u-1)m_j} \cdot O_t(\tau, m_j) \right] \diff m  = \mathcal{O}_p(\Delta m).
    \end{align}

    The result~\eqref{zeta3_bounds}, however, needs to be adapted for the joint asymptotic scheme, where $\overline{m}$, $\underline{m}$ and $\Delta m$ all depend on $n$, with $n \to \infty$. 
    To this end, we first note that
    \begin{align*}
         \Big\rvert \sum_{j=2}^n &\frac{1}{F_t} \int_{m_{j-1}}^{m_j} \left[ e^{(\mathrm{i}u-1) m } \cdot O_t(\tau, m) - e^{(\mathrm{i}u-1)m_j} \cdot O_t(\tau, m_j) \right] \diff m \Big\rvert\\
         &\leq C_t \sum_{j=2}^{n}  e^{-[(p+1)m_{j-1} \vee -q m_{j-1}]} (\Delta m_j)^2 \\
         &\leq C_t \Delta m \sum_{j=2}^{n}  e^{-[(p+1)m_{j-1} \vee -q m_{j-1}]} \Delta m_j.
    \end{align*}
    The sum on the far right-hand side of the inequality is a Riemann approximation that converges to the following integral, as $n \to \infty$:
    \begin{align*}
        \sum_{j=2}^{n}  e^{-[(p+1)m_{j-1} \vee -q m_{j-1}]} \Delta m_j \longrightarrow& \int_{\underline{m}}^{\overline{m}} e^{-[(p+1)m \vee -q m]} \diff m 
        = \int_{\underline{m}}^{0} e^{q m} \diff m + \int_{0}^{\overline{m}} e^{-(p+1)m} \diff m \\
        &= \mathcal{O}_p\left( n^{-q \underline{\alpha}} \right) + \mathcal{O}_p\left( n^{-(p+1) \overline{\alpha}} \right)\\
        &= \mathcal{O}_p\left( n ^{- (q\underline{\alpha} \wedge (p+1) \overline{\alpha})} \right).
    \end{align*}
    Next, given Assumption~\ref{ass-moments}\textit{(ii)} on the log-moneyness grid, we can bound $\Delta m$ as 
    \begin{align*}
        \frac{\overline{m} - \underline{m}}{\eta n} \geq \Delta m \geq \frac{\overline{m} - \underline{m}}{n}.
    \end{align*}
    Hence, $\Delta m = \mathcal{O}_p\left(n^{-1}\log n \right)$.
    Thus, the order of magnitude of the discretization errors under the joint asymptotic scheme is given by
    \begin{align*}
        \zeta_t^{(3)}(u, \tau) = \mathcal{O}_p\left( n ^{- (q\underline{\alpha} \wedge (p+1) \overline{\alpha})} \right) \mathcal{O}_p\left(\frac{\log n}{n} \right) = \mathcal{O}_p\left(\frac{\log n}{n^{1 + q\underline{\alpha} \wedge (p+1) \overline{\alpha}}} \right).
    \end{align*}
    \hfill $\square$

    \subsection{Proof of Proposition 1}

    Using Lemma~\ref{lemma2}, Assumption~\ref{ass-moments} on the moments of the underlying process and observation error Assumption~\ref{ass:option_errors}, we can decompose the measurement errors in the CCF approximation as 
    \begin{align}
        \widehat{\phi}_t(u, \tau) - \phi_t(u, \tau) 
        &= \zeta_t^{(1)}(u, \tau) + \mathcal{O}_p\left(\frac{\log n}{n^{1 + 2(\underline{\alpha} \wedge \overline{\alpha})}} \vee n^{-2(\underline{\alpha} \wedge \overline{\alpha})}\right)\nonumber \\
        &=\zeta_t^{(1)}(u, \tau) + \mathcal{O}_p\left(n^{-2(\underline{\alpha} \wedge \overline{\alpha})}\right),\label{eq:CCFerrdec}
    \end{align}
    with 
    \begin{align*}
        \zeta_t^{(1)}(u, \tau) = - u_t \sum_{j=2}^{n} e^{(\mathrm{i}u-1)m_j} \cdot \zeta_t(\tau, m_j) \Delta m_j. 
    \end{align*}

    We now show that $\zeta_t^{(1)}(u, \tau) = \mathcal{O}_p\left( \sqrt{ n^{-1}\log n } \right)$. 
    In fact, the standard deviation of the observation errors is proportional to the Black-Scholes vega, which decreases with $|m| \to \infty$. 
    More specifically, the vega is given by
    \begin{align*}
        \nu_t(\tau, m) &= F_t \sqrt{\tau} \varphi(d_+), \\
        d_+ &= -m \varpi^{-1/2}(m) + \frac{1}{2}\varpi^{1/2}(m),
    \end{align*}
    where $\varphi(x)$ is the standard normal pdf and $\varpi(m):= \kappa^2(\tau, m) \tau$ is the total implied variance.\footnote{The total implied variance $\varpi(m)$ is a function of both the moneyness level $m$ and the time-to-maturity $\tau$. 
    For ease of notation, we write it as a function of moneyness only since the time-to-maturity $\tau$ is fixed in our exposition.} Hence, 
    \begin{align*}
        \nu^2_t(\tau, m) = F_t^2 \tau  \frac{1}{2 \pi} e^{-d_+^2} 
        = F_t^2 \tau \frac{1}{2 \pi} e^{-(\varpi^{-1}(m)m^2 - m + \frac{1}{4}\varpi(m))}.
    \end{align*}
    
    Therefore, given Assumption~\ref{ass:option_errors}, we obtain
    \begin{align*}
        \E\left[\Big\rvert \zeta^{(1)}_t(u, \tau) \Big\rvert^2 \Big\rvert \F_t \right] 
        &\leq  |u_t|^2 \sum_{j=2}^n e^{- 2 m_j} \cdot \E \left[ \zeta_t(\tau, m_j)^2 \rvert \F_t \right] (\Delta m_j)^2 \\
        &\leq  |u_t|^2 \sum_{j=2}^n e^{- 2 m_j} \cdot \sigma_{\varkappa}^2 \kappa^2(\tau, m_j)  F_t^2 \tau \frac{1}{2 \pi} e^{-(\varpi_j^{-1}m_j^2 - m_j + \frac{1}{4}\varpi_j)} (\Delta m_j)^2 \\
        &\leq  C_t \Delta m \sum_{j=2}^n \varpi_j  e^{-\varpi_j^{-1}m_j^2 - m_j - \frac{1}{4}\varpi_j} \Delta m_j\\
        &\leq  C_t \Delta m \sum_{j=2}^n \varpi_j  e^{-d_{-}^2(m_j)} \Delta m_j,
    \end{align*}
    where $d_{-}(m) := -m \varpi^{-1/2}(m) - \frac{1}{2}\varpi^{1/2}(m)$ and $\varpi_j := \varpi(m_j)$. 
    Then, as $n\to\infty$, the right-hand side summation converges to 
    \begin{align*}
        \int_{-\infty}^{\infty}   \varpi(m)\exp\left(-d_{-}^2(m)\right) \diff m =: \int_{-\infty}^{\infty} h(m) \diff m,
    \end{align*}
    provided that the function $h(m)$ is integrable. 
    
    To show the latter, we focus on the tail behavior, since $h$ is continuous and hence bounded on the bounded interval $[\underline{m}, \overline{m}]$.
    For that, we will make use of the following asymptotic results of \citeA{lee2004moment}: 
    \begin{align*}
        &\limsup_{m \to -\infty} \frac{\varpi(m) }{|m|} = \underline{\beta}^* \mbox{ with } \underline{\beta}^* \in [0,2], \quad\mbox{ and }\quad \frac{1}{2\underline{\beta}^*} + \frac{\underline{\beta}^*}{8} - \frac{1}{2} = \sup\{q:\E[F_{t+\tau}^{-q}|\F_t] < \infty\}, \\
         &\limsup_{m \to \infty} \frac{\varpi(m)}{|m|} = \overline{\beta}^* \mbox{ with } \overline{\beta}^* \in [0,2], \quad\mbox{ and }\quad \frac{1}{2\overline{\beta}^*} + \frac{\overline{\beta}^*}{8} - \frac{1}{2} = \sup\{p:\E[F_{t+\tau}^{1+p}|\F_t] < \infty\}.
    \end{align*}
    That is, for $m<0$, the total implied variance $\varpi(m)$ grows at most as fast as $-\underline{\beta}^* m$ for some $\underline{\beta}^* \in [0,2]$. 
    Given Assumption~\ref{ass-moments} on the moments of the reciprocal process, we further have $\underline{\beta}^* < 1$, which implies that $h$ is integrable over the negative domain, as for sufficiently small $m$,
    \begin{align*}
        0\leq h(m) = \varpi(m) \exp\left(-\varpi^{-1}(m)m^2 - m - \frac{1}{4}\varpi(m) \right) \leq -\underline{\beta}^*m \exp\left(\frac{m}{\underline{\beta}^*} - m\right).
    \end{align*}
    Integrability over the positive domain is achieved even without exploiting moment conditions.
    Therefore, since the right-hand side summation converges under the joint asymptotic scheme to the integral above, we have that $\zeta_t^{(1)}(u, \tau) = \mathcal{O}_p\left( \sqrt{\Delta m} \right) = \mathcal{O}_p\left( \sqrt{n^{-1} \log n}\right)$.

    Furthermore, from Assumption~\ref{ass:option_errors}, it also follows that $\E[\zeta_t^{(1)}(u, \tau)|\F_t] = 0$, while the discretization and truncation errors $ \zeta^{(2)}_t(u, \tau)$ and $ \zeta^{(3)}_t(u, \tau)$ are $\F_t$-measurable.
    Hence, the covariance and pseudo-covariance terms of the CCF approximation are given by the second moments of the observation errors $\zeta^{(1)}_t(u, \tau)$, that is,
    \begin{align*}
        \mbox{Cov}(\zeta^{\phi}_t(u_i, \tau), \zeta^{\phi}_t(u_j, \tau)) 
        :&= \E\left[ \left(\zeta^{\phi}_t(u_i, \tau) - \E[\zeta^{\phi}_t(u_i, \tau)]\right) \overline{\left(\zeta^{\phi}_t(u_j, \tau) - \E[\zeta^{\phi}_t(u_j, \tau)]\right) } \Bigg\rvert \F_t \right] \\
        & = \E\left[\zeta^{(1)}_t(u_i, \tau) \zeta^{(1)}_t(-u_j, \tau)\Big\rvert \F_t \right] \\
        & = u_{i,t} \overline{u_{j,t}} \sum_{j=2}^n e^{(\mathrm{i}(u_i - u_j)-2)m_j} \cdot \sigma^2_{t}(\tau, m_j) (\Delta m_j)^2\\
        & = \sigma_{\varkappa}^2 \cdot \underbrace{u_{i,t} \overline{u_{j,t}} \sum_{j=2}^n e^{(\mathrm{i}(u_i - u_j)-2)m_j} \cdot \kappa^2_{t}(\tau, m_j) \nu^2_{t}(\tau, m_j) (\Delta m_j)^2}_{=:\gamma_t(u_i, u_j, \tau)}\\
        & = \sigma_{\varkappa}^2 \cdot \gamma_t(u_i, u_j, \tau),
    \end{align*} 
    and
    \begin{align*}
        \mbox{PCov}(\zeta^{\phi}_t(u_i, \tau), \zeta^{\phi}_t(u_j, \tau))
        :&= \E\left[ \left(\zeta^{\phi}_t(u_i, \tau) - \E[\zeta^{\phi}_t(u_i, \tau)]\right) \left(\zeta^{\phi}_t(u_j, \tau) - \E[\zeta^{\phi}_t(u_j, \tau)]\right) \Bigg\rvert \F_t \right] \\
        & = \E\left[\zeta^{(1)}_t(u_i, \tau) \zeta^{(1)}_t(u_j, \tau) \Big \rvert \F_t \right] \\
        & = \sigma_{\varkappa}^2 \cdot \underbrace{u_{i,t} u_{j,t} \sum_{j=2}^n e^{(\mathrm{i}(u_i + u_j)-2)m_j} \cdot \kappa^2_{t}(\tau, m_j) \nu^2_{t}(\tau, m_j) (\Delta m_j)^2}_{=: c_t(u_i, u_j, \tau)}\\
        & = \sigma_{\varkappa}^2 \cdot c_t(u_i, u_j, \tau),
    \end{align*}
    for any $u_i, u_j \in \mathcal{U}$, where $\overline{z}$ denotes the complex conjugate of a complex number $z \in \mathbb{C}$. 
    In other words, the covariances of the total measurement errors in the CCF approximation are determined by the properties of the observation errors in option prices only.
    Note that the terms $\gamma_t(u_i, u_j, \tau)$ and $c_t(u_i, u_j, \tau)$ depend only on option's characteristics such as BSIV, BS vega and moneyness levels. 
    That is, the covariance terms are parametrized using only a single parameter $\sigma_\varkappa$ that reflects the variance of the observation errors in option prices.

    The measurement equation for the filtering problem is given in terms of the log CCF. 
    Therefore, by applying a Taylor-series expansion to the difference of the logs and using the error decomposition of the CCF approximation (in particular, \eqref{eq:CCFerrdec}), we have
    \begin{align*}
        \xi_t(u, \tau) := \log \widehat{\phi}_t(u, \tau) - \log \phi_t(u, \tau) 
        &= \log \left(1 + \frac{\zeta^{(1)}_t(u, \tau) + \zeta^{(2)}_t(u, \tau) + \zeta^{(3)}_t(u, \tau)}{\phi_t(u, \tau)}\right)\\ 
        &= \xi_t^{(1)}(u, \tau)  + r_t(u, \tau),
    \end{align*}
    where 
    \begin{align*}
        \xi_t^{(1)}(u, \tau) := \frac{\zeta^{(1)}_t(u, \tau) }{\phi_t(u, \tau)} = \mathcal{O}_p\left( \sqrt{\frac{\log n}{n}}\right),\quad \mbox{ and } \quad
        r_t(u, \tau) = \mathcal{O}_p\left(n^{-2(\underline{\alpha} \wedge \overline{\alpha})} \vee \frac{\log n}{n} \right).
    \end{align*}
    We note that the remainder term collects the log-linearization of the truncation and discretization errors and higher-order terms from a Taylor-series expansion.

    After stacking each component of the measurement equation~\eqref{key_relation2} as well as the observation errors $\xi_t^{(1)}(u, \tau)$ and remainder term $r_t(u, \tau)$ along arguments, real and imaginary parts, and maturity, we obtain the state space measurement equation~\eqref{ss-observ} in Proposition~\ref{prop1}.

    To derive the covariance matrix of the measurement errors, we first consider the covariance and pseudo-covariance matrices of the stacked vector $\xi_{t, \tau}^{(1)} = \left(\xi_t^{(1)}(u_1, \tau), \dots, \xi_t^{(1)}(u_q, \tau)\right)'$ for a fixed time $t$ and time-to-maturity $\tau$.
    They are given by 
    \begin{align*}
        \Gamma_{t, \tau} 
        :&= \E\left[\xi_{t, \tau}^{(1)} \overline{\xi_{t, \tau}^{(1)}}'\right]
        = \left[ \E[\xi_t^{(1)}(u_i, \tau) \xi_t^{(1)}(-u_j, \tau)] \right]_{1\leq i,j \leq q}\\
        &= \sigma_{\varkappa}^2 \cdot \left[ \frac{\gamma_t(u_i, u_j, \tau)}{\phi_t(u_i, \tau) \phi_t(-u_j, \tau)} \right]_{1\leq i,j \leq q}
        =: \sigma_{\varkappa}^2 \cdot \widetilde{\Gamma}_{t, \tau}, \notag\\
        %
        C_{t, \tau} 
        :&= \E\left[\xi_{t, \tau}^{(1)} {\xi_{t, \tau}^{(1)}}'\right]
        = \left[ \E[\xi_t^{(1)}(u_i, \tau) \xi_t^{(1)}(u_j, \tau)] \right]_{1\leq i,j \leq q}\\
        &= \sigma_{\varkappa}^2 \cdot \left[ \frac{c_t(u_i, u_j, \tau)}{\phi_t(u_i, \tau) \phi_t(u_j, \tau)} \right]_{1\leq i,j \leq q}
        =: \sigma_{\varkappa}^2 \cdot \widetilde{C}_{t, \tau}. \notag
    \end{align*}
    Next, since $\xi_{t, \tau}^{(1)}$ is a complex-valued random vector, the covariance matrix of the stacked real and imaginary parts of $\xi^{(1)}_{t, \tau}$ is of the following form:
    \begin{align*}
        H_{t, \tau}:=
        \mbox{Var}\left[ 
            \begin{pmatrix}
                \Re(\xi_{t, \tau}^{(1)})\\
                \Im(\xi_{t, \tau}^{(1)})
            \end{pmatrix}
        \right]
        &=
        \begin{pmatrix}
            \frac{1}{2} \Re(\Gamma_{t, \tau} +  C_{t, \tau}) & \frac{1}{2} \Im(-\Gamma_{t, \tau} +  C_{t, \tau})\\
            \frac{1}{2} \Im(\Gamma_{t, \tau} +  C_{t, \tau}) & \frac{1}{2} \Re(\Gamma_{t, \tau} - C_{t, \tau})
        \end{pmatrix}\\
        &= \sigma_\varkappa^2 \cdot
        \begin{pmatrix}
            \frac{1}{2} \Re(\widetilde\Gamma_{t, \tau} + \tilde C_{t, \tau}) & \frac{1}{2} \Im(-\widetilde\Gamma_{t, \tau} + \widetilde C_{t, \tau})\\
            \frac{1}{2} \Im(\widetilde\Gamma_{t, \tau} + \widetilde C_{t, \tau}) & \frac{1}{2} \Re(\widetilde\Gamma_{t, \tau} - \widetilde C_{t, \tau})
        \end{pmatrix}\\
        &=: \sigma_\varkappa^2 \cdot \widetilde{H}_{t,\tau}.
    \end{align*}
This establishes~\eqref{complex_covar}.

Given Assumption~\ref{ass:option_errors}, the error terms $\zeta^{(1)}_{t, \tau}$ and $\xi^{(1)}_{t, \tau}$ are conditionally independent along maturity and time. 
This implies that the measurement errors $\varepsilon_t$ stacked along maturities are also conditionally independent, thus $\E[\varepsilon_t \varepsilon_s'] = 0$ for $s\neq t$, and their covariance matrix has a block-diagonal form: $H_t = \mbox{blkdiag}\{H_{t,1},\dots,H_{t,k}\}$. 

The disturbance term in the state updating equation is given by $\eta_{t+1} = x_{t+1} - \E[x_{t+1}|\F_t]$. 
Therefore, $\eta_t$ constitutes a martingale difference sequence, thus $\E[\eta_t \eta_s'] = 0$ for $s\neq t = 1, \dots, T$.

Since the measurement errors $\varepsilon_t$ have zero mean conditional on the filtration $\F_t$, we also have that $\E[\varepsilon_t x_t] = 0$. 
Given that the state process is stationary and the initial condition is the unconditional mean, $\E[\varepsilon_t x_1'] = 0$ and $\E[\eta_{t+1} x_1'] = 0$. 
This implies that $\E[\varepsilon_t \eta_s'] = 0$ for all $s,t=1,\dots,T$.
Thus, the proof is established.
\hfill $\square$

\section{Conditional Moments}
\label{sec:Appendix-CM}

In this appendix, we describe how the conditional mean and variance can be computed for the AJD class of models. 
In particular, we derive closed-form expressions for the conditional mean and variance in the univariate case, and briefly discuss how these moments can be obtained numerically in the multivariate setting at low computational costs. 
While semi-closed-form expressions are also available in the multivariate setting, they are more cumbersome to work with in practice since they typically require matrix exponentials and integrations.

We start with considering the univariate version of the AJD process in~\eqref{AJD-x}, using shorthand notation as follows:
    \begin{align}\label{AJD1}
        \diff x_t = \mu(x_t) \dt + \sigma(x_t) \diff W_t + J_t \diff N_t,
    \end{align}
    with $\mu(x) = k_0 + k_1 x,\ \sigma^2(x) = h_0 + h_1x,\ \lambda(x) = l_0 + l_1 x$, where all coefficients are real-valued numbers and the standard Brownian motion $W_t$ and the counting process $N_t$ are univariate processes. 
The jump size distribution $\nu$ on $\R$ is independent of time and of any form of randomness in the model. 
We further assume that the SDE~\eqref{AJD1} has a unique strong solution and the first two moments are well defined. 
For more details, see Section~\ref{sec:Framework2} and~\citeA{DPS2000}. 
For notational simplicity, let 
    \begin{align*}
        \mu_J := \E[J],\quad \mu_{J2}:= \E[J^2],\quad
        g_0 := k_0 + l_0 \mu_J, \quad 
        g_1 := k_1 + l_1 \mu_J.
    \end{align*}

The associated infinitesimal generator $\mathcal{D}$, defined at a bounded $C^2$ function $f{:}\ D\to \R$, with bounded first and second derivatives $f_x$ and $f_{xx}$, is given by
    \begin{align*}
        \mathcal{D}f(x) = f_x(x) \mu(x) + \frac{1}{2} f_{xx}(x) \sigma(x)^2 + \lambda(x) \int_{\R}[f(x+z) - f(x)] \diff \nu(z).
    \end{align*}
Dynkin's formula yields that
    \begin{align*}
        \E[f(x_{T})| \F_t] = f(x_t) + \E\left[ \int_t^{T} \mathcal{D}f(x_s) \diff s \big \rvert \F_t \right].
    \end{align*}
Therefore, we can find the conditional moments by applying Dynkin's formula for $f(x) = x$:
    \begin{align*}
        \E[x_{T}|\F_t] 
        &= x_t + \E\left[ \int_t^{T} \left( \mu(x_s) + \lambda(x_s) \int_{\R}z \diff \nu(z) \right) \diff s \big \rvert \F_t \right]\\
        &= x_t + \E\left[ \int_t^{T} \left( k_0 + k_1 x_s + (l_0 + l_1 x_s) \mu_{J} \right)  \diff s \big \rvert \F_t \right]\\
        &= x_t + \int_t^{T} \left( k_0 + l_0 \mu_{J} +  ( k_1  +  l_1 \mu_{J} )  \E [x_s \big \rvert \F_t] \right) \diff s\\
        &= x_t + \int_t^{T} \left( g_0 +  g_1  \E [x_s \big \rvert \F_t] \right) \diff s,
    \end{align*}
where Fubini's theorem is used in the third line.
Hence, we can obtain the first conditional moment by solving the following ODE:
    \begin{align*}
        \diff \E[x_s|\F_t] = \left(g_0 + g_1 \E [x_s \big \rvert \F_t] \right) \ds,
    \end{align*} 
with initial condition $\E[x_t|\F_t] = x_t$. 
Thus, the conditional expectation is given by
    \begin{align}\label{Mean1}
        m_t(T):= \E[x_T| \F_t] = e^{g_1(T-t)} x_t + \frac{g_0}{g_1}\left(e^{g_1(T-t)} - 1 \right).
    \end{align}

Next, we are interested in deriving the conditional variance:
    \begin{align*}
        \mbox{Var}(x_T | \F_t) = \E[(x_T - \E[x_T| \F_t])^2 | \F_t].
    \end{align*}
Note that 
    \begin{align*}
        x_T - \E[x_T| \F_t] = \E[x_T| \F_T] - \E[x_T| \F_t] = \int_t^T \diff \E[x_T| \F_s] = \int_t^T \diff m_s(T).
    \end{align*}
The dynamics of the conditional mean for fixed $T > t$ can be obtained by using It\^o's lemma:
    \begin{align*}
        \diff m_t(T) &= \left[ -g_1 e^{g_1(T-t)}x_t - g_0 e^{g_1(T-t)} \right] \dt + e^{g_1 (T-t)}(\mu(x_t) \dt + \sigma(x_t) \diff W_t) + e^{g_1 (T-t)}J_t \diff N_t\\
        &= e^{g_1(T-t)} \left[ -(l_0 + l_1 x_t) \mu_{J} \dt + \sigma(x_t) \diff W_t + J_t \diff N_t \right].
    \end{align*} 

Note that the process $m_t(T)$ for fixed $T$ is a local martingale. 
Thus, we can use the It\^o isometry to obtain the conditional variance:
    \begin{align*}
        \mbox{Var}(x_T | \F_t) &= \E\left[ \left( \int_t^T \diff m_s(T) \right)^2 \Big\rvert \F_t \right]\\
        &= \E\left[ \int_t^T e^{2g_1(T-s)} \sigma^2(x_s) \ds \Big\rvert \F_t \right] + \mu_{J2} \cdot \E \left[ \int_t^T e^{2g_1(T-s)} \lambda(x_s) \ds \Big\rvert \F_t \right]\\
        &=  \int_t^T e^{2g_1(T-s)} (h_0 + h_1 \E[x_s|\F_t]) \ds + \mu_{J2} \cdot \int_t^T e^{2g_1(T-s)} (l_0 + l_1 \E[x_s|\F_t]) \ds\\
        &=  (h_0 + l_0 \mu_{J2}) \int_t^T e^{2g_1(T-s)}  \ds + (h_1 + l_1 \mu_{J2}) \cdot \int_t^T e^{2g_1(T-s)} \E[x_s|\F_t] \ds,
    \end{align*}
where we have again used Fubini's theorem in the third line. 
Given the conditional expectation, the second integral on the far right-hand side can be simplified further:
    \begin{align*}
        \int_t^T &e^{2g_1(T-s)} \E[x_s|\F_t] \ds = \int_t^T e^{2g_1(T-s)} \left[ e^{g_1(s-t)} x_t + \frac{g_0}{g_1}\left(e^{g_1(s-t)} - 1 \right) \right] \ds \\
        &= e^{2g_1T} \left[ \int_t^T e^{-g_1(s+t)} x_t + \frac{g_0}{g_1}\left(e^{-g_1(s+t)} - e^{-2g_1 s} \right) \ds \right]\\
        &= e^{2g_1T} \left[ -\frac{1}{g_1} \left(e^{-g_1(T+t)} - e^{-2g_1t} \right) x_t 
        - \frac{g_0}{g_1^2}\left(e^{-g_1(T+t)} - e^{-2g_1t}\right)  + \frac{g_0}{2g_1^2} \left(e^{-2g_1 T} - e^{-2g_1 t} \right) \right]\\
        &= -\frac{1}{g_1} \left(e^{g_1(T-t)} - e^{2g_1(T-t)} \right) x_t + \frac{g_0}{2 g_1^2}\left( 1 - e^{g_1(T-t)} \right)^2.
    \end{align*}
Thus, the conditional variance in the univariate case is given by 
    \begin{align}\label{Var1}
        \mbox{Var}(x_T | \F_t) = & -\frac{1}{2g_1} (h_0 + l_0 \mu_{J2}) \left( 1 - e^{2g_1(T-t)}  \right) \notag\\ 
        &- \frac{1}{2g_1^2}  (h_1 + l_1 \mu_{J2} )  \left[ 2 g_1\left(e^{g_1(T-t)} - e^{2g_1(T-t)} \right) x_t - g_0\left( 1 - e^{g_1(T-t)} \right)^2\right].
    \end{align}

Equations~\eqref{Mean1} and~\eqref{Var1} serve as the basis for the formulation of the transition equation~\eqref{ss-state} as discussed in Section~\ref{sec:Estimation-ss}.
It is crucial for our application to note that the conditional mean~\eqref{Mean1} and conditional variance~\eqref{Var1} of the univariate AJD process $x_{T}$ at time $T > t$, conditional on information at time $t$, are affine functions in $x_t$. 
The affinity of the conditional moments yields the linear state updating equation, which, in turn, allows us to use the linear Kalman filtering technique.

Using the same reasoning, it is in principle also possible to derive the analogues of equations~\eqref{Mean1} and~\eqref{Var1} for the multivariate AJD process. 
However, these expressions typically involve matrix exponentials and integrals thereof, which makes them burdensome to work with. 
Fortunately, the conditional moments can easily be obtained numerically by differentiating the CCF, which, as discussed in Section~\ref{sec:Framework2} is known in semi-closed form for the AJD class. 
Indeed, finite difference approximations of the first and second derivatives around the origin yield the moments with high precision and little additional computational costs.
One can also easily verify that the affine property of the conditional moments holds in the multivariate case by differentiating the exponentially-affine CCF.

\section{Interpolation-Extrapolation Scheme\\ and CCF Replication}\label{sec:Appendix-InterExtraSpanning}


In this appendix, we discuss in detail the option interpolation-extrapolation scheme we adopt and illustrate the impact of the different measurement errors on the option-implied CCF `payoff' replication. 

\subsection{Interpolation-extrapolation scheme}
   \label{sec:Appendix-Extrapolation}
   
  \subsubsection{Interpolation} 
    
For each trading day and for each tenor, we interpolate option prices between moneyness levels using cubic splines. 
For interpolation, we consider option data expressed in terms of their total implied variance, defined as $\varpi(m, \tau) = \kappa^2(m, \tau) \cdot \tau$, where $\kappa(m, \tau)$ is the Black-Scholes implied volatility for an option with log-moneyness $m$ and tenor $\tau$. 
This is similar to interpolating on the implied volatility domain, but it will provide us further advantages when we proceed to the extrapolation scheme, discussed in the next subsection.

Cubic splines provide a useful tool for the interpolation of options data and are commonly employed for this purpose in the literature; see, for instance, \citeA{jiang2007extracting}, \citeA{malz2014simple} among many others. 
Furthermore, they are also used as an approximation method that allows to smooth out noise in the data; see, for instance, \citeA{bliss2002testing}, \citeA{fengler2009arbitrage}. 
For the latter, it is common to penalize the squared second derivative of the spline. 
This might, however, induce a loss of flexibility of the spline leading to larger approximation errors, especially for short-dated options, which are of pivotal importance in our analysis.
In this paper, we therefore use a standard cubic spline, but instead of providing all data as knot points for spline interpolation, we explicitly specify which data points shall be used as knots. 
This allows us to interpolate in some domains and smooth out in others, taking the `best' out of the spline interpolation and approximation schemes. 

\begin{figure}
        \centering
        \caption{Spline interpolation-extrapolation example: April 1, 2021, 15 days to maturity}
        \begin{subfigure}{0.45\textwidth}
            \hspace*{-1.5cm}
            \includegraphics[scale=0.4]{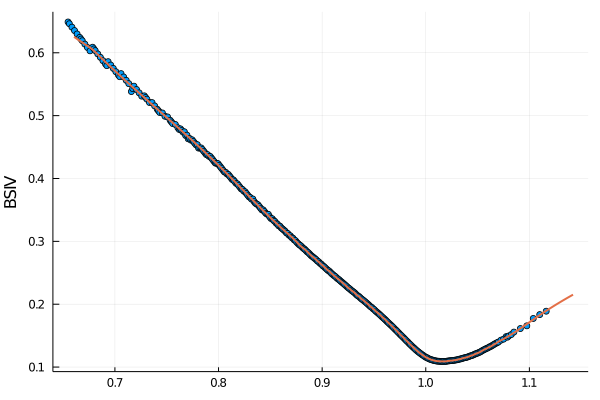}
        \end{subfigure}
        \begin{subfigure}{0.45\textwidth}
            \includegraphics[scale=0.4]{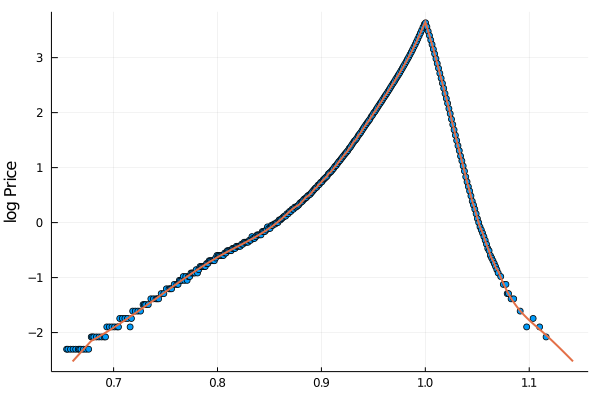}
        \end{subfigure}
        \label{fig:spl_example}

        \medskip
        \begin{minipage}{\textwidth}\scriptsize
            Note: This figure plots an example of the interpolation-extrapolation scheme for options traded on April 1, 2021 with 15 days to maturity. 
            The option data (blue dots) are interpolated using a cubic spline (orange line). 
            Interpolation is conducted on the total implied variance domain. 
            The left panel plots the data in terms of Black-Scholes implied volatility, whereas the right panel plots the data in terms of log prices. 
            Moneyness $K/F_t$ is on the horizontal axis.
        \end{minipage}
    \end{figure}

Close to ATM options are more liquid than very deep OTM counterparts. 
Thus, intuitively, information in the former options is more reliable, and we would not like to distort this information by imposing smoothing constraints. 
Very deep OTM options, on the other hand, may be quite illiquid. 
Furthermore, the tick size for deep OTM options becomes large relative to their value. 
This might lead to observing a sequence of the same midpoint quote prices in the data. 
Figure~\ref{fig:spl_example} provides an example of such `flat' prices for put options, visible in the right panel for very deep OTM options (i.e., small $k$). 
These prices clearly violate no-arbitrage assumptions. 
However, throwing them away would reduce available information, needed to extract the CCF; these prices are not uninformative, but the tick size distorts their information. 
Therefore, instead of eliminating `flat' prices, we will just not include them as knot points in our spline interpolation scheme. 
In other words, we do not require the spline function to go through all data points for deep OTM options, but rather let it approximate the information in them.\footnote{Recall that we interpolate/approximate data on the total implied variance domain, not in terms of implied volatility, option prices or log prices as considered in Figure~\ref{fig:spl_example}.}

More formally, we first include the closest to ATM put option, $m_{n^*}$, in the knot sequence and then iteratively include put options with smaller moneyness level $m_i$ for $i=n^*{-}1,\dots,2$ such that all of the following conditions are satisfied: ($i$) $P(m_i) < P(m_{i-1})$ and $C(m_i) > C(m_{i-1})$; ($ii$) $P(m_{i+1}) < P(m_{i})$ and $C(m_{i+1}) > C(m_{i})$; ($iii$) daily trading volume for $P(m_i)$ is larger than one. 
The first two criteria check for no-arbitrage conditions. 
The third one filters out possible stale prices from being a knot point. 
Similar mirrored conditions are applied to OTM call prices. 
The knot sequence thus constructed will likely contain more close to ATM options and fewer deep OTM options, resulting in more interpolation in the former range and more approximation in the latter one.
    
We emphasize again that we do not filter out option data that violate no-arbitrage conditions, which would reduce available information for CCF extraction. 
Instead, we do not include these points into the knot sequence, thus we do not require the spline to go exactly through these points. 
Furthermore, another reason not to filter out options that violate no-arbitrage conditions, is that we use option-implied CCFs rather than option prices themselves as inputs in our estimation procedure. 
Similarly, the CBOE does not impose any no-arbitrage filters in the calculation of the VIX index, except for eliminating zero-bid quotes \cite{CBOE}.
Figure~\ref{fig:spl_example} provides an example of the interpolation-extrapolation scheme for an option slice traded on April 1, 2020, with 15 days to maturity.
    
\subsubsection{Extrapolation}

Truncation errors are, in a sense, more challenging to address than discretization errors, since they require to make assumptions about the dynamics of option prices (either in dollar or volatility terms) beyond the observable range of strikes. 
On the other hand, as prices of OTM options decrease with $|m|$, the impact of the truncation errors is expected to be small for highly liquid options that cover a wide range of strike prices (such as index options).
Nevertheless, truncation might deteriorate the CCF approximation even for small argument values. 
This can especially be a relevant issue after a sudden market shock, since options with smaller or larger strikes might not be issued immediately to cover a new range of strikes.  

It is common in the literature to use flat extrapolation; see again e.g.,  \citeA{bliss2002testing}, \citeA{jiang2005model} and \citeA{malz2014simple}.  
Under a flat extrapolation scheme, the implied volatility beyond the observable
range of strikes is simply set equal to the volatility of the observable extreme-strike options, i.e., $\kappa(\underline{m},\tau)$ for the left-hand side of the volatility smile and $\kappa(\overline{m},\tau)$ for the right-hand side. 
This approach is very easy to implement. 
However, the main caveat of flat extrapolation is that it assumes the Black-Scholes log-normal model to apply in the tails, for the extrapolated range of strikes. 

Instead, we extrapolate the total implied variance $\varpi(m, \tau)$ linearly in log-moneyness $m$ beyond the observable range of strikes. 
This particular linear parametrization is motivated by the asymptotic results of \citeA{lee2004moment}, who analyzed the behavior of the implied volatility smile as strikes tend to infinity. 
Another example of a parametrization that satisfies Lee's asymptotic results is the SVI model, commonly used among practitioners \cite{gatheral2014arbitrage}. 
However, it is well known that the SVI approach may not provide a good fit for short-dated options. 
Thus, different from SVI, we use the more flexible cubic spline for interpolation within the observable range of strikes, as detailed above, and, similar to SVI, extrapolate implied variance linearly in log-moneyness.

The asymptotic results of \citeA{lee2004moment}, exploited also in Appendix~\ref{sec:Appendix-Proofs} and recalled here for convenience, entail that the implied volatility wings should not grow faster than $|m|^{1/2}$ and, unless the underlying asset has finite moments of all orders, should not grow slower than $|m|^{1/2}$. 
More specifically, \citeA{lee2004moment} first shows that 
    \begin{align}
        \label{limsup_left}
        &\limsup_{m \to -\infty} \frac{\kappa^2 (\tau, m) \tau}{|m|} = \underline{\beta}^* \mbox{ with } \underline{\beta}^* \in [0,2], \quad \mbox{ and }\\
        \label{limsup_right}
         &\limsup_{m \to \infty} \frac{\kappa^2 (\tau, m) \tau}{|m|} = \overline{\beta}^* \mbox{ with } \overline{\beta}^* \in [0,2].
    \end{align}
Furthermore, he establishes that there is a one-to-one correspondence between $\overline{\beta}^*$ and the number of finite moments of the underlying process $F_{\tau}$, and between $\underline{\beta}^*$ and the number of finite moments of $1/F_{\tau}$. 
For instance, for the right tail, the moment formula for implied volatility is given by
    \begin{align*}
        \frac{1}{2 \overline{\beta}^*} + \frac{\overline{\beta}^*}{8} - \frac{1}{2}
        = \sup\{p: \E[F_\tau^{1+p}] < \infty\}.
    \end{align*}

These results allow us to conjecture the asymptotically valid parametrization to extrapolate implied volatility beyond the observable range of strikes. 
Hence, we assume that the total variance $\varpi(m) = \kappa^2( \tau, m) \tau$ is an affine function of log-moneyness:
    \begin{align*}
        \varpi(m) = c + \beta  m.
    \end{align*}
An intercept coefficient is introduced to guarantee continuity between the interpolation and extrapolation domains.
The intercept coefficients for the left and right tails, denoted by $\underline{c}$ and $\overline{c}$, are exactly determined by the smallest and largest observable strike prices (or corresponding log-moneyness levels) given the slopes $\underline{\beta}$ and $\overline{\beta}$ for the left and right tails, respectively: 
    \begin{align*}
        \underline{c} = \varpi(\underline{m}) - \underline{\beta}  \underline{m} 
        \quad \mbox{and} \quad 
        \overline{c} = \varpi(\overline{m}) - \overline{\beta} \overline{m}.
    \end{align*}

Therefore, what is left to be done is to establish the choice of the slope coefficients $\underline{\beta}$ and $\overline{\beta}$. 
Note that the formulas~\eqref{limsup_left} and~\eqref{limsup_right} provide asymptotic bounds for the slope coefficients.
Furthermore, finding the number of finite moments of the underlying, and exploiting the respective moment formulas, would require parametrizing the dynamics of $F_\tau$.\footnote{Note that flat extrapolation assumes log-normality of the underlying asset in the tails. 
Since all moments of the log-normal distribution exist, it means that the slope indeed has to be zero in this case.} The latter is not desirable in our application, since we want to fit another parametric model afterwards. 
Instead, we simply use the derivatives of the fitted cubic splines at the last observable strikes to determine the slope coefficients: 
    \begin{align*}
        \underline{\beta} = -\frac{\partial \varpi(m)}{\partial m} \Bigg \rvert_{m=\underline{m}} \quad
        \mbox{ and }\quad
        \overline{\beta} = \frac{\partial \varpi(m)}{\partial m} \Bigg \rvert_{m=\overline{m}}.
    \end{align*}

Lee's bounds for the slopes constitute an asymptotic result. 
The chosen slopes $\underline{\beta}$ and $\overline{\beta}$ should satisfy these bounds. (Note that, due to the adopted sign convention in the extrapolation formula, this translates into $\underline{\beta} \in [-2, 0]$ for the left slope.) 
However, no-arbitrage conditions for our parametrization can be tighter, since we are in a setting with finite log-moneyness levels. 
To obtain these conditions, we  follow the derivation sketched in \citeA{jackel2014clamping}. 
This yields the following no-arbitrage bounds for the right-tail slope $\overline{\beta}$ (our detailed derivations are available upon request; they are suppressed to save space):
    \begin{align*}
        0 \leq \overline{\beta} < \min(\beta_{max}, 2),
    \end{align*} 
    where
    \begin{align*}
        \beta_{max} = \begin{cases}
            \max \left(\frac{\overline{m}(\overline{\varpi} - 2) + \sqrt{\overline{\Delta}}}{\overline{m}^2 + 1}, \frac{-2\overline{m} + 2\sqrt{\overline{m}^2 + 2 \overline{\varpi}^2 + 4 \overline{\varpi}}}{\overline{\varpi} + 2}\right), &\quad \mbox{ if } \overline{\Delta} > 0;\\
            \frac{-2\overline{m} + 2\sqrt{\overline{m}^2 + 2 \overline{\varpi}^2 + 4 \overline{\varpi}}}{\overline{\varpi} + 2}, &\quad \mbox{ if } \overline{\Delta} \leq 0;
        \end{cases}
    \end{align*}
    with $\overline{\varpi}:= \varpi(\overline{m})$ and $\overline{\Delta}:= 4\overline{m}^2 - \overline{\varpi}^2 + 4\overline{\varpi}$.

Similarly, for the left-tail slope $\underline{\beta}$,
    \begin{align*}
        \max(\beta_{min}, -2) < \underline{\beta} \leq 0,
    \end{align*} 
    where
    \begin{align*}
        \beta_{min} = \begin{cases}
            \max \left(\frac{\underline{m}(\underline{\varpi} - 2) - \sqrt{\underline{\Delta}}}{\underline{m}^2 + 1}, \frac{-2\underline{m} - 2\sqrt{\underline{m}^2 + 2 \underline{\varpi}^2 + 4 \underline{\varpi}}}{\underline{\varpi} + 2}\right), &\quad \mbox{ if } \underline{\Delta} > 0;\\
            \frac{-2\underline{m} + 2\sqrt{\underline{m}^2 + 2 \underline{\varpi}^2 + 4 \underline{\varpi}}}{\underline{\varpi} + 2}, &\quad \mbox{ if } \underline{\Delta} \leq 0;
        \end{cases}
    \end{align*}
    with $\underline{\varpi}:= \varpi(\underline{m})$ and $\underline{\Delta}:= 4\underline{m}^2 - \underline{\varpi}^2 + 4\underline{\varpi}$.

\subsection{CCF replication}
\label{sec:Appendix-CCFReplication}

\begin{figure}[!ht]
            \centering
            \caption{The three types of measurement errors}
            \begin{subfigure}{0.45\textwidth}
                \hspace*{-1.0cm}
                \includegraphics[scale=0.07]{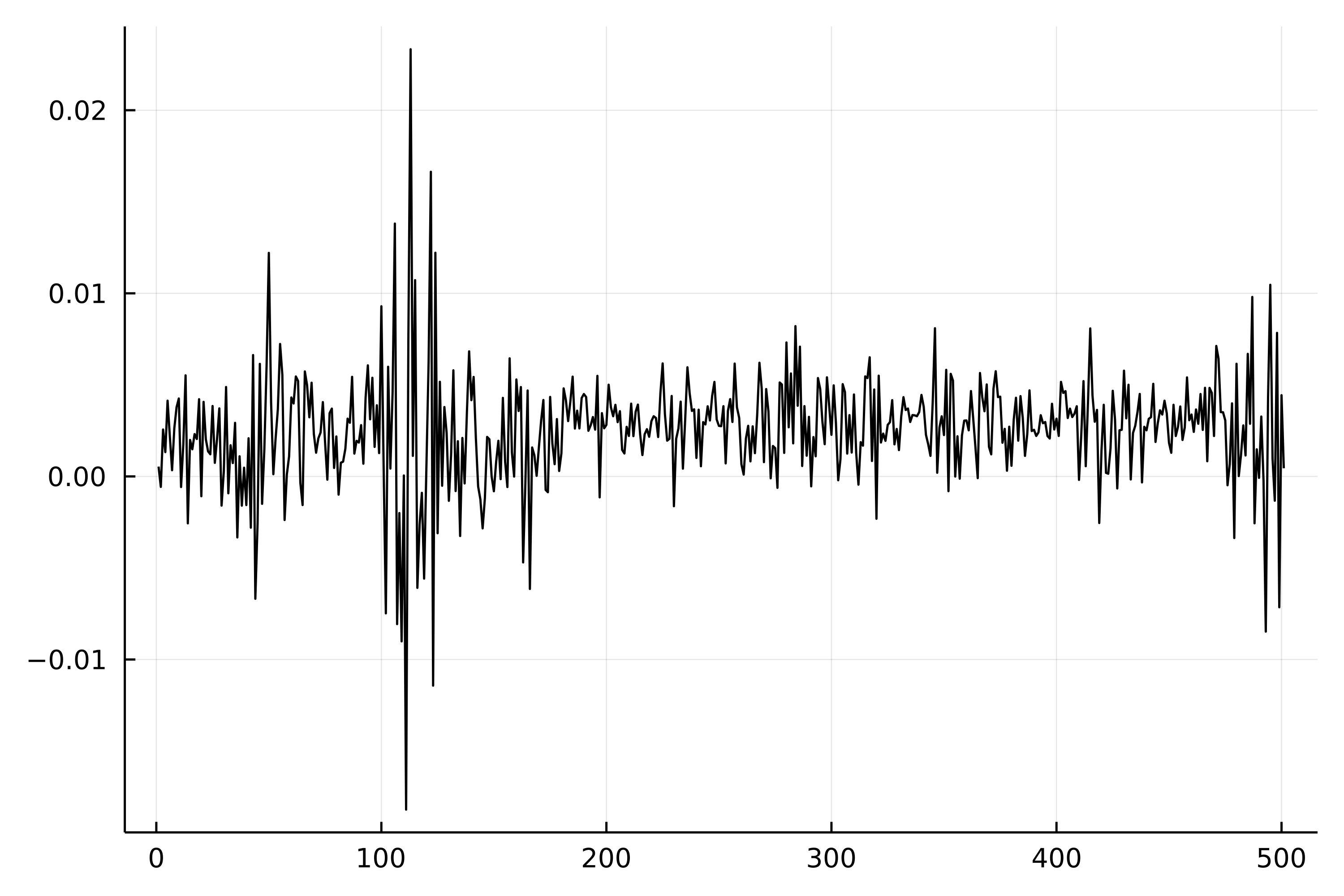}
                \caption{$\zeta^{(1)}_t + \zeta^{(2)}_t + \zeta^{(3)}_t$}
            \end{subfigure}
            \begin{subfigure}{0.45\textwidth}
                \includegraphics[scale=0.07]{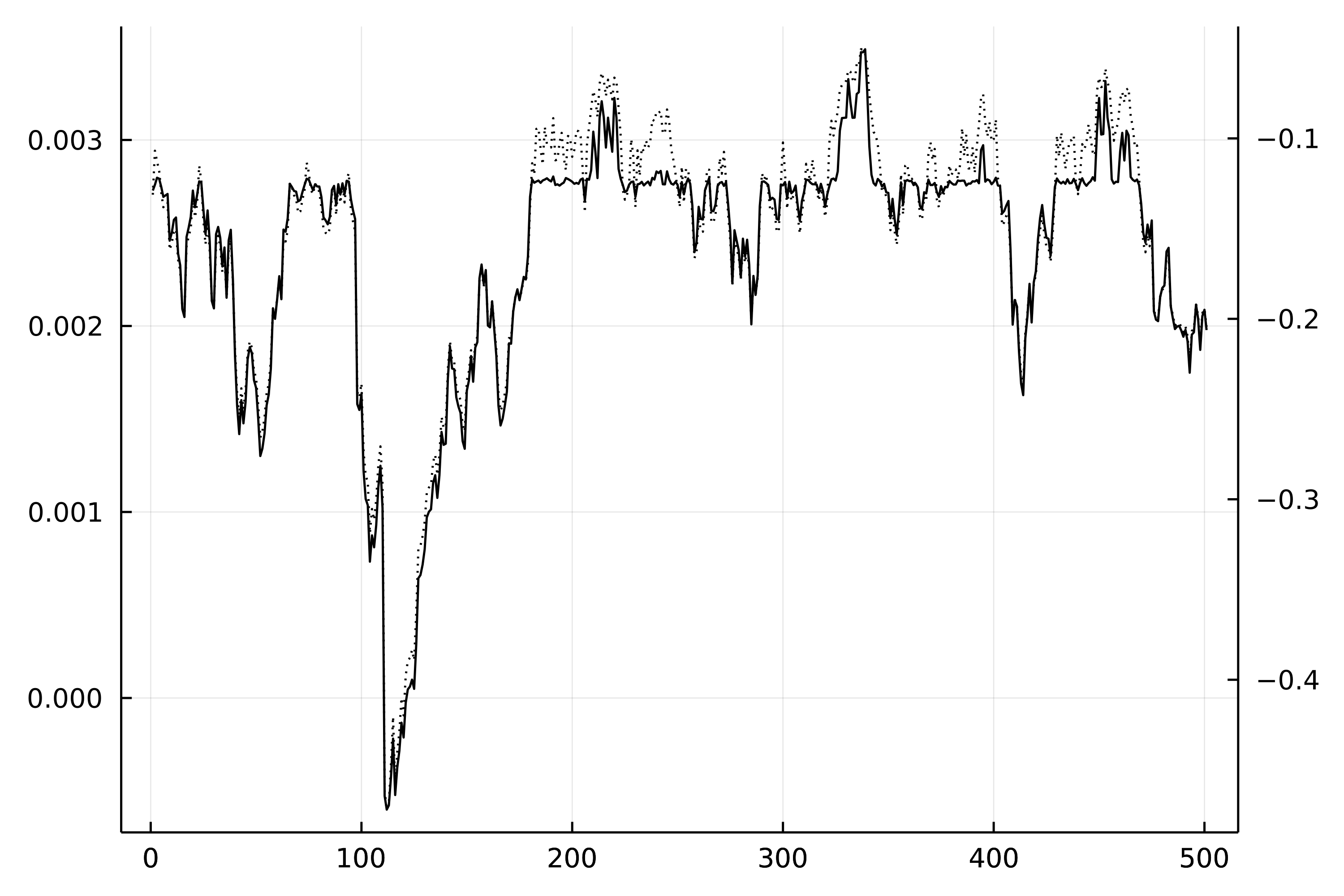}          
                \caption{$\zeta^{(2)}_t+\zeta^{(3)}_t$ (solid) and $-$ATM BSIV (dots)}
            \end{subfigure}
            \begin{subfigure}{0.45\textwidth}
            \hspace*{-1.0cm}
                \includegraphics[scale=0.07]{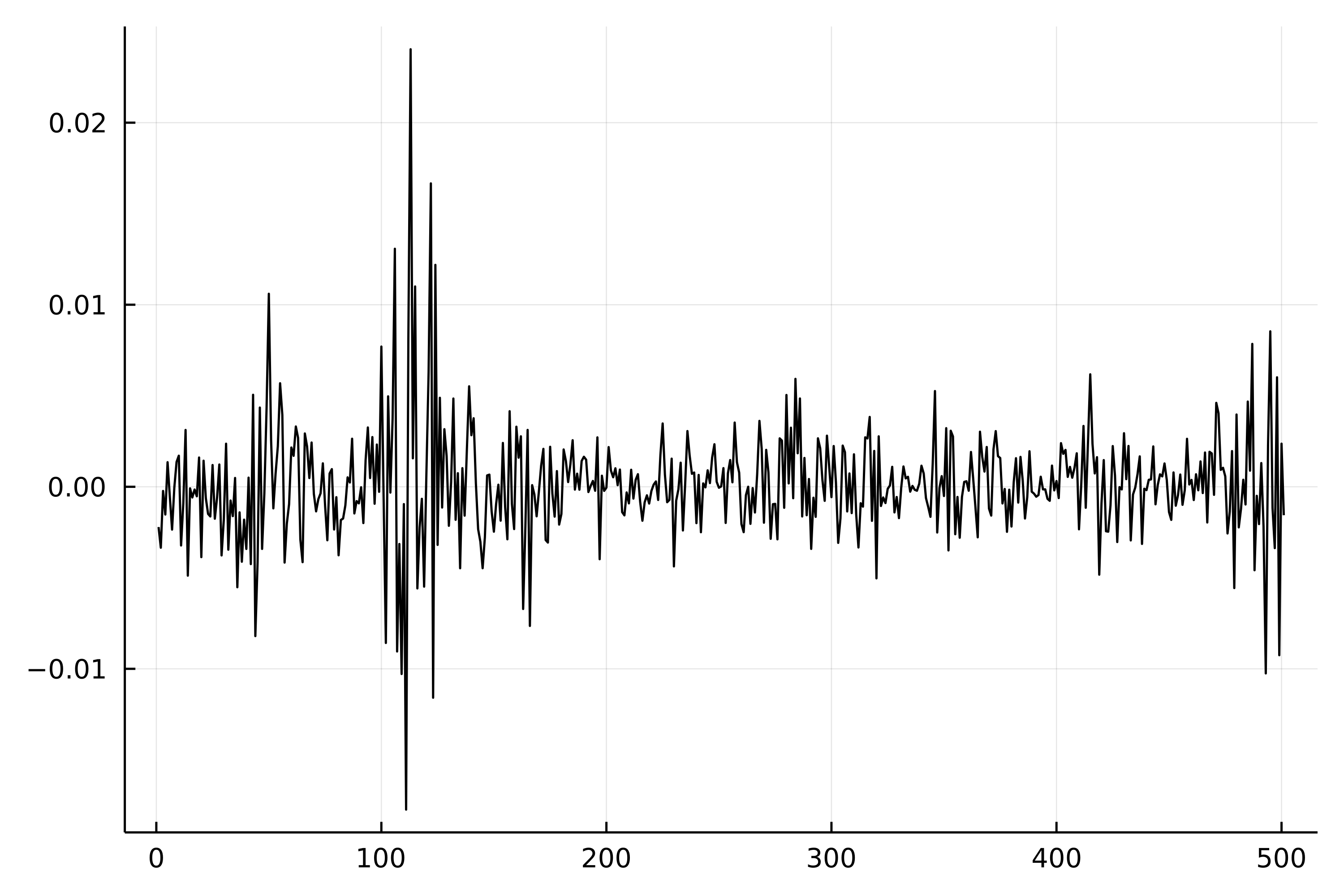}
            \caption{$\phi(u,\tau) - \phi(spl(\widehat{O}))$\\ \ \ }
            \end{subfigure}
            \begin{subfigure}{0.45\textwidth}
                \includegraphics[scale=0.07]{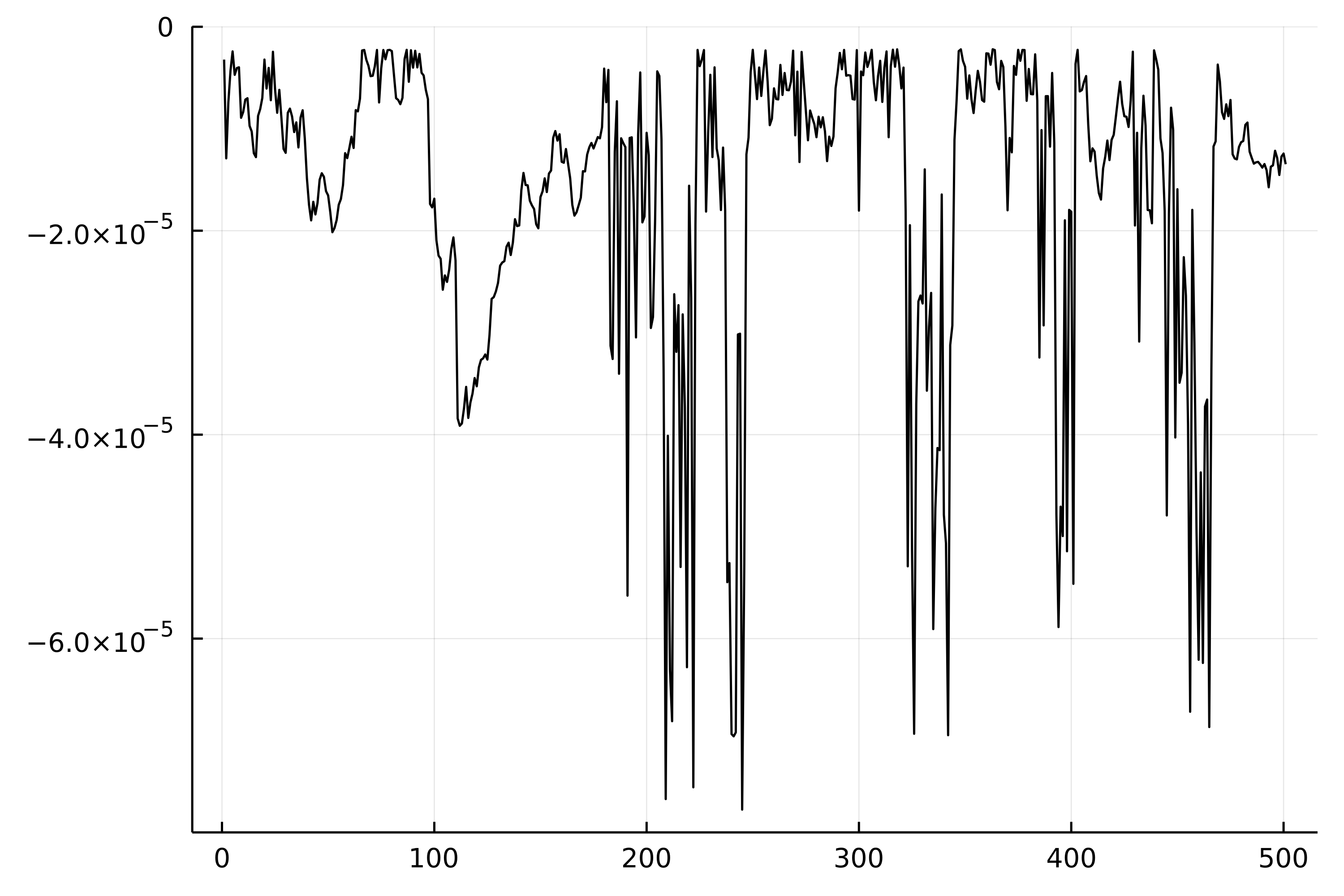}
                \caption{$\phi(u,\tau) - \phi(spl(O))$}
            \end{subfigure}
        \label{fig:illustration}
        \medskip
        \begin{minipage}{\textwidth}\scriptsize
            Note: This figure plots the impact of the three types of measurement errors on the option-implied CCF. 
            The figures illustrate the approximation for the real part of the CCF with $\tau=10,\ u=20$.
            The same Monte Carlo simulation setup as described in Section~\ref{sec:Simulation} is used here to simulate data from the SVCDEJ model.
        \end{minipage}
    \end{figure}

As discussed in Section~\ref{sec:Estimation-ss}, we replicate the CCF `payoff' using a Riemann sum approximation, and employ the interpolation-extrapolation scheme detailed in the previous subsection, applied to the set of observable option prices, to reduce the discretization and truncation errors. 
Figure~\ref{fig:illustration} illustrates the impact of the different measurement errors on the option-implied CCF. 
For the illustration, we simulate option prices from the SVCDEJ model using a similar setup as described in Section~\ref{sec:Simulation}. 
In particular, at each time point we have a discrete set of strikes and additive observation errors in the observed option prices. 
We fix the time-to-maturity to $\tau=10$ days, take $u=20$ and focus only on the real part of the CCF. 
These values are chosen to emphasize the impact of the measurement errors. 
The impact of the discretization errors, for instance, is typically smaller for larger maturities and smaller argument values.

Panel~(a) of Figure~\ref{fig:illustration} plots the measurement errors when we use a finite set of observed option prices $\widehat{O}_t(\tau,m)$. 
That is, it shows the full measurement errors $\zeta^{\phi}_t(u, \tau)$, given by the sum of the observation errors $\zeta^{(1)}_t(u, \tau)$, the truncation errors $\zeta^{(2)}_t(u, \tau)$ and the discretization errors $\zeta^{(3)}_t(u, \tau)$, formally defined in Appendix~\ref{sec:Appendix-Proofs}. 
We observe that the errors are not exactly centered at zero, implying a small bias in the CCF approximation. From panel (b), which eliminates the impact of the observation errors $\zeta^{(1)}_t(u, \tau)$ (by using a finite set of \textit{true} option prices $O_t(\tau,m)$), we observe the same small non-zero mean in $\zeta^{(2)}_t(u, \tau) + \zeta^{(3)}_t(u, \tau)$. We overlay this plot with the ATM BSIV to illustrate that the sum of the truncation and discretization errors is strongly negatively correlated with the implied volatility, and hence driven by the volatility dynamics.

For panels~(c) and~(d), we use a cubic spline interpolation and extrapolate option prices outside of the observed range of strikes, as described in Appendix~\ref{sec:Appendix-Extrapolation}. 
Panel~(d) plots the errors in the CCF approximation when we use the true finite set of option prices.
As we can see, the discretization and truncation errors are largely reduced by the interpolation-extrapolation scheme (note the scale of the  vertical axis). 
Finally, panel~(c) illustrates the errors in the CCF approximation when we apply the same interpolation and extrapolation to option prices observed with error. 
We observe that the observation errors $\zeta^{(1)}_t(u, \tau)$ (which dominate both panels~(a) and~(c)) are largely unaffected by the interpolation-extrapolation scheme, but the (already small) bias from the impact of the discretization and truncation errors has been further reduced, leading to errors that are virtually centered around zero.

\section{Additional Simulation and Empirical Results}
\label{sec:Appendix-SimulationEmpirical}

In this appendix, we first provide additional simulation results for two related alternative option pricing models, to supplement Section~\ref{sec:Simulation}. 
We also consider a model specification that includes a variance risk premium. 
Next, we provide some additional empirical results to analyze the robustness of our empirical findings reported in Section~\ref{sec:Empirics}. 

\subsection{Additional simulation results}
\subsubsection{SVCJ}

We additionally illustrate the developed estimation procedure using the `double-jump' stochastic volatility model of \citeA{DPS2000} with a Gaussian jump size distribution. 
In particular, we assume the following process, referred to in shorthand as `SVCJ', for the log forward price under both the $\P$ and $\Q$ probability measures:
    \begin{align}
        \label{SVCJ-y}
        \diff \log F_t &= ( -\tfrac{1}{2}v_t - \mu\lambda_t) \dt + \sqrt{v_t} \diff W_{1,t} + J_t \diff N_t,\\
        \label{SVCJ-v}
        \diff v_t &= \kappa(\bar{v} - v_t) \dt + \sigma \sqrt{v_t} \diff W_{2,t} + J_t^v \diff N_t,
    \end{align}
where nearly all ingredients are the same as in the SVCDEJ specification in Section~\ref{sec:Simulation-svcdej}, except for the distribution of the jump sizes. 
In particular, we assume here that the jump sizes in returns are Gaussian, $J \sim \mathcal{N}(\mu_J, \sigma_J^2)$, and the jump sizes in volatility are independent from the jump sizes in returns with $J^v \sim \exp(1/\mu_v)$.

Similar to the specification in the main text, this model belongs to the AJD class and the log of the option-spanned CCF is linear in the latent state process $v_t$.
The conditional mean and variance of the latent stochastic volatility process are given by
    \begin{align}\label{mean_svcj}
        \E[v_{t+1}| \F_t] &= e^{g_1\Delta t} v_t + \frac{g_0}{g_1}\left(e^{g_1\Delta t} - 1 \right),\\
        \label{var_svcj}
        \mbox{Var}(v_{t+1} | \F_t) &= - \frac{\sigma^2 + 2 \delta \mu_v^2}{2g_1^2}  \left[ 2 g_1\left(e^{g_1\Delta t} - e^{2g_1\Delta t} \right) v_t - g_0\left( 1 - e^{g_1\Delta t} \right)^2\right],
    \end{align}
    with $g_0 = \kappa \bar{v}$ and $g_1 = -\kappa + \delta \mu_v$. 
Equations~\eqref{mean_svcj} and~\eqref{var_svcj} are used to define the state updating equation:
    \begin{align}\label{ss_svcj}
        v_{t+1} = c_t + T_t v_t + \eta_{t+1},
    \end{align}
    where $c_t =\frac{g_0}{g_1}\left(e^{g_1\Delta t} - 1 \right),\ T_t = e^{g_1\Delta t}$ and $\mbox{Var}(\eta_{t+1}| \F_t) =\mbox{Var}(v_{t+1}|\F_t)$. 
We also impose the Feller condition $2 \kappa \bar{v} > \sigma^2$ and the covariance stationarity condition $\kappa > \delta \mu_v$.

    \begin{table}[h]
        \centering
        \scriptsize
        \caption{Monte Carlo results for the SVCJ model}
          \begin{tabular}{lccccccccc}
            \toprule
            parameter & $\sigma$ & $\kappa$ & $\bar{v}$ & $\rho$ & $\delta$ & $\mu_J$ & $\sigma_J$ & $\mu_v$ & $\sigma_{\varkappa}$ \\ [1ex]
            \multicolumn{10}{c}{$u=1,\ldots,15$} \\
            \midrule
            true value & 0.400 & 5.000 & 0.02  & -0.95 & 20.000 & -0.100 & 0.04  & 0.05  & 0.02 \\
            mean  & 0.410 & 4.869 & 0.0207 & -0.9382 & 17.013 & -0.110 & 0.0343 & 0.0520 & 0.0215 \\
            std dev   & 0.012 & 0.115 & 0.0007 & 0.0151 & 3.171 & 0.014 & 0.0124 & 0.0026 & 0.0044 \\
            q10   & 0.400 & 4.681 & 0.0201 & -0.9516 & 11.209 & -0.136 & 0.0100 & 0.0501 & 0.0166 \\
            q50   & 0.405 & 4.905 & 0.0204 & -0.9445 & 18.537 & -0.103 & 0.0404 & 0.0507 & 0.0206 \\
            q90   & 0.433 & 4.988 & 0.0219 & -0.9114 & 19.573 & -0.100 & 0.0431 & 0.0570 & 0.0274 \\ [1ex]
            \multicolumn{10}{c}{$u=1,\ldots,20$} \\
            \midrule
            true value & 0.400 & 5.000 & 0.02  & -0.95 & 20.000 & -0.100 & 0.04  & 0.05  & 0.02 \\
            mean  & 0.403 & 4.913 & 0.0202 & -0.9444 & 18.866 & -0.103 & 0.0397 & 0.0502 & 0.0198 \\
            std dev  & 0.006 & 0.091 & 0.0003 & 0.0085 & 1.420 & 0.006 & 0.0050 & 0.0010 & 0.0062 \\
            q10   & 0.396 & 4.830 & 0.0199 & -0.9524 & 17.903 & -0.105 & 0.0375 & 0.0496 & 0.0156 \\
            q50   & 0.403 & 4.930 & 0.0202 & -0.9456 & 19.048 & -0.102 & 0.0408 & 0.0501 & 0.0185 \\
            q90   & 0.410 & 5.000 & 0.0206 & -0.9373 & 20.088 & -0.100 & 0.0426 & 0.0506 & 0.0234 \\ [1ex]
            \multicolumn{10}{c}{$u=1,\ldots,25$} \\
            \midrule
            true value & 0.400 & 5.000 & 0.02  & -0.95 & 20.000 & -0.100 & 0.04  & 0.05  & 0.02 \\
            mean  & 0.395 & 4.907 & 0.0200 & -0.9555 & 20.424 & -0.097 & 0.0435 & 0.0495 & 0.0207 \\
            std dev  & 0.009 & 0.129 & 0.0003 & 0.0146 & 1.297 & 0.004 & 0.0029 & 0.0005 & 0.0097 \\
            q10   & 0.384 & 4.776 & 0.0196 & -0.9710 & 19.209 & -0.101 & 0.0405 & 0.0489 & 0.0152 \\
            q50   & 0.395 & 4.934 & 0.0200 & -0.9564 & 20.392 & -0.097 & 0.0436 & 0.0495 & 0.0178 \\
            q90   & 0.404 & 5.010 & 0.0203 & -0.9409 & 22.029 & -0.093 & 0.0473 & 0.0500 & 0.0284 \\
          \bottomrule
          \end{tabular}%
        \label{tab:svcj_mc}%

        \medskip
        \begin{minipage}{0.9\textwidth}\scriptsize
            Note: This table provides Monte Carlo simulation results for the SVCJ model, based on 500 replications. 
            Three settings with different ranges of the argument $u$ are considered. 
            Each panel lists, for each parameter,
            the true value, the Monte Carlo mean and standard deviation, and
            the 10th, 50th and 90th Monte Carlo percentiles, respectively.
            We use $T=500$ time points with $\Delta t =1/250$.
            The initial values are set to $F_0 = 100$ and $v_0=0.02$.
            The threshold for singular values is set to $\bar{s} = 10^{-7}$.
        \end{minipage}
    \end{table}%

We use the same simulation setting as in Section~\ref{sec:Simulation-svcdej}, \textit{mutatis mutandis}. 
The simulation results are provided in Table~\ref{tab:svcj_mc}. 
Just like for the SVCDEJ model specification of Section~\ref{sec:Simulation-svcdej}, the results for the SVCJ model also display high-quality finite-sample properties. 
We also note that the `double-jump' specification includes other widely used option pricing models as special cases, such as the stochastic volatility model of \citeA{heston1993}.

\subsubsection{SVCJ with a variance risk premium}
\label{sec:svcej_vrp}
    
Since the transition equation in the state space representation reflects the $\P$-dynamics of the latent components, it is, in principle, possible to conduct inference on the risk premia associated with this latent process. 
In this subsection, we provide Monte Carlo simulation results for the SVCJ model with a variance risk premium (VRP). 
(To facilitate identification, we focus on the slightly more parsimonious SVCJ model rather than the SVCDEJ model; it will turn out that already in the more parsimonious model, the VRP is weakly identified.) 
In particular, we model the VRP $\pi_v$ as the difference between the mean-reversion parameters under the $\P$ and $\Q$ measures, that is, in the state transition equation~\eqref{ss_svcj} we change the mean-reversion parameter to $\kappa^{\P}= \kappa + \pi_v$.
    
    \begin{table}[h]
        \centering
        \scriptsize
        \caption{Monte Carlo results for the SVCJ model with variance risk premium}
          \begin{tabular}{lcccccccccc}
            \toprule
            parameter & $\sigma$ & $\kappa$ & $\bar{v}$ & $\rho$ & $\delta$ & $\mu_J$ & $\sigma_J$ & $\mu_v$ & $\sigma_{\varkappa}$ & $\pi_v$ \\ [1ex]
            \multicolumn{11}{c}{constrained, $\pi_v = 1.0$} \\
            \midrule
    true value & 0.400 & 5.000 & 0.0200 & -0.950 & 20.000 & -0.100 & 0.0400 & 0.0500 & 0.0200 &  1.000\\
    mean  & 0.402 & 4.924 & 0.0202 & -0.946 & 18.882 & -0.103 & 0.0396 & 0.0503 & 0.0158 &  -\\
    std dev  & 0.007 & 0.073 & 0.0003 & 0.008 & 1.515 & 0.006 & 0.0051 & 0.0010 & 0.0031 &  -\\
    q10   & 0.396 & 4.841 & 0.0199 & -0.953 & 18.012 & -0.105 & 0.0374 & 0.0496 & 0.0127 & - \\
    q50   & 0.401 & 4.931 & 0.0202 & -0.947 & 19.136 & -0.101 & 0.0408 & 0.0501 & 0.0155 & - \\
    q90   & 0.409 & 5.003 & 0.0205 & -0.939 & 20.182 & -0.099 & 0.0427 & 0.0508 & 0.0189 & - \\ [1ex]
            \multicolumn{11}{c}{constrained, $\pi_v = 0.0$} \\
            \midrule
            true value & 0.400 & 5.000 & 0.0200 & -0.950 & 20.000 & -0.100 & 0.0400 & 0.0500 & 0.0200 &  1.000\\
    mean  & 0.402 & 4.924 & 0.0202 & -0.946 & 18.882 & -0.103 & 0.0396 & 0.0503 & 0.0158 & - \\
    std dev  & 0.007 & 0.073 & 0.0003 & 0.008 & 1.515 & 0.006 & 0.0051 & 0.0010 & 0.0031 & - \\
    q10   & 0.396 & 4.841 & 0.0199 & -0.953 & 18.012 & -0.105 & 0.0374 & 0.0496 & 0.0127 & - \\
    q50   & 0.401 & 4.931 & 0.0202 & -0.947 & 19.136 & -0.101 & 0.0408 & 0.0501 & 0.0155 & - \\
    q90   & 0.409 & 5.003 & 0.0205 & -0.939 & 20.182 & -0.099 & 0.0427 & 0.0508 & 0.0189 & - \\ [1ex]
            \multicolumn{11}{c}{unconstrained} \\
            \midrule
    true value & 0.400 & 5.000 & 0.0200 & -0.950 & 20.000 & -0.100 & 0.0400 & 0.0500 & 0.0200 & 1.000 \\
    mean  & 0.402 & 4.923 & 0.0202 & -0.945 & 18.865 & -0.103 & 0.0395 & 0.0503 & 0.0158 & 4.029 \\
    std dev  & 0.007 & 0.074 & 0.0003 & 0.008 & 1.546 & 0.006 & 0.0052 & 0.0011 & 0.0031 & 4.942 \\
    q10   & 0.396 & 4.835 & 0.0199 & -0.953 & 17.985 & -0.105 & 0.0373 & 0.0496 & 0.0127 & -4.338 \\
    q50   & 0.401 & 4.931 & 0.0202 & -0.947 & 19.136 & -0.101 & 0.0408 & 0.0502 & 0.0155 & 4.558 \\
    q90   & 0.409 & 5.003 & 0.0206 & -0.939 & 20.182 & -0.099 & 0.0427 & 0.0508 & 0.0189 & 9.883 \\
          \bottomrule
          \end{tabular}%
        \label{tab:svcj_vrp_mc}%

        \medskip
        \begin{minipage}{0.9\textwidth}\scriptsize
            Note: This table provides Monte Carlo simulation results for the SVCJ model with a variance risk premium, based on 500 replications. Each panel lists, for each parameter,
            the true value, the Monte Carlo mean and standard deviation, and
            the 10th, 50th and 90th Monte Carlo percentiles, respectively. 
            We use $T=500$ time points with $\Delta t =1/250$.
            The range of arguments  is set to $u=1,\ldots,20$ and the threshold to $\bar{s} = 10^{-7}$. 
            The initial values are set to $F_0 = 100$ and $v_0=0.02$.
            
        \end{minipage}
    \end{table}%
    
Table~\ref{tab:svcj_vrp_mc} provides Monte Carlo simulation results for the SVCJ model with a VRP. 
We consider three estimation strategies. 
First, we fix the VRP parameter to its true value $\pi_v = 1$. 
Second, we assume no VRP when estimating the model, although the true model is simulated with a non-zero VRP, that is, we fix $\pi_v = 0$ in the estimation procedure. 
Finally, we estimate the VRP along with all model parameters. 
    
As the results suggest, it is hard to identify the VRP in this setting (see the third panel in Table~\ref{tab:svcj_vrp_mc}). 
It appears that the $\Q$-information in option prices largely dominates the $\P$-information, making the identification of the VRP relatively weak.
A similar issue arises in the term structure literature, where calibrated bond prices often imply unrealistic $\P$-dynamics (see, e.g., the discussion in \citeNP{kim2012term}). 
However, we also notice that under all three estimation strategies, the identification of the $\Q$-parameters barely changes. 
That is, even in the misspecified model with the VRP parameter fixed to zero, the parameter estimates display good finite-sample properties (see the second panel in Table~\ref{tab:svcj_vrp_mc}). 
Consistent with this, we also find (in results not provided here) that if we were to introduce a VRP parameter in the SVCDEJ model in the empirical application of Section~\ref{sec:Empirics} (which, supported by the Monte Carlo results, we do not), it would not have a significant effect on the estimates of the model's $\Q$-parameters. 
For more focused VRP estimation, one can use, e.g., a non-parametric approach based on high-frequency data, as in \citeA{BT2011tails} and \citeA{andersen2015risk}.
    
\subsubsection{SVCEJ}
\label{sec:svcej}

Instead of the double-exponential jump size distribution considered in Section~\ref{sec:Simulation-svcdej}, or the Gaussian distribution considered above, one may also consider separate exponential distributions for positive and negative jumps. 
Following \citeA{bardgett2019inferring}, we consider two separate counting processes $N_t^{-}$ and $N_t^{+}$ for negative and positive jumps, respectively, 
and modify the SVCJ specification to obtain the `SVCEJ' model as follows:
\begin{align}
    \label{SVCEJ-y}
    \diff \log F_t &= ( -\frac{1}{2}v_t - \mu^{-} \lambda_t^{+} - \mu^{-} \lambda_t^{+}) \dt + \sqrt{v_t} \diff W_{1,t} + J_t^{-} \diff N_t^{-} + J_t^{+} \diff N_t^{+},\\
    \label{SVCEJ-v}
    \diff v_t &= \kappa(\bar{v} - v_t) \dt + \sigma \sqrt{v_t} \diff W_{2,t} + J_t^v \diff N_t^{-},
\end{align}
where $\lambda_t^{-}$ and $\lambda_t^{+}$ are the corresponding jump intensities for negative and positive jumps, and $-J_t^{-}$ and $J_t^{+}$ are exponentially distributed negative and positive jump sizes in log returns with means $\eta^{-}$ and $\eta^{+}$, respectively. 
Note that the negative jump sizes have negative support, that is, $J_t^{-}$ is negative exponential. 
Given the jump size distributions, the expected relative jump sizes in returns are $\mu^{-} = \E[e^{J^{-}} {-} 1] =  -\eta^{-}/(1+\eta^{-})$ and $\mu^{+} = \E[e^{J^{+}} {-} 1] =  \eta^{+}/(1-\eta^{+})$.
We further let the intensities be affine functions of the stochastic volatility, that is, $\lambda_t^{-} = \delta_0^{-} + \delta_1^{-} v_t$ and $\lambda_t^{+} = \delta_0^{+} + \delta_1^{+} v_t$. 
However, to keep a moderate number of parameters, we set $\delta_1^+ =0$ and $\delta_0^-=0$. 

\begin{table}[h]
    \centering
    \scriptsize
    \caption{Monte Carlo results for the SVCEJ model}
      \begin{tabular}{lcccccccccc}
        \toprule
        parameter & $\sigma$ & $\kappa$ & $\bar{v}$ & $\rho$ & $\delta_0^+$ & $\delta_1^-$ & $\eta^+$ & $\eta^-$ & $\mu_v$ & $\sigma_{\varkappa}$ \\ [1ex]
        \multicolumn{11}{c}{$u=1,\ldots,15$} \\
        \midrule
        true value & 0.450 & 8.000 & 0.015 & -0.95 & 2.000 & 100.000 & 0.01  & 0.05  & 0.05  & 0.02 \\
        mean  & 0.483 & 8.063 & 0.0159 & -0.9242 & 0.512 & 99.475 & 0.0355 & 0.0527 & 0.0529 & 0.0626 \\
        std dev  & 0.069 & 1.506 & 0.0026 & 0.0796 & 1.690 & 21.104 & 0.0211 & 0.0196 & 0.0351 & 0.1820 \\
        q10   & 0.459 & 7.642 & 0.0150 & -0.9577 & 0.028 & 90.701 & 0.0191 & 0.0485 & 0.0473 & 0.0170 \\
        q50   & 0.486 & 7.914 & 0.0157 & -0.9261 & 0.125 & 99.718 & 0.0320 & 0.0495 & 0.0482 & 0.0229 \\
        q90   & 0.511 & 8.306 & 0.0168 & -0.9049 & 0.578 & 106.121 & 0.0524 & 0.0505 & 0.0504 & 0.0302 \\ [1ex]
        \multicolumn{11}{c}{$u=1,\ldots,20$} \\
        \midrule
        true value & 0.450 & 8.000 & 0.015 & -0.95 & 2.000 & 100.000 & 0.01  & 0.05  & 0.05  & 0.02 \\
        mean  & 0.462 & 8.178 & 0.0149 & -0.9551 & 1.309 & 110.631 & 0.0166 & 0.0510 & 0.0464 & 0.0482 \\
        std dev  & 0.042 & 0.804 & 0.0020 & 0.0266 & 1.566 & 18.690 & 0.0130 & 0.0156 & 0.0056 & 0.1425 \\
        q10   & 0.443 & 7.849 & 0.0144 & -0.9851 & 0.531 & 101.950 & 0.0121 & 0.0478 & 0.0452 & 0.0143 \\
        q50   & 0.454 & 8.099 & 0.0148 & -0.9588 & 0.964 & 107.556 & 0.0155 & 0.0489 & 0.0468 & 0.0184 \\
        q90   & 0.488 & 8.410 & 0.0154 & -0.9165 & 1.742 & 115.366 & 0.0190 & 0.0498 & 0.0484 & 0.0368 \\ [1ex]
        \multicolumn{11}{c}{$u=1,\ldots,25$} \\
        \midrule
        true value & 0.450 & 8.000 & 0.015 & -0.95 & 2.000 & 100.000 & 0.01  & 0.05  & 0.05  & 0.02 \\
        mean  & 0.466 & 7.966 & 0.0154 & -0.9417 & 1.526 & 104.783 & 0.0157 & 0.0515 & 0.0479 & 0.0381 \\
        std dev  & 0.048 & 0.615 & 0.0022 & 0.0321 & 1.322 & 17.732 & 0.0203 & 0.0140 & 0.0114 & 0.1101 \\
        q10   & 0.442 & 7.671 & 0.0146 & -0.9726 & 0.602 & 97.221 & 0.0106 & 0.0488 & 0.0462 & 0.0129 \\
        q50   & 0.454 & 7.963 & 0.0150 & -0.9484 & 1.394 & 103.412 & 0.0126 & 0.0496 & 0.0476 & 0.0152 \\
        q90   & 0.494 & 8.171 & 0.0159 & -0.8935 & 2.130 & 109.523 & 0.0167 & 0.0505 & 0.0492 & 0.0417 \\
      \bottomrule
      \end{tabular}%
      \label{tab:svcej_mc}%

      \medskip
      \begin{minipage}{0.9\textwidth}\scriptsize
        Note: This table provides Monte Carlo simulation results for the SVCEJ model, based on 300 replications. 
        Three settings with different ranges of the argument $u$ are considered. 
        Each panel lists, for each parameter, the true value, the Monte Carlo mean and standard deviation, and the 10th, 50th and 90th Monte Carlo percentiles, respectively.
        We use $T=500$ time points with $\Delta t =1/250$.
        The initial values are set to $F_0 = 100$ and $v_0=0.015$.
        The threshold for singular values is set to $\bar{s} = 10^{-7}$.
      \end{minipage}
\end{table}%

This specification is somewhat richer than the SVCDEJ considered in Section~\ref{sec:Simulation-svcdej} since  positive jumps are modeled by a separate counting process with its own jump intensity process $\lambda_t^+$. 
Nevertheless, this specification also belongs to the AJD class and the CCF of log forward prices has a semi-closed form. 
The state updating equation is defined in a similar way as for the other specifications.

The Monte Carlo simulation results for the SVCEJ model are provided in Table~\ref{tab:svcej_mc}.
We notice that most of the parameters exhibit good finite-sample performance. 
However, the parameters related to the positive jumps are biased and have a large standard deviation.

\subsection{Additional empirical results}\label{sec:Appendix-Empirical}

Table~\ref{tab:svcdej2} provides additional empirical results for the model specification of Section~\ref{sec:Empirics}. 
Next to the empirical results with fixed $p^{-}=0.7$ reported in Section~\ref{sec:Empirics}, we provide the estimates for an unrestricted probability of negative jumps and for different fixed values $p^{-} = 0.65$ and $p^{-} = 0.75$. 
Overall, the results indicate similar parameter estimates as in Table~\ref{tab:svcdej_res}, which is reassuring for the robustness of our empirical results. 
We also note larger standard errors of the parameter estimates in the unrestricted model, specifically for the parameter $\delta$, which enters the model as a multiple of $p^-$. 
This is in line with our simulation results for the unrestricted model (not provided here), which show the limits to identification in the considered unrestricted model. 
Therefore, in the empirical application in the main text, we focus on the restricted model.

    \begin{table}[h!]
        \centering
        \caption{SVCDEJ estimation results}
        \footnotesize
        \begin{tabular}{lrrrrrrrrrr}
            \toprule
            & $\sigma$ & $\kappa$ & $\bar{v}$ & $\rho$ &  $\delta$ & $p^-$ & $\eta^+$ & $\eta^-$ & $\mu_v$ & $\sigma_{\varkappa}$ \\
            \midrule
            \multicolumn{11}{c}{unconstrained} \\
            $\widehat\theta$ & 0.505 & 8.368 & 0.0152 & -1.000 & 167.68 & 0.6619 & 0.0195 & 0.0424 & 0.0516 & 0.253 \\
            s.e.  & 0.071 & 0.762 & 0.0018 & 0.054 & 17.04 & 0.0155 & 0.0014 & 0.0007 & 0.0039 & 0.004 \\
            \midrule
            \multicolumn{11}{c}{constrained, $p^- = 0.75$} \\
            $\widehat\theta$  & 0.503 & 8.259 & 0.0153 & -1.000 & 148.01 & 0.75  & 0.0218 & 0.0422 & 0.0517 & 0.253 \\
            s.e.  & 0.006 & 0.091 & 0.0004 & 0.011 & 2.66  &       & 0.0005 & 0.0006 & 0.0004 & 0.004 \\
            \midrule
            \multicolumn{11}{c}{constrained, $p^- = 0.65$} \\
            $\widehat\theta$  & 0.506 & 8.007 & 0.0160 & -1.000 & 162.09 & 0.65  & 0.0196 & 0.0432 & 0.0523 & 0.253 \\
            s.e.  & 0.006 & 0.178 & 0.0003 & 0.021 & 1.36  &       & 0.0003 & 0.0006 & 0.0009 & 0.004 \\
            \bottomrule
          \end{tabular}%
          \label{tab:svcdej2}%

          \medskip
          \begin{minipage}{0.8\textwidth}\scriptsize
              Note: This table provides the parameter estimates and standard errors for the SVCDEJ model. 
              The model is estimated based on $u=1,\dots,20$ and $\bar{s} = 10^{-7}$.
          \end{minipage}
    \end{table}%
    
Table~\ref{tab:svcej_res} provides empirical results for the alternative model specification SVCEJ detailed in Subsection~\ref{sec:svcej}, without and with external state variables. 
Positive and negative jumps are modeled by separate counting processes with their own jump intensities $\lambda_t^+$ and $\lambda_t^-$, possibly depending on the external state variable with coefficients $\gamma^+$ and $\gamma^-$, respectively. 
We observe a similar magnitude as in Table~\ref{tab:svcdej_res} (and Table~\ref{tab:svcdej2}) for most of the parameter estimates, corroborating again the robustness of our empirical results.

    \begin{table}
      \centering
      \caption{SVCEJ estimation results}
      \scriptsize
        \begin{tabular}{lccccccccccccc}
          \toprule
            & $\sigma$ & $\kappa$ & $\bar{v}$ & $\rho$ & $\delta_0^+$ & $\delta_1^-$ & $\eta^+$ & $\eta^-$ & $\mu_v$ & $\gamma^+$ & $\gamma^-$ & $q$ & $\sigma_{\varkappa}$ \\ 
          \midrule
            \multicolumn{14}{c}{no external factors} \\
          $\widehat{\theta}$ &    0.481 & 8.31 & 0.0139 & -1.00 & 3.76 & 107.3 & 0.0100 & 0.0445 & 0.061 & - & - & - & 0.221 \\
          s.e.    & 0.007 & 0.26 & 0.0002 & 0.01 & 0.08 & 3.99 & 0.0002 & 0.0006 & 0.002 & & & & 0.004 \\
          \midrule
            \multicolumn{14}{c}{$R_0$} \\
            $\widehat{\theta}$ &   0.547 & 11.28 & 0.0133 & -1.00 & 0.999 & 83.11 & 0.0150 & 0.0439 & 0.079 & 0.036 & 1.587 & 0.016 & 0.213 \\
            s.e.    &     0.005 & 0.25 & 0.0003 & 0.01 & 0.045 & 2.97  & 0.0004 & 0.0006 & 0.002 & 0.300 & 0.142 & 0.015 & 0.004 \\
         \bottomrule
        \end{tabular}%
      \label{tab:svcej_res}
      
          \medskip
          \begin{minipage}{\textwidth}\scriptsize
              Note: This table provides the parameter estimates and standard errors for the SVCEJ model. 
              The model is estimated based on $u=1,\dots,20$ and $\bar{s} = 10^{-7}$.
          \end{minipage}
    \end{table}%

\noindent  

    \end{appendices}

    \clearpage
    \bibliography{lit}
    \bibliographystyle{apacite}

\end{document}